\documentstyle{report}
\catcode`@=11
\def\thechapterhead{\arabic{chapter}}
\def\thesectionhead{\thechapterhead.\arabic{section}}

\def\@sect#1#2#3#4#5#6[#7]#8{\ifnum #2>\c@secnumdepth
     \def\@svsec{}\else
     \refstepcounter{#1}\edef\@svsec{\csname the#1head\endcsname\hskip 1em }\fi
     \@tempskipa #5\relax
      \ifdim \@tempskipa>\z@
        \begingroup #6\relax
          \@hangfrom{\hskip #3\relax\@svsec}{\interlinepenalty \@M #8\par}
        \endgroup
       \csname #1mark\endcsname{#7}\addcontentsline
         {toc}{#1}{\ifnum #2>\c@secnumdepth \else
                      \protect\numberline{\csname the#1head\endcsname}\fi
                    #7}\else
        \def\@svsechd{#6\hskip #3\@svsec #8\csname #1mark\endcsname
                      {#7}\addcontentsline
                           {toc}{#1}{\ifnum #2>\c@secnumdepth \else
                             \protect\numberline{\csname
the#1head\endcsname}\fi
                       #7}}\fi
     \@xsect{#5}}

\def\appendixname{Appendix}     
\def\appendix{\par
 \setcounter{chapter}{0}
 \setcounter{section}{0}
 \def\@chapapp{\appendixname} 
 \def\thechapter{\Alph{chapter}}
 \def\thechapterhead{\Alph{chapter}}
 \def\thesection{\thechapter.\arabic{section}}
 \def\thesectionhead{\thechapterhead.\arabic{section}}}
\def\@makechapterhead#1{ {\setbox0=\hbox{#1} \parindent 0pt
 \begin{raggedright}
 \Large \bf
 \ifnum \c@secnumdepth >\m@ne \ifdim \wd0=0pt \@chapapp{} \thechapterhead
   \else \thechapterhead \hspace{1ex} \ #1\fi \end{raggedright}\par
 \fi \nobreak \vskip 2ex }}
\def\@makeschapterhead#1{ { \parindent 0pt \begin{raggedright}
 \Large \bf  #1 \end{raggedright}\par
 \nobreak \vskip 2ex } }
\def\thebibliography#1{\chapter*{\bibname\@mkboth
 {\uppercase{\bibname}}{\uppercase{\bibname}}}\list
 {[\arabic{enumi}]}{\settowidth\labelwidth{[#1]}\leftmargin\labelwidth
 \advance\leftmargin\labelsep
 \usecounter{enumi}}
 \def\newblock{\hskip .11em plus .33em minus -.07em}
 \sloppy
 \sfcode`\.=1000\relax}
\def\bibname{References}
\def\chapter{\par \vspace{3.5ex} \thispagestyle{plain} \global\@topnum\z@
\@afterindentfalse \secdef\@chapter\@schapter}
\def\section{\@startsection {section}{1}{\z@}{-2.0ex plus -1ex minus
 -.2ex}{1.0ex plus .2ex}{\Large\bf}}
\def\subsection{\@startsection{subsection}{2}{\z@}{-2.0ex plus -1ex minus
 -.2ex}{0.5ex plus .2ex}{\large\bf}}
\def\citenum{\@ifnextchar [{\@tempswatrue\@zcitex}{\@tempswafalse\@zcitex[]}}
\def\@zcitex[#1]#2{\if@filesw\immediate\write\@auxout{\string\citation{#2}}\fi
  \def\@citea{}\@zcite{\@for\@citeb:=#2\do
    {\@citea\def\@citea{,\penalty\@m\ }\@ifundefined
       {b@\@citeb}{{\bf ?}\@warning
       {Citation `\@citeb' on page \thepage \space undefined}}%
\hbox{\csname b@\@citeb\endcsname}}}{#1}}
\def\@zcite#1#2{{#1\if@tempswa , #2\fi}}

\def\citevb{\@ifnextchar [{\@tempswatrue\@citexvb}{\@tempswafalse\@citexvb[]}}
\def\@citexvb[#1]#2{\if@filesw\immediate\write\@auxout{\string\citation{#2}}\fi
  \def\@citea{}\@cite{\@for\@citeb:=#2\do
    {\@citea\def\@citea{--\penalty\@m}\@ifundefined
       {b@\@citeb}{{\bf ? \@citeb}\@warning
       {Citation `\@citeb' on page \thepage \space undefined}}%
\hbox{\csname b@\@citeb\endcsname}}}{#1}}
\newbox\uphookbox
\setbox\uphookbox=\hbox{\vrule width 1.2pt depth 3.0pt height 1.18pt}
\dp\uphookbox=0\p@\def\uphook{\copy\uphookbox}

\def\uphookbracefill#1#2{%
\setbox1=\hbox{$\m@th\displaystyle#1$\hskip-1.2pt}%
\setbox2=\hbox{$\m@th\displaystyle#2$\hskip-1.2pt}%
$\m@th\hskip0.5\wd1\uphook\leaders\vrule\hfill\uphook\hskip0.5\wd2$}
\def\uphookbrace#1#2#3{\mathop
   {\vbox
      {\ialign{##\crcr
         \noalign{\kern3\p@}%
         \uphookbracefill{#1}{#3}\crcr
         \noalign{\kern3\p@\nointerlineskip}%
         $\hfil\displaystyle{#1#2#3}\hfil$\crcr}
      }%
   }%
\limits}
\def\sqr#1#2{{\vcenter{\hrule height.#2pt
\hbox{\vrule width.#2pt height#1pt \kern#1pt
\vrule width.#2pt}
\hrule height.#2pt}}}

\newif\ifseparateabstract\separateabstractfalse
\def\abstract#1{\gdef\@abstract{#1}}
\def\pacs#1{\gdef\@pacs{#1}}
\def\maketitle{\begin{titlepage}%
 \let\footnotesize\small      
 \setcounter{page}{0}%
 \null
 \vfil
 \vskip 30pt                  
 \vbox
  {\begin{center}
   {\Large\bf \@title \par}        
   \vskip 3em                  
   {
     \lineskip .75em
     \begin{tabular}[t]{c}\@author
     \end{tabular}\par}
    \vskip 1.5em               
   { \@date \par}
  \end{center} \par}\nopagebreak 
  \@thanks
 \setcounter{footnote}{0}       
 \ifseparateabstract
   \vfil\null
   \end{titlepage}
   \begin{titlepage}
   \null\vfil
 \else
   \vskip 0pt plus 0.6fil\penalty5000
 \fi
 \vbox
   {\begin{center}
     {\large\bf Abstract}       
   \end{center}\nopagebreak
   \@abstract
   \par\vskip4ex \noindent\@pacs}
 \vfil\null\end{titlepage}
\let\thanks\relax
\gdef\@thanks{}\gdef\@author{}\gdef\@title{}\let\maketitle\relax
\gdef\@abstract{}\gdef\@pacs{}}
\gdef\@thanks{}\gdef\@author{}\gdef\@title{}
\gdef\@abstract{}\gdef\@pacs{}
\catcode`@=12
\newcommand{\sml}{
\raisebox{-1.0mm}{$\mbox{}\stackrel{\textstyle <}{\sim}\mbox{}$}}

\newcommand{\tr}{\mathop{\rm tr} \nolimits}
\newcommand{\Tr}{\mathop{\rm Tr} \nolimits}

\newcommand{\trg}{\mathop{\rm trg} \nolimits}

\newcommand{\asymx}{\mathop{\sim}}
\newcommand{\asym}[1]{\mathrel{\asymx_{#1}}}

\newcommand{\half}{{\textstyle \frac{1}{2}}}

\newdimen\savebaselineskip
\savebaselineskip=\baselineskip

\newcommand{\opf}{\bf}
\newcommand{\artan}{\mathop{\rm artan} \nolimits}

\newcommand{\diag}{\mathop{\rm diag} \nolimits}

\newcommand{\Trg}{\mathop{\rm Trg}}
\renewcommand{\Tr}{\mathop{\rm Tr}}
\newcommand{\Detg}{\mathop{\rm Detg} \nolimits}
\newcommand{\detg}{\mathop{\rm Detg} \nolimits}
\newcommand{\re}{\mathop{\rm Re} \nolimits}
\newcommand{\im}{\mathop{\rm Im} \nolimits}

\newcommand{\s}{t}

\begin{document}
\separateabstracttrue
\title{Crosssover from Orthogonal to Unitary Symmetry for
       Ballistic Electron Transport in Chaotic Microstructures}
\author{{\bf Z.~Pluha\v{r}}\thanks{On leave from Charles University, Prague,
     Czech Republic}, {\bf H.A.~Weidenm\"uller} \\
     {\em Max--Planck--Institut f\"ur Kernphysik, D-69029 Heidelberg,
     Germany} \\
and \\
     {\bf J.A.~Zuk} \\
     {\em Max--Planck--Institut f\"ur Kernphysik, D-69029 Heidelberg,
     Germany} \\
     \& \\
     {\em Department of Physics, University of Manitoba,
     Winnipeg, R3T 2N2, Canada}\thanks{Permanent address} \\
and \\
     {\bf C.H.~Lewenkopf} \\
     {\em Depertment of Physics, University of Washington, Seattle,
     WA 98192, U.S.A.}}

\date{}
\pacs{PACS numbers: 72.20.My, 73.20.Fz, 05.45.$+$b, 72.20.Dp
     \hfill October 1994}
\abstract{
We study the ensemble-averaged conductance as a function of applied magnetic
field for ballistic electron transport across few-channel microstructures
constructed in the shape of classically chaotic billiards. We analyse the
results of recent experiments, which show suppression of weak localization due
to magnetic field, in the framework of random-matrix theory.
By analysing a random-matrix Hamiltonian for the billiard-lead system with the
aid of Landauer's formula and Efetov's supersymmetry technique, we derive a
universal expression for the weak-localization contribution to the mean
conductance that depends only on the number of channels and the magnetic flux.
We consequently gain a theoretical understanding of the continuous crossover
from orthogonal symmetry to unitary symmetry arising from the violation of
time-reversal invariance for generic chaotic systems.
}
\maketitle
%
\newcommand{\opfone}{\bf 1}
\chapter{Introduction}
Random-matrix theory has proven to be a very successful tool for the
understanding of localization phenomena in disordered diffusive systems,
for example, universal conductance fluctuations \cite{IWZ,MMZ},
Altshuler-Aronov-Spivak oscillations \cite{JZ,AMG} and persistent currents
\cite{AIMW}.
In a number of recent experiments, weak-localization effects have been
observed in ballistic cavities \citevb{Marcus,Keller} in which electron
scattering from the boundaries is classically chaotic.
It is interesting to also investigate the application of random-matrix theory
to such problems
as one can expect it to be able to address questions not accessible to other
methods, such as semiclassical analysis.

In this work, we look at ballistic electron transport through a microstructure
that has the shape of a classically chaotic billiard, in the presence of an
applied magnetic field $B$.
The typical set-up for a stadium is shown in Figure~1.
The stadium, whose length
$L$ is less than the mean free path $\ell$ and the dephasing length $L_\phi$,
has attached to it two
asymmetrically positioned ideal quasi-one-dimensional leads which carry the
incoming and outgoing electron transverse modes.
We assume an equal number, $M/2$, of incoming channels (labelled by index $a$)
and outgoing channels (labelled by index $b$).
Thus $M$ counts the total number of open channels coupled to the stadium.

We use this physical system to study the breaking of time-reversal symmetry,
as measured by the conductance
\mbox{$G(B) = (e^2/h)g(B)$},
as the magnetic field $B$ is turned on.
The relevant experimental observation \citevb{Marcus,Keller}
is that the average dimensionless conductance $\overline{g}$ dips at zero
magnetic field
$B=0$. The $\overline{g}$ versus $B$ curve is illustrated schematically in
Figure~2. The conductance interpolates between a minimum value, which can be
associated with the Gaussian orthogonal ensemble (GOE), at
\mbox{$B = 0$}
and an asymptotic upper value, which can be associated with the Gaussian
unitary ensemble (GUE), for large $B$. The difference $\delta g$
between the upper GUE limit and the
mean dimensionless conductance $\overline{g}$
represents the weak-localization contribution to $\overline{g}$.
The suppression of weak localization due to non-zero magnetic field, i.e.\ the
fact that
\mbox{$\delta g \to 0$} as
\mbox{$B \to \infty$},
corresponds to a GOE~$\to$~GUE transition.

The appearance of the orthogonal and unitary random-matrix ensembles
in this discussion arises from the fact that the fluctuation properties of the
quantum spectra for closed chaotic billiards are the same as those of
(i) the GOE for
\mbox{$B=0$},
and (ii) the GUE for $B$ sufficiently different from zero. The former, being an
ensemble of Gaussion distributed real-symmetric matrices, respects
time-reversal invariance, while the latter, being an ensemble of Gaussian
distributed Hermitian matrices, does not.

The presence and magnitude of the weak-localization term in
\mbox{$\overline{g}(B)$} at
\mbox{$B=0$}
can easily be understood in the framework of random-matrix theory
(without detailed calculation)
as a consequence of (i) elastic enhancement (or equivalently, coherent
back-scattering), derived in Section~5.1, and (ii) unitarity of the S-matrix.
Our starting point is the connection between the conductance $g$ and the
S-matrix given by Landauer's formula:
\begin{equation}
g = \sum_{a=1}^{M/2}\sum_{b=M/2+1}^M\left\{|S_{ab}|^2 +
     |S_{ba}|^2\right\} \;,
\label{landauer}
\end{equation}
where a factor of two for spin is effectively included in this expression.
Here, the index $a$ sums over the $M/2$ incoming channels and $b$ sums over the
$M/2$ outgoing channels, so that the summation in
Eq.~(\ref{landauer}) runs over
two non-overlapping sets of channels
\mbox{$\{a\} \neq \{b\}$}.
If we assume all channels to be equivalent, so that all mean values
\mbox{$\overline{|S_{ab}|^2}$}
are equal for
\mbox{$a \neq b$},
then we obtain
\begin{equation}
\overline{g} = 2\left(\frac{M}{2}\right)\left(\frac{M}{2}\right)
     \overline{|S_{ab}|^2} =
     \frac{M^2}{2}\overline{|S_{ab}|^2} \;.
\label{avg}
\end{equation}
Now, according to (i), we have
\begin{equation}
\overline{|S_{cc}|^2} = \kappa\overline{|S_{cc'}|^2}
     \quad\mbox{for}\quad c \neq c'
\label{elastic}
\end{equation}
with
\renewcommand{\arraystretch}{0.5}
\begin{equation}
\kappa = \left\{
\begin{array}{ll}
2 & \quad\mbox{for GOE} \\
1 & \quad\mbox{for GUE}.
\end{array}
\right.
\end{equation}
As a consequence of (ii), we have
\renewcommand{\arraystretch}{1.0}
\begin{equation}
\sum_{c,c'=1}^M\overline{|S_{cc'}|^2} = \tr{\opfone}_M = M \;.
\label{unitarity}
\end{equation}
Combination of the two equations (\ref{elastic}) and
(\ref{unitarity}) yields
\begin{eqnarray}
M & = & \sum_{c=1}^M\overline{|S_{cc}|^2} +
     \sum_{{c,c'=1 \atop c\neq c'}}^M \overline{|S_{cc'}|^2}
\nonumber \\
& = & M\kappa\overline{|S_{ab}|^2} +
     M(M-1)\overline{|S_{ab}|^2}
\end{eqnarray}
for any
\mbox{$a \neq b$}.
Therefore, we see that
\renewcommand{\arraystretch}{0.5}
\begin{equation}
\overline{|S_{ab}|^2} = \frac{1}{M-1+\kappa} = \left\{
\begin{array}{lll}
1/M     & \quad\mbox{for GUE} & \quad (\kappa=1) \\
1/(M+1) & \quad\mbox{for GOE} & \quad (\kappa=2) \;.
\end{array}
\right.
\label{wls}
\end{equation}
We can now use Eq.~(\ref{wls}) to look at the expansions of the corresponding
values of $\overline{g}$ in powers of $M^{-1}$.
For the pure GUE (large magnetic field), we obtain
\renewcommand{\arraystretch}{1.0}
\begin{equation}
\overline{g}_{\rm GUE} = M/2 \;,
\label{igue}
\end{equation}
while for the pure GOE
(\mbox{$B=0$}),
we see that
\begin{eqnarray}
\overline{g}_{\rm GOE} & = & \frac{M^2}{2(M+1)}
\nonumber \\
& = & \overline{g}_{\rm GUE} - \frac{1}{2} +
     O\left(\frac{1}{M}\right) \;.
\label{igoe}
\end{eqnarray}
The contribution $-1/2$ to the RHS of the equation above represents the
weak-localization term in the conductance, and gives the magnitude of
the gap between the upper and lower asymptotic values in Figure~2 in the limit
of large channel number $M$. Weak localization is absent for the GUE, evidently
being suppressed by the magnetic field.

One should also observe that, for general values of $M$, the weak localization
term has the value $M/(2(M+1))$.
It increases from $1/3$ for
\mbox{$M = 2$}
to $1/2$ for
\mbox{$M \rightarrow \infty$}.
Nonetheless, the weak localization correction decreases relative to
${\overline g}$ which is linear in $M$.
All this agrees with recent results obtained \cite{BM} for
the circular unitary (CUE) and circular orthogonal (COE) ensembles.
The agreement is not surprising: It has been shown
\cite{LW} that the GOE and the COE distributions for the $S$-matrix
elements coincide when
the coupling with the leads is maximized.
Similar arguments apply to the unitary case.

We now consider the question: How does a finite magnetic field
interpolate between $\overline{g}_{\rm GOE}$
(\mbox{$B=0$})
and $\overline{g}_{\rm GUE}$ ($B$ sufficiently large)?
Semiclassical analysis \cite{BJS} implies that
\mbox{$\delta g(B) = \overline{g}_{\rm GUE} - \overline{g}(B)$}
is a Lorentzian. However, this work was only able to include the diagonal
contributions (i.e.\ taking into account only interference between
symmetry-related classical paths),
and was not able to achieve quantitative agreement with fully quantal
numerical simulations.
In view of this, one should ask whether the full result remains Lorentzian.

Our aim in this investigation is to understand the generic features
of the continuous
GOE~$\to$~GUE crossover as a function of magnetic field $B$ for {\em any}
classically chaotic microstructure in the framework of random-matrix theory.
{}From the semiclassical perspective, our random-matrix approach may be viewed
as
taking into account the generic properties described by the long classical
trajectories, in which system-specific information is washed out.
Therefore, we shall need to restrict attention to billiard/lead
configurations that are dominated by long classical trajectories.
Our results will not take account of effects associated with the specific
geometry of a billiard, which would be probed by short trajectories.
Moreover, the density of long trajectories is large --- growing
exponentially with the period --- and therefore virtually impossible to
account for in the framework of semiclassical theory.

On an energy (rather than a time) scale, such generic features are
displayed on intervals measured in units of the mean level spacing $d$. The
suppression of weak localization becomes effective whenever the matrix
elements of the interaction of an electron with the magnetic field become
comparable with $d$. This is why we believe that our approach should yield
results which are relevant for the problem. There exists a time scale --- the
mixing time $\tau_{\rm mix}$, defined below ---
which relates to the strength of
these matrix elements. The suppression of weak localization will be seen to
depend actually on the competition between this time scale and a second time
scale $\tau_{\rm dec}$, the mean life-time of the levels in the billiard. The
associated width
\mbox{$\Gamma_{\rm dec} = \hbar/\tau_{\rm dec}$}
grows linearly with the number of channels. Our final
result essentially depends on the ratio of these two time scales.
%

The paper is set out as follows:
In Section~2, we define the model and discuss the values of its parameters.
Section~3 deals with relating the parameter that controls the
GOE~$\to$~GUE crossover to the strength of the external magnetic field.
In Section~4, we outline the supersymmetry formalism used to extract an
exact result for the ensemble-averaged conductance, and explicitly construct
the generating function that is central to this approach.
We also summarize the steps leading to the final analytical expression for
$\overline{g}$,
which is then displayed and discussed in Section~5.
Results of the numerical evaluation of the magnetic field and channel number
dependence of the weak-localization term are presented in Section~6, along with
a discussion of the Lorentzian approximation.
Section~7 contains some interpretation of our results, as well as a comparison
with experimental data.
Finally, we present our conclusions in Section~8. A few appendices contain
various technical details.

\chapter{The Model}
We first consider an electron moving in a closed billiard, whose
Hamiltonian $H$ we model in terms of an ensemble of random matrices.
We then take account of the presence of external channels by introducing
a coupling between $H$ and the leads. This allows us to define a scattering
matrix which we relate to the conductance via Landauer's formula, viz.\
Eq.~(\ref{landauer}).
We shall maximize the coupling between $H$
and the leads. Then, the (ensemble-averaged) conductance
$\overline{g}(\s,M)$ depends only on the strength of the magnetic field $B$
(expressed in terms of a dimensionless parameter $t$), and on the total number
$M$ of channels. Using ergodicity, we compare this quantity with experiment.
We neglect the Coulomb interaction between electrons as well as all other
dephasing effects.

For the Hamiltonian of the closed billiard, we consider an ensemble of random
matrices $H_{\mu\nu}$ of dimension $N$, given by \cite{PM2}
\begin{equation}
H_{\mu\nu} = H_{\mu\nu}^{\rm (S)} + i\sqrt{\frac{\s}{N}}
     H_{\mu\nu}^{\rm (A)} \;.
\label{billiard}
\end{equation}
The ensemble $H^{\rm (S)}$ represents a GOE with zero mean and whose
non-diagonal elements have variance $\lambda^2/N$, while $H^{\rm (A)}$ denotes
an ensemble of Gaussian distributed real antisymmetric matrices whose
non-vanishing elements have zero mean, the same variance $\lambda^2/N$ and are
uncorrelated for
\mbox{$\mu > \nu$}.
Explicitly,
\begin{eqnarray}
\overline{H_{\mu\nu}^{\rm (S)}H_{\mu'\nu'}^{(S)}} & = & \frac{\lambda^2}{N}
     (\delta_{\mu\mu'}\delta_{\nu\nu'} +
      \delta_{\mu\nu'}\delta_{\nu\mu'}) \;,
\nonumber \\
\overline{H_{\mu\nu}^{\rm (A)}H_{\mu'\nu'}^{(A)}} & = & \frac{\lambda^2}{N}
     (\delta_{\mu\mu'}\delta_{\nu\nu'} -
      \delta_{\mu\nu'}\delta_{\nu\mu'}) \;.
\label{var}
\end{eqnarray}
The parameter $\lambda$ fixes the mean level spacing $d$ of the GOE and must,
in principle, be adjusted to the local mean level spacing at the Fermi energy
of the billiard at hand.
In the end, we shall take the limit
\mbox{$N\to\infty$},
in which case the average level density of the ensemble (\ref{billiard})
will take the form of Wigner's semicircle.
We shall also normalize all energies such that the Fermi energy is zero,
\mbox{$E_{\rm F} = 0$},
in which case it sits at the centre of the Wigner semicircle. The mean level
spacing at the Fermi energy is then given by
\mbox{$d = \pi\lambda/N$}.

To estimate the scale of the GOE~$\to$~GUE crossover, we note that this
transition will take place when the typical matrix element of the time-reversal
violating perturbation becomes comparable with the mean GOE level spacing:
\begin{equation}
\sqrt{\frac{\s}{N}}{\cdot}\sqrt{\frac{\lambda^2}{N}} \sim
     \frac{\lambda}{N} \;,
\end{equation}
which implies
\mbox{$\s \sim 1$},
i.e.\ for a value of $\s$ of order unity on the scale of $N$.
This is why we chose
to include a factor $N^{-1/2}$ in front of $H^{\rm (A)}$ in
Eq.~(\ref{billiard}).

The coefficient of
\mbox{$iH^{\rm (A)}$}
gives the strength of the time-reversal violating perturbation and is expected
to be proportional to the magnetic field, since the variation of the levels of
the stadium with magnetic field is linear in the flux
\mbox{$\phi = BA$}
through the area $A$ of the stadium.
It is convenient to express
\mbox{$\sqrt{\s} = k\phi/\phi_0$},
where
\mbox{$\phi_0 = h/e$}
is the fundamental flux quantum, and $k$ is then a number of order unity.
An estimate of $k$ can be provided by computing the $\Delta_3$--statistic of a
closed stadium for various values of the applied magnetic field, and then
fitting the results using the matrix ensemble $H_{\mu\nu}$ of
Eq.~(\ref{billiard}). More details of this procedure will be given later.

Next, we must consider the coupling of the closed stadium with two ideal
external leads.
The Hamiltonian operator for the total system (comprising stadium and leads)
is given by
\begin{eqnarray}
H & = & \sum_{\mu,\nu=1}^{N}|\mu\rangle H_{\mu\nu}\langle\nu| +
     \sum_{c=1}^{M}\int d\mu(E,c)\, |E,c\rangle E \langle E,c|
\nonumber \\
& & {}+ \sum_{\mu,c}\int d\mu(E,c)\,
     \left\{|E,c\rangle W_{c\mu}(E) \langle\mu| + {\rm h.c.}\right\} \;.
\end{eqnarray}
The first term represents the stadium Hamiltonian as discussed above.
The second term describes the $M$ free electron channels in both of the leads.
Each channel $c$ is defined by a propagating transverse mode with an energy $E$
below the Fermi energy.
Thus
\mbox{$|E,c\rangle$}
is the plane wave mode of transverse energy $\epsilon_c$ and longitudinal
momentum
\mbox{$\hbar k$},
such that the total energy of the state is
\mbox{$E = \epsilon_c + \hbar^2k^2/(2m_e^*)$},
with $m_e^*$ being the effective mass of the electrons, and
\mbox{$d\mu(E,c)$}
is the corresponding energy measure:
\begin{equation}
d\mu(E,c) = \sqrt{\frac{m_e^*}{8\pi^2\hbar^2}}\Theta(E-\epsilon_c)
     \frac{dE}{\sqrt{E-\epsilon_c}} \;,
\end{equation}
where $\Theta(E)$ denotes the step function.
The final term contains the lead-stadium couplings.
We have denoted by $W_{c\mu}$ the matrix element which connects the basis state
$|\mu\rangle$ in the stadium's internal region with channel $c$, where
\mbox{$\mu = 1,2,\ldots,N$} and
\mbox{$c = 1,2,\ldots,M$}.
We assume that the breaking of time-reversal symmetry by the magnetic field $B$
need only be taken into account in the Hamiltonian (\ref{billiard})
of the stadium and not in the matrix elements $W_{c\mu}$
describing the passage of the electron
between one of the channels and the stadium interior. Accordingly, we set
\mbox{$W_{c\mu} = W_{c\mu}^*$}.

General quantum scattering theory allows us to write down the scattering
matrix $S_{cd}$ for the problem at hand \cite{MW}. It has the form
\begin{equation}
S_{cd} = \delta_{cd} - 2i\pi\sum_{\mu,\nu}W_{c\mu}[D^{-1}]_{\mu\nu}
     W_{d \nu} \;,
\label{S}
\end{equation}
where the inverse propagator $D_{\mu\nu}$ is given by
\begin{equation}
D_{\mu\nu} = E\delta_{\mu\nu} - H_{\mu\nu} + i\pi\sum_c W_{c\mu}
     W_{c\nu} \;.
\label{D}
\end{equation}
Here, the energy $E$ is taken to be the Fermi energy $E_F$, and thus will
ultimately be set to zero.
The scattering matrix defined in Eqs.~(\ref{S},\ref{D}) is
manifestly unitary but not symmetric (unless
\mbox{$\s = 0$}).
In writing eq.~(\ref{D}), we have assumed that the energy dependence of
the matrix elements $W_{c\mu}$ is negligible over an energy interval of
the order of $d$, and we have therefore omitted a principal-value integral.
It was shown in Ref.~\citenum{VWZ} that taking such an integral along
complicates the calculation but does not affect the GOE scattering result,
except for a redefinition of the transmission coefficients introduced below.
Insertion of Eqs.~(\ref{S}) and (\ref{D}) into the Landauer formula
(\ref{landauer}) furnishes us with a formal expression for the conductance $g$
in terms of the Hamiltonian, which must now undergo ensemble averaging.
For this purpose, it is useful to note the compact representation
\begin{equation}
S_{ab}S_{ab}^* = 4\Tr\Omega^a D^{-1}\Omega^b [D^{-1}]^\dagger \;,
\label{ss*}
\end{equation}
where the trace `Tr' runs over the level index $\mu$ and we have introduced the
quantities
\begin{equation}
\Omega^c_{\mu\nu} = \pi W_{c\mu}W_{c\nu} \;.
\end{equation}

It is worth pointing out at this stage that
the parameters $\{W_{c\mu}\}$ will not appear explicitly in our final
expression for $\overline{g}$, for the following reason: The random-matrix
ensemble (\ref{billiard}) is invariant under unitary transformations.
Therefore, only unitary invariants constructed from the $W_{c\mu}$'s
can appear in averages like $\overline{g}$.
The only such invariants are the bilinear forms
\begin{equation}
v_{cd}^2 = N^{-1}\sum_{\mu} W_{c \mu} W_{d \mu} \;.
\label{ortho1}
\end{equation}
Because of the summations in Eq.~(\ref{landauer}), we may assume without
loss of generality that
\begin{equation}
v_{cd}^2 = v_c^2 \delta_{cd}
\label{ortho2}
\end{equation}
is diagonal in the
channel indices if $c$ and $d$ are channels in the {\it same} lead. In
assuming that this relation holds in general, we disregard short classical
trajectories connecting the two leads.
It turns out \cite{VWZ} that the $v_c$'s
enter into the final expression for ${\overline g}$ only in the form of the
dimensionless coefficients
\begin{equation}
T_c = \frac{4\lambda X_c}{(\lambda + X_c)^2} \;,
\label{T}
\end{equation}
where
\mbox{$X_c = \pi N v_c^2$}.
Likewise, the parameter $\lambda$ specifying the local level density
appears {\it only} in the combination $X_c$. Obviously, we
have
\mbox{$0 \le T_c \le 1$}
for all $c$. The quantities $T_c$ are interpreted as transmission coefficients
which measure
the strength of the coupling between the cavity and the leads. The absence
of barriers between cavity and leads motivates us to put
\mbox{$T_c = 1$}
for all $c$, thereby maximizing the coupling to the leads.
Then, ${\overline g}$ depends only on $B$ and on $M$.
All dependence on $\lambda$ and on the Fermi energy has disappeared,
and we are left with a calculation that contains {\em no} free parameters.

We emphasize that in our approach, all channels are strictly equivalent.
There is no distinction made between channels in lead one and those in lead
two. Hence, the sums in Eq.~(\ref{landauer}) are restricted only in the sense
that they must be carried out over different sets
\mbox{$\{a\} \neq \{b\}$}
of channels. Clearly, this equivalenve assumption is justified only for long
delay times of the electron inside the billiard. The experiments
\citevb{Marcus,Keller}
are arranged in such a way that it is unlikely for electrons to pass directly
from one lead to the other, lending credence to this assumption.

A crucial constraint  applies in our calculation of $\overline{g}$.
We cannot use the usual perturbative expansion in powers of $1/M$ as done in
previous work \cite{IWZ}.
First, experiments typically have a small number of open channels,
\mbox{$M = 2,4$} or $6$. For example,
in the experiments of Refs.~\citenum{Marcus} and \citenum{Berry},
the stadium billiard is connected to two almost
one-dimensional contacts in each of which the number of channels $M/2$
depends
on the gate voltage and varies between one and three. For $M \gg 1$, the
semiclassical approximation is expected to work. For small $M$, on
the other hand, we
enter a different regime which is not accessible to perturbative methods.
Physically speaking, the levels in the billiard are broadened by the
coupling to the leads. For small $M$, this broadening is of the order of
the mean level spacing $d$. This is the regime we mainly address in this
paper, although our results are generic and apply to all values of $M$.
Second, and more importantly, we shall see that the dependence of the mean
conductance enters in the combination $\s/M$. This implies that the
perturbative
results will be good only as long as
\mbox{$\s \ll M$},
but this will be shown to fall short of the crossover region.
Consequently, we must resort to a computation that is exact in the number of
channels $M$. This we shall achieve by using the method of the supersymmetric
generating function \cite{Efetov,VWZ} to find an expression for
\mbox{$\overline{g}(\s,M)$}
valid for all values of $\s$ and $M$.

\chapter{Magnetic-Field Parameter}
In order to completely specify our model in terms of physical quantities,
it remains to discuss how one can relate $\s$ to the strength of the magnetic
field $B$.
A previous investigation of the GOE~$\to$~GUE transiton in terms of the
same magnetic-field parameter $\s$ by Pandey \& Mehta \cite{PM83} offers
the possibilty of relating $\s$ and $B$ by direct inspection of the spectral
fluctuations of the stadium for different values of the external magnetic
field.
Also, in more recent work, Oz\'orio~de~Almeida \cite{Alm91} derived
semiclassically the dependence of $\s$ on $B$ for a chaotic system
in terms of purely classical correlators.
Although these classical quantities are difficult to compute with
very good accuracy, an efficient average over the phase space is
sufficient to provide another independent quantitative
estimate of $\s$ in terms of physical quantities. In this section,
both approaches are discussed.

We shall start with the first method, and consider the Schr\"odinger
equation for an electron moving in a billiard in the presence of an external
magnetic field $B$. Working in the gauge
${\bf A}(x,y) = (-By/2, Bx/2, 0)$
and introducing rescaled coordinates
\mbox{${x'} = \eta^{1/2}x$},
\mbox{${y'} = \eta^{1/2}y$}
with
\mbox{$\eta = 4\pi/\cal{A}$},
${\cal A}$ being the area of the stadium,
we arrive at the Schr\"odinger equation in the form
\begin{equation}
\label{Schr2}
\left[ \nabla^2 - \frac{i}{2}\frac{\phi}{\phi_0}
 \left(x' \frac{\partial}{\partial y'} -
 y'\frac{\partial}{\partial x'}\right) -
 \frac{\pi}{8} \left(\frac{\phi}{\phi_0}\right)^2 ({x'}^2 + {y'}^2) +
 \frac{{\cal{A}}}{4\pi}\frac{2m}{\hbar^2} E \right]\psi(x', y') = 0 \;,
\end{equation}
where $\phi_0 = h/e$ is the magnetic flux quantum.
The boundary conditions are given by
\mbox{$\psi(x',y') = 0$}
for
\mbox{$(x',y') \in \partial D$},
where $\partial D$ is the border of the stadium.
This is the eigenvalue equation that we solve numerically.
We subsequently drop the primes on the coordinates.
Recalling the Weyl formula for the average level density
$\rho^{\rm av}$ for Dirichlet
boundary conditions in the absence of magnetic field,
\begin{equation}
\rho^{\rm av}(E) = \frac{{\cal{A}}}{4\pi}\frac{2m}{\hbar^2} -
          \frac{{\cal{P}}}{4\pi}\sqrt{\frac{2m}{\hbar^2E}} + ... \;,
\label{Weylrho}
\end{equation}
where $\cal{P}$ denotes the perimeter of the stadium,
we see that, up to a correction of order $E^{-1/2}$,
the rescaling leads to to an
energy spectrum of average density one in units of $2m/\hbar$.

The linear term in $\phi$ in Eq.~(\ref{Schr2}) breaks time-reversal
symmetry. The term quadratic in $\phi$ brings the system to the
integrable limit as one increases $\phi$.
This term is of little importance for the range of $B$ fields
relevant to the experiment under consideration.
The structure of Eq.~(\ref{Schr2}) already suggests a linear dependence
between the parameter $\s$ and ${\phi}/{\phi_0}$, provided that the
operator
\mbox{$\hat{O} = i(x \partial/\partial y - y\partial/\partial x)$}
locally connects different states efficiently,
thereby mimicking a random matrix.
Although the latter assumption seems to be rather natural, in a
dynamical system, the variance of the matrix elements associated
with $\hat{O}$ will typically vary with the energy.
A qualitative prediction of the trend is possible by invoking
the following arguments:
For $B=0$, Berry conjectured that the wavefunction of an eigenstate
of energy
\mbox{$E=\hbar^2 k^2/2m$}
for a classically chaotic billiard is
typically given by a superposition of plane waves
\mbox{$a_i\exp(i{\bf k}_i{\cdot}{\bf r})$}
with ${\bf k}_i$ randomly oriented,
\mbox{$k = |{\bf k}_i|$}
and Gaussian amplitudes $a_i$.
Since according to the previous  conjecture,
\mbox{$\hat{O} = -\hbar ({\bf r\times \hat{k}})_z$},
it follows that for eigenstates $\mu$ and $\nu$ with
\mbox{$k_\nu \approx k_\mu \approx k$},
\mbox{$\hat{O} \approx -\hbar
     k\langle \nu |({\bf r \times \hat{k}})_z |\mu\rangle =
     k v_{\nu\mu}$},
where
\mbox{${\bf \hat{k}} = {\bf k}/k$}
and $v_{\nu\mu}$ is a random variable.
In order to compute the wavefunctions of the stadium with good accuracy,
Heller typically uses
\mbox{$k{\cal P}$}
plane-wave components. Taking into consideration the number of components
and the randomness of $a_i$, one can estimate that
\mbox{$v_{\mu\nu} \propto k^{-1}$}.
{}From this naive argument, one concludes that
\mbox{$\hat{O}^2 \propto k$}.
Therefore, any numerical study of the GOE~$\to$~GUE transition in a
dynamical system will be quantitative within a given energy range.
A precise correction for the energy dependence of our results can be
made via the semiclassical approach, and will be discussed later.

Analytical results for the GOE~$\to$~GUE transition are available for
the fluctuating part of the level-level density correlator, defined as
\begin{equation}
Y_2(\epsilon/d) =
     d^2\,\overline{\rho^{\rm fl}(E)\,\rho^{\rm fl}(E + \epsilon)}
\end{equation}
where $d$ is the mean level density.
For the model defined in Eq.~(\ref{billiard}), one finds in
Ref.~\citenum{PM83} the following analytical expression for $Y_2$
as a function of $x =\epsilon/d$ and $\s$:
\begin{equation}
\label{Y2PM}
Y_2(x,\s) = \Big(\frac{\sin \pi x}{\pi x}\Big)^2 -
     \frac{1}{\pi^2}\int_0^{\pi} dr\, r\, e^{2\s r^2/\pi^2}\sin rx
     \,\int_\pi^\infty dq\, \frac{\sin qx}{q} e^{-2\s q^2/\pi^2} \;.
\end{equation}
Unfortunately, this statistical measure is not very efficient
for quantifying $\s$ in practice. Many numerical studies show that
a more adequate one is the $\Delta_3(x)$ statistic, which
can be related to $Y_2(x)$ straightforwardly \cite{Meh91}.
We have proceeded as follows: For different values of $\phi/\phi_0$,
we computed the $\Delta_3$ level statistics and adjusted $\s$
in Eq.~(\ref{Y2PM}) to obtain the best fit.
The results for the quarter stadium, with radius
\mbox{$R = 1$}
and length
\mbox{$\ell = 0.5$}
are presented in Figs.~3a--3c.
The statistical analysis of the spectra was performed over an energy
interval where the level density stays almost constant. (Only the leading
term in Eq.~(\ref{Weylrho}) is relevant.)
Since we are interested in a quantitative determination of the transition
parameter, care must be taken with respect to the following:
(a) Berry's saturation of the $\Delta$-statistics, which would jeopardize
the method. By restricting our analysis to
\mbox{$x < 10$},
this problem is circumvented.
(b) The lowest eigenstates of the stadium show some marked regularities.
In our analysis, we considered levels only above the $50^{\rm th}$.
In so doing, the agreement for
\mbox{$\phi = 0$}
with the GOE results (for levels up to $300$) is very good.
(c) Spectral modulations due to `bouncing ball' orbits would make
comparison of the model (i.e.\ Eq.~(\ref{billiard})) with the billiard
difficult. For our choice of a relatively small $\ell/R$, similar to the
experimental situation, there are no sizeable modulations.
The numerical results indicate that
\mbox{$\sqrt{\s} = 3.87\, \phi/\phi_0$}.
The uncertainty in the fitting of the best $Y_2(x,\s)$ is about 5\% due to
statistical error bars.
Here, we determined $\s$ for a system containing
approximately
\mbox{$n = 400$}
electrons.
The physical situation of interest is more likely to deal with
\mbox{$n \approx O(10^3)$},
and the question is whether our result will be changed for this
region of the spectrum.

A partial answer to this question was given by Berry and
Robnik \cite{BR86}, who analysed the semiclassical transition
parameter for a billiard threaded by an Aharonov-Bohm flux line.
They found that for a fixed $\phi/\phi_0$, $\s$ increases as $n^{1/2}$.
A complete semiclassical treatment can be found in
Ref.~\citenum{Alm91}, where $\s$ is expressed in terms of
the density of states and purely classical quantities:
\begin{equation}
\sqrt{\s} = \frac{K(E) \, \rho^{\rm av}(E)}{4\hbar} \;,
\end{equation}
where $K(E)$ is a classical quantity which measures the
average magnetic flux squared that is enclosed by a typical
trajectory.
We should like to stress that it is not the purpose of this discussion to
further improve the semiclassical approach, but rather apply the findings
of Ref.~\citenum{Alm91} to obtain some quantitative results.
For this purpose, we shall present the semiclassical results in a
nutshell, motivate their origins, discuss briefly $K(E)$ and rescale
it in a convenient way. Finally, we compute $K(E)$ for a certain
geometry and determine
the $\s$-dependence in a way independent from the previous.

After some manipulations on the Gutzwiller trace formula,
Ref.~\citenum{Alm91} gives $Y_2(\epsilon)$ as
\begin{equation}
\label{Y2sc}
Y_2(\epsilon) = \Big(\frac{d}{2\pi\hbar}\Big)^2\int_{-\infty}^{\infty}
    d\tau\, |\tau| e^{(i\epsilon \tau - 2 \eta |\tau|)/\hbar^2}
    \Big( 1 + \Big\langle e^{-\langle
    \Delta S^2(\tau)\rangle_{\rm p.s.}/\hbar^2}\Big\rangle_E \Big) \;,
\end{equation}
where the average
\mbox{$\langle ... \rangle_E$}
should be taken over an
energy interval large compared with $\epsilon$ and the smoothing
parameter $\eta$ should be small compared with $\epsilon$.
The central dynamical quantity to be computed is the phase space average
\mbox{$\langle ... \rangle_{\rm p.s.}$}
of $\Delta S^2$ as a function of time $\tau$,
which replaces the usual average over periodic orbits.
It is interesting to note that all the physics due to the presence
of the magnetic field is contained in the latter quantity.
Introducing the dynamical variable
\mbox{$
\widetilde{\varphi}(\tau) = {\bf A}({\bf r}) \cdot {\bf \dot{r}} =
     eB\varphi(\tau)$},
one can write
\begin{equation}
\langle \Delta S^2(\tau) \rangle_{\rm p.s.} = 8 (eB)^2
     \int_0^\tau dt\,\int_0^{\tau - t}
     dt'\, \langle \varphi(t)\varphi(t+t')\rangle_{\rm p.s.} \;.
\end{equation}
There is considerable work in the literature dealing with
long power law tails
\mbox{$t' \rightarrow \infty$}
of
\mbox{$\langle \varphi(t)\varphi(t+t')\rangle_{\rm p.s.}$}
and similar correlators.
These works and our numerical studies assure that
\mbox{$\langle \varphi(t)\varphi(t+t')\rangle_{\rm p.s.}$}
decays sufficiently fast so that one can define
\begin{equation}
K(E,\tau) = \int_0^{\tau} dt'\,
       \langle \varphi(0) \varphi(t') \rangle_{\rm p.s.}
\end{equation}
and
\mbox{$\langle \Delta S^2(\tau) \rangle_{\rm p.s.} \approx
     8 (eB)^2 K(E,\tau\to\infty)\tau$}
for sufficiently large $\tau$.
The quantity $K(E)$ has trivial $E$ and $\cal A$ dependence,
and it is convenient
to single out these variables so as to obtain a correlator $\kappa$
defined for a unit area and unit velocity, which depends
solely on the shape of the billiard. This can be easily done
by a proper rescaling of the coordinates and the time; as a
result, one has
\begin{equation}
K(E,\tau) = \sqrt{\frac{2E{\cal A}^3}{m}}\,\kappa(\tau) \;.
\label{kappa}
\end{equation}

Inserting this into the semiclassical expression for the transition
parameter and recalling Eq.~(\ref{Weylrho}), one arrives at the relation
\begin{equation}
\label{tsc}
\s = \left(\frac{\phi}{\phi_0}\right)^2 \Big( N^{\rm av}(E)\Big)^{1/2}\,
     \pi^{5/2}\, \kappa(\tau\to\infty) \;,
\end{equation}
where $N^{\rm av}(E)$ is the cumulative level density, which
counts the number of single-particle states in the billiard
up to the energy $E$. This result gives us the correct energy scaling of the
problem, implying that for a fixed magnetic field the GOE~$\to$~GUE
transition will become faster as the number of electrons $n$
increases (in agreement with Ref.~\citenum{BR86} and our naive estimate).
Moreover, it indicates that $\sqrt{\s}$ depends linearly on
$\phi/\phi_0$ as a naive inspection of Eq.~(\ref{Schr2})
would suggest.
Eq.~(\ref{tsc}) also allows one to obtain quantitative results
by determining $\kappa$ for the the specific system under
consideration. In Fig.~4, we present our result for $\kappa$,
obtained by a numerical average over the phase space.
This was done by computing $10^7$ trajectories with randomly
chosen initial points.
(It is of interest to mention that such an average has to be performed over
closed paths; otherwise the results depend on where the origin of the
coordinate system is taken.)
Our numerical result is
\mbox{$\kappa(\tau\to\infty) \simeq 0.06$}.
It follows that
\mbox{$\sqrt{\s} = \alpha[N^{\rm av}(E)]^{1/4}\,\phi/\phi_0$},
with
\mbox{$\alpha \simeq 1.05$}
in good agreement with
\mbox{$1.03 = 3.87/(200)^{1/4}$}
obtained by the other method (where we should
note that $N^{\rm av}(E)$, being the number of levels, is half as small as
$n$ due to spin degeneracy).

\chapter{Supersymmetry Formalism}
To calculate the ensemble average ${\overline g}$, we use Efetov's
supersymmetry approach in the version described in Ref.~\citenum{VWZ}.
In that paper, the scattering problem formulated in Section~2 was worked
out {\it without} the GOE-symmetry-breaking term in Eq.~(\ref{billiard}).
The present calculation follows very closely the lines of Ref.~\citenum{VWZ}.

The appearance of the symmetry-breaking term causes the scattering matrix not
to be symmetric. This leads to a set of source terms not considered in
Ref.~\citenum{VWZ}. Moreover, the appearance of the
symmetry-breaking term in the effective Lagrangian of the zero-dimensional
non-linear sigma model derived in the
\mbox{$N \rightarrow \infty$}
limit makes it much harder to calculate the terms of highest (eighth) order
in the anticommuting integration variables. These are the terms that yield the
answer to our problem. (All terms of lower order yield vanishing
contributions. This statement, well known in the absence of
symmetry-breaking, applies also in the case considered here \cite{MRZ}.)
There are two ways to overcome this difficulty: (i) One may use a
parametrization of the $Q$-matrix developed by Altland, Efetov and Iida
\cite{AIE} which is particularly taylored to time-reversal symmetry breaking,
or (ii) one may use Efetov's original parametrization \cite{Efetov} of the
$Q$-matrix in conjunction with the method developed in Ref.~\citenum{AIMW} to
simplify the form of the symmetry-breaking term.

We have tried both approaches. It turns out that method (ii) is very much
simpler: The terms of highest order can be obtained by hand, i.e.\ without
recourse to a computer-based algorithm. The reason is that in comparison
with the level correlation function, the source terms needed to obtain the
elements of the scattering matrix depend on $Q$ in a more complicated way:
The matrix $Q$ appears in the denominator. In such a case, the method of
Ref.~\citenum{AIE} becomes very unwieldy.
This is why we have adopted method (ii).

In what follows, we have attempted to render a reasonably complete and
systematic account of the derivation of the explicit formal expression
for $\overline{g}$ as given within the supersymmetry formalism.
Though much of this material has appeared previously in various guises,
these sources are rather fragmentary, and so we consider it worthwhile to
present here a pedagogical treatment that is hopefully comprehensible to the
relatively uninitiated reader who has a little background in Grassmann algebra
and general random-matrix theory. At the same time, we aim to achieve
consistency of notation and conventions.
On the other hand, the expert reader should simply take note of our generating
function as given by Eqs.~(\ref{gdouble})--(\ref{hdouble}),
which incorporates the new source matrix that is employed in the present
calculation. This is given in Eq.~(\ref{jdouble}).
{}From here, one can proceed directly to the final integral expression over
supermatrices for
\mbox{$\overline{|S_{ab}|^2}$},
which is presented in Eq.~(\ref{qint}).

\section{Generating Function}
Let us consider a generating function given by
\begin{eqnarray}
Z(\varepsilon) & = & \Detg^{-1}[D+J(\varepsilon)]
\nonumber \\
& = & \exp\left\{-\Trg\ln [D+J(\varepsilon)]\right\} \;,
\label{genfn}
\end{eqnarray}
where the symbols `Detg' and `Trg' denote the graded determinant and graded
trace as defined in the supersymmetric formalism of Ref.~\citenum{VWZ}.
Here, the inverse propagator $D$ has been extended to a $4N\times4N$
supermatrix,
\begin{equation}
D = (E-H){\opfone}_4 +\half\omega L +i\Omega L \;,
\label{dsuper}
\end{equation}
where
\mbox{$\Omega = \sum_c\Omega^c$}
and
\mbox{$L_{pp'}^{\alpha\alpha'} = (-1)^{p+1}\delta_{pp'}
     \delta^{\alpha\alpha'}$}
is the diagonal supermatrix that distinguishes between advanced and retarded
parts of $D$.
We have
\mbox{$D_{pp'} = [\diag (D,D^\dagger)]_{pp'}$}
where
\mbox{$p,p'=1,2$},
with
\mbox{$p=1$} referring to the retarded block and
\mbox{$p=2$} to the advanced block.
The indices
\mbox{$\alpha,\alpha' = 0,1$}
determine the grading (with
\mbox{$\alpha=0$}
for the commuting (bosonic) components and
\mbox{$\alpha=1$}
for the anticommuting (fermionic) components).
We have also allowed for the presence of a difference
\mbox{$\omega = E_1 - E_2$}
between the energy arguments of the two S-matrix elements that multiply
together to form $g$. According to Eq.~(\ref{dsuper}),
$D_{11}$ corresponds to energy
\mbox{$E_1 = E + \half\omega$}
while $D_{22}$ corresponds to energy
\mbox{$E_2 = E - \half\omega$}.
We shall ultimately set
\mbox{$\omega \to 0$},
but keep it in the interim to facilitate comparison with other related work
\cite{AIE}.

The source supermatrix $J(\varepsilon)$ depends on a set of parameters
$\varepsilon_m$ labelled by by some (multi)-index $m$, and is taken to have the
general form
\begin{equation}
J(\varepsilon) = \varepsilon_m M_m \;.
\end{equation}
for some set of $4N\times4N$ supermatrices $M_m(\mu,\nu)$.
It then follows that
\begin{equation}
\left. \frac{\partial^2}{\partial\varepsilon_m\partial\varepsilon_n}
     Z(\varepsilon)\right|_{\varepsilon=0} =
     \Trg M_mD^{-1}{\cdot}\Trg M_nD^{-1} +
     \Trg M_mD^{-1}M_nD^{-1} \;.
\label{diffj}
\end{equation}
For the problem at hand, we shall make the choice of source matrix
\begin{equation}
J_{\mu\nu}(\{\varepsilon_{ab}^j\}) = \pi\sum_{a,b}\sum_{j=1}^2
     I(j)W_{a\nu}\varepsilon_{ab}^j W_{b\mu} \;,
\label{source}
\end{equation}
where
\begin{equation}
I_{pp'}^{\alpha\alpha'}(j) = -k^{\alpha\alpha'}\delta_{pp'}\delta_{pj}
\label{proj}
\end{equation}
is a projector onto the
\mbox{$p = j$}
block with
\mbox{$k^{\alpha\alpha'} = (-1)^\alpha\delta^{\alpha\alpha'}$}.
In this case, the second term on the RHS of Eq.~(\ref{diffj}) vanishes if we
differentiate with respect to
\mbox{$\varepsilon_m = \varepsilon^1_{ab}$}
and
\mbox{$\varepsilon_n = \varepsilon^2_{ba}$}.
Comparison with Eq.~({\ref{ss*}) then leads to the identification
\begin{equation}
|S_{ab}|^2 = \left. \frac{\partial^2}{\partial\varepsilon^1_{ab}
     \partial\varepsilon^2_{ba}}Z(\varepsilon)\right|_{\varepsilon=0} \;.
\end{equation}

\section{Ensemble Average}
As a Gaussian superintegral, the generating function
\mbox{$Z(\varepsilon)$}
reads
\begin{equation}
Z(\varepsilon) = \int{\cal D}\varphi{\cal D}\overline{\varphi}\,
     \exp\biggl\{i\sum_{p,\alpha}\langle\overline{\varphi}_p^\alpha,
     [(D+J)\varphi]_p^\alpha\rangle\biggr\} \;,
\label{gsuper}
\end{equation}
where we employ the notation
\begin{equation}
\langle F(\mu),G(\mu)\rangle \equiv \sum_\mu F(\mu)G(\mu) \;,
\end{equation}
and
\mbox{$\varphi_p^\alpha(\mu)$}
is a four-component supervector field.
The adjoint supervector is defined by
\mbox{$\overline{\varphi} = \varphi^\dagger s$}
with
\begin{equation}
s_{pp'}^{\alpha\alpha'} = s_p^\alpha\delta^{\alpha\alpha'}\delta_{pp'}\;,
     \quad\quad s^\alpha_p = (-1)^{(1-\alpha)(1+p)} \;.
\end{equation}
The presence of the supermatrix $s$ in the definition of the adjoint ensures
the convergence of the final integral representation of the ensemble-averaged
generating function as a supermatrix non-linear $\sigma$-model and ensures the
correct combination of compact and non-compact symmetries therein \cite{VZ}.

With the definitions
\renewcommand{\arraystretch}{0.5}
\begin{equation}
\Phi = \left(
\begin{array}{c}
\varphi \\
s\varphi^*
\end{array}
\right) \;, \quad\quad
\overline{\Phi} = \Phi^\dagger s \;,
\label{Phi}
\end{equation}
we have the property
\begin{equation}
\overline{\varphi}A\varphi = \half\overline{\Phi}{\opf A}\Phi \;,
\label{double}
\end{equation}
where
\begin{equation}
{\opf A} = \left(
\begin{array}{ll}
A & 0   \\
0 & A^{\rm T}
\end{array}
\right)
\label{dmatrix}
\end{equation}
for {\em any} supermatrix $A$ that is diagonal in the graded indices
\mbox{$\alpha,\alpha'$}.
The eight-dimensional supervector $\Phi$ possesses the `charge-conjugation'
or reality property
\mbox{$\Phi^* = C\Phi$}
which can equivalently be expressed as a condition of Majorana type, viz.\
\mbox{$\overline{\Phi} = (\Lambda\Phi)^{\rm T}$},
where the matrices $C,\Lambda$ are defined in Appendix~A.
In view of Eq.~(\ref{double}), Eq.~(\ref{gsuper}) may be cast in a form that
renders the GOE ensemble average particularly simple, namely
\renewcommand{\arraystretch}{1.0}
\begin{equation}
Z(\varepsilon) = \int{\cal D}\Phi\,
     \exp\biggl\{\frac{i}{2}\sum_{p,\alpha,r}
     \langle\overline{\Phi}_{pr}^\alpha,
     [({\opf D}+{\opf J})\Phi]_{pr}^\alpha\rangle\biggr\} \;,
\label{gdouble}
\end{equation}
The indices $r,r'$  span the additional matrix structure due to the doubling of
dimensionality implied by Eq.~(\ref{Phi}). We shall say that the index $r$
labels the `GOE blocks'.
Here, we have
\begin{equation}
{\opf D} = E{\opfone}_8 - {\opf H} + \half\omega L + i\Omega L \;,
\end{equation}
and the definitions of ${\opf H}$ and ${\opf J}$ are those implied by
Eq.~(\ref{dmatrix}). In particular,
\renewcommand{\arraystretch}{0.5}
\begin{equation}
{\opf H} = \left(
\begin{array}{ll}
H & 0 \\
0 & H^{\rm T}
\end{array}
\right)
= H^{(S)} \otimes {\opfone}_8 + i\sqrt{\frac{\s}{N}}H^{(A)} \otimes \tau^3 \;,
\label{hdouble}
\end{equation}
where the matrix $\tau^3$ is defined in Appendix~A.
We should note the following symmetry properties:
(i) Transformations
\mbox{$T: \Phi \mapsto T\Phi$}
that preserve the relation
\mbox{$\Phi^* = C\Phi$}
obey
\renewcommand{\arraystretch}{1.0}
\begin{equation}
T^* = CTC^{-1} \;,
\label{ctc}
\end{equation}
and (ii) transformations
\mbox{$T: \Phi \mapsto T\Phi$}
that leave the bilinear
\mbox{$\overline{\Phi}\Phi$}
invariant are those which satisfy
\begin{equation}
T^{-1} = sT^\dagger s \;.
\label{sts}
\end{equation}
The set of $8\times8$ supermatrices respecting conditions (\ref{ctc})
{\em and} (\ref{sts})
forms a supergroup with compact and non-compact bosonic subgroups.
In the absence of symmetry-breaking and source terms,
the generating function in
Eq.~(\ref{gdouble}) is invariant under this supergroup.

Special  attention should be paid to the induced $r$-space structure of
the source supermatrix ${\opf J}$:
\begin{equation}
{\opf J}_{\mu\nu} = \pi\sum_{a,b}\sum_{j=1}^2{\opf I}_1(j)
     W_{a\nu}\varepsilon^{(s)j}_{ab}W_{b\mu} +
     \pi\sum_{a,b}\sum_{j=1}^2{\opf I}_2(j)
     W_{a\nu}\varepsilon^{(a)j}_{ab}W_{b\mu} \;,
\label{jdouble}
\end{equation}
where $\varepsilon^{(s)j}_{ab}$ and $\varepsilon^{(a)j}_{ab}$ are the parts
of $\varepsilon^{j}_{ab}$ symmetric and antisymmetric in the indices
\mbox{$a,b$}
respectively, and we have introduced
\mbox{${\opf I}_1(j) = I(j) \otimes {\opfone}_2$},
\mbox{${\opf I}_2(j) = I(j) \otimes \tau^3$}.
Consequently, the symmetrized form of
\mbox{$|S_{ab}|^2$}
can be written as the sum of two terms:
\begin{equation}
\half\left\{|S_{ab}|^2 + |S_{ba}|^2\right\} = {\textstyle \frac{1}{4}}
     \left. \frac{\partial^2}{\partial\varepsilon^{(s)1}_{ab}
     \partial\varepsilon^{(s)2}_{ba}}Z(\varepsilon^{(s)},0)
     \right|_{\varepsilon=0} +
     {\textstyle \frac{1}{4}}
     \left. \frac{\partial^2}{\partial\varepsilon^{(a)1}_{ab}
     \partial\varepsilon^{(a)2}_{ba}}Z(0,\varepsilon^{(a)})
     \right|_{\varepsilon=0}
\end{equation}
for
\mbox{$a \neq b$}.
Cast in this form, the contribution to
\mbox{$|S_{ab}|^2$}
from the first term on the RHS of the equation above is generated by the
essentially same source matrix as used in Ref.~\citenum{VWZ} for the pure
GOE problem; the second term is
generated by a new source matrix. This latter term must vanish for
\mbox{$\s=0$}
and accounts for the asymmetry of the S-matrix.

The factor in the integral of Eq.~(\ref{gdouble}) which undergoes ensemble
averaging can be decomposed as
\begin{equation}
\overline{\exp\Bigl\{-\frac{i}{2}\sum_{p,\alpha,r}
     \langle\overline{\Phi}_{pr}^\alpha, ({\opf H}\Phi)_{pr}^\alpha\rangle
     \Bigr\}}^{\rm H} =
     \overline{\exp\bigl\{-i\Tr_{\mu}H^{(S)}{\cal S}^{(0)}\bigr\}}^{\rm S}
     \cdot
     \overline{\exp\bigl\{\sqrt{\s/N}\Tr_{\mu}H^{(A)}
     {\cal S}^{(3)}\bigr\}}^{\rm A} \;,
\end{equation}
where we have introduced the ordinary $N\times N$ matrices in the level
indices $\mu,\nu$,
\begin{equation}
{\cal S}^{(j)}(\nu,\mu) = \half\sum_{p,\alpha \atop r,r'}
     \overline{\Phi}^\alpha_{pr}(\mu)\tau^j_{rr'}\Phi^\alpha_{pr'}(\nu) \;,
\end{equation}
and we take
\mbox{$\tau^0 = {\opfone}$}.
Gaussian ensemble averaging leads to the relations
\begin{eqnarray}
\overline{\exp\Bigl\{-i\Tr_{\mu}H^{(S)}{\cal S}^{(0)}\Bigr\}}^{\rm S}
     & = & \exp\left\{-\frac{\lambda^2}{4N}\trg S^2\right\} \;,
\nonumber \\
\overline{\exp\Bigl\{\sqrt{\s/N}\Tr_{\mu}H^{(A)}{\cal S}^{(3)}\Bigr\}}^{\rm A}
     & = & \exp\left\{-\frac{\s}{N}\frac{\lambda^2}{4N}
     \trg \tau^3S\tau^3S\right\} \;,
\end{eqnarray}
where
\begin{equation}
S^{\alpha,\alpha'}_{pr,p'r'} \equiv \sum_\mu\Phi^\alpha_{pr}(\mu)
     \overline{\Phi}^{\alpha'}_{p'r'}(\mu)
\end{equation}
is a supermatrix.
These results are easily derived with the aid of the relations
\begin{eqnarray}
\sum_{\mu,\nu}{\cal S}^{(0)}(\nu,\mu){\cal S}^{(0)}(\mu,\nu) & = &
     {\textstyle \frac{1}{4}}\trg S^2 \;,
\nonumber \\
\sum_{\mu,\nu}{\cal S}^{(0)}(\nu,\mu){\cal S}^{(0)}(\nu,\mu) & = &
     {\textstyle \frac{1}{4}}\trg S^2 \;,
\end{eqnarray}
and
\begin{eqnarray}
\sum_{\mu,\nu}{\cal S}^{(3)}(\nu,\mu){\cal S}^{(3)}(\mu,\nu) & = &
     +{\textstyle \frac{1}{4}}\trg \tau^3S\tau^3S \;,
\nonumber \\
\sum_{\mu,\nu}{\cal S}^{(3)}(\nu,\mu){\cal S}^{(3)}(\nu,\mu) & = &
     -{\textstyle \frac{1}{4}}\trg \tau^3S\tau^3S \;.
\end{eqnarray}
The ensemble-averaged generating function then becomes
\begin{equation}
\overline{Z(\varepsilon)} = \int{\cal D}\Phi\,
     e^{i{\cal L}^\#(S)} \exp\biggl\{\frac{i}{2}\sum_{p,\alpha,r}
     \langle\overline{\Phi}^\alpha_{pr} ,
     [(E{\opfone} + \half\omega L + i\Omega L +
     {\opf J})\Phi]^\alpha_{pr}\rangle
     \biggr\} \;,
\end{equation}
where
\begin{equation}
i{\cal L}^\#(S) = -\frac{\lambda^2}{4N}\left[\trg S^2 + \frac{\s}{N}
     \trg\tau^3 S\tau^3 S \right] \;.
\label{hash}
\end{equation}

\section{Hubbard-Stratonovich Transformation}
Following Ref.~\citenum{PS}, let us introduce unity under the $\Phi$-integral
in the form
\begin{equation}
1 = \frac{1}{{\cal N}(S)}\int {\cal D}\sigma\, e^{iW(\sigma,S)} \;,
     \quad\quad {\cal N}(S) = \int {\cal D}\sigma\, e^{iW(\sigma,S)} \;,
\label{unity}
\end{equation}
followed with an interchange of the $\sigma$ and $\Phi$ integrations,
where $\sigma$ spans some appropriate space of supermatrices, not yet
specified.
The quantity
\mbox{$W(\sigma,S)$}
can be any function provided the integral in Eq.~(\ref{unity}) is convergent.
The choice
\begin{equation}
W(\sigma,S) = -{\cal L}^\#\left(S-\frac{iN}{\lambda}\sigma\right)
\end{equation}
leads to
\begin{equation}
i{\cal L}^\#(S) + iW(\sigma,S) = -\frac{N}{4}\left[\trg \sigma^2 +
     \frac{\s}{N}\trg \tau^3 \sigma\tau^3 \sigma\right]
     -\frac{i\lambda}{2}\sum_{p,\alpha,r}\langle\overline{\Phi}^\alpha_{pr},
     [(\Sigma + \frac{\s}{N}\tau^3\Sigma\tau^3)\Phi]^\alpha_{pr}\rangle \;,
\end{equation}
where
\mbox{$\Sigma_{\mu\nu} = \sigma\delta_{\mu\nu}$}.
The quartic dependence on $\Phi$ in the exponent of the generating function
implied in
\mbox{${\cal L}^\#(S)$}
is thus eliminated, and if the space of $\sigma$'s is chosen such that
\mbox{${\cal N}(S) = {\cal N}(0)$}
(i.e.\ shifts are allowed), then the remaining quadratic dependence on $\Phi$
is amenable to exact integration, which yields
\begin{eqnarray}
\overline{Z(\varepsilon)} & = & \int{\cal D}\sigma\,
     \exp\biggl\{-\frac{N}{4}\left[\trg \sigma^2 +
     \frac{\s}{N}\trg \tau^3\sigma\tau^3\sigma\right]\biggr\}
\nonumber \\
& &  \cdot\exp\left\{-\half\Trg_\mu \ln
     \Bigl[E{\opfone} + \half\omega L + i\Omega L + {\opf J}
     -\lambda(\Sigma + \frac{\s}{N}\tau^3\Sigma\tau^3)\Bigr]\right\} \;.
\label{sigma}
\end{eqnarray}
The appropriate space of matrices $\sigma$ is given by those which
assume the form
\begin{equation}
\sigma = T^{-1}(P-i\eta L)T \;,
\label{tpt}
\end{equation}
where the $P$, which represent massive modes, are essentially Hermitian and
block-diagonal in the $p$-index,
and the $T$ span the set of matrices defined by Eqs.~(\ref{ctc}) and
Eq.~(\ref{sts}) modulo the subgroup that commutes with $L$.
The positive constant $\eta$ is required for convergence, and its precise value
will be determined from the saddle-point equation.
Next, we write
\mbox{$P - i\eta L = \sigma_0 + \delta P$},
where $\sigma_0$ is the unique diagonal saddle-point that lies within the
integration manifold, and $\delta P$ are the fluctuations around it.
Thus we have
\mbox{$\sigma = \sigma_{\rm G} + T^{-1}\delta PT$}
with
\mbox{$\sigma_{\rm G} = T^{-1}\sigma_0T$}.
The massive fluctuations can be integrated out in the limit
\mbox{$N\to\infty$},
which leads to an expression for
\mbox{$\overline{Z(\varepsilon)}$}
identical with Eq.~(\ref{sigma}) except for the replacement
\mbox{$\sigma \to \sigma_{\rm G}$}
everywhere.

With
\mbox{$E=0$}
and neglecting $\omega$ (which is assumed to be of order
\mbox{${\rm O}(N^{-1})$}),
the saddle-point equation reads
\mbox{$\sigma_0 + \sigma_0^{-1} = 0$}.
Setting
\mbox{$\sigma_0 = x{\opfone} + yL$}
leads to
\mbox{$x=0$},
\mbox{$y=\pm i$}.
According to Eq.~(\ref{tpt}), the correct choice is
\mbox{$\sigma_0 = -iL$}.
So we write
\mbox{$\sigma_{\rm G} = -iQ$}
where
\mbox{$Q \equiv T^{-1}LT$}.
Having set
\mbox{$E=0$},
we can write
\begin{equation}
\overline{Z(\varepsilon)} = \int {\cal D}Q\,
     e^{i{\cal L}_{\rm eff}(Q)}
     e^{i{\cal L}_{\rm source}(Q;J)} \;,
\end{equation}
where
\begin{equation}
i{\cal L}_{\rm eff}(Q) = \frac{\s}{4}\trg\tau^3Q\tau^3Q - \frac{1}{2}
     \Trg_\mu\ln (\half\omega L +i\Omega L +i\lambda {\tilde Q})
\end{equation}
and
\begin{equation}
i{\cal L}_{\rm source}(Q;J) = -\frac{1}{2}\Trg_\mu\ln\biggl[{\opfone} +
     \frac{1}{\half\omega L +i\Omega L +i\lambda{\tilde Q}}{\opf J}
     \biggr] \;,
\end{equation}
with
\begin{equation}
{\tilde Q} \equiv Q + \frac{\s}{N}\tau^3Q\tau^3 \;.
\end{equation}
The summation over the level index $\mu$ (implied in the trace `Trg') can be
performed by virtue of the orthogonality relation of
Eqs.~(\ref{ortho1}, {\ref{ortho2}),
from which one can derive the result
\begin{equation}
\Tr_\mu (\bar{\omega}+\Omega)^n = (N-M)\bar{\omega}^n +
     \sum_c(\bar{\omega}+X_c)^n \;,
\end{equation}
for
\mbox{$n=1,2,\ldots$}
and any constant $\bar{\omega}$. (In our case,
\mbox{$\bar{\omega} = -i\omega/2$}.)
Consequently,
\mbox{${\cal L}_{\rm eff}(Q)$}
can be decomposed according to
\mbox{${\cal L}_{\rm eff}(Q) = {\cal L}_{\rm free}(Q) +
     {\cal L}_{\rm ch}(Q)$}
with
\begin{eqnarray}
i{\cal L}_{\rm free}(Q) & = & \frac{\s}{4}\trg\tau^3Q\tau^3Q -
     \frac{N}{2}\trg\ln(\half\omega L + i\lambda {\tilde Q}) \;,
\nonumber \\
i{\cal L}_{\rm ch}(Q) & = & -\frac{1}{2} \sum_c \trg\ln \biggl[
     1 + \frac{1}{\half\omega L + i\lambda{\tilde Q}}iX_cL\biggr] \;.
\end{eqnarray}
Next, we expand ${\cal L}_{\rm free}(Q)$ and ${\cal L}_{\rm ch}(Q)$
in the small quantities $\omega$ and $\s/N$ to leading order.
Using the fact that
\mbox{$Q^2={\opfone}$},
this yields
\begin{eqnarray}
i{\cal L}_{\rm free}(Q) & = & -\frac{\s}{4}\trg\tau^3Q\tau^3Q +
     \frac{iN\omega}{4\lambda}\trg LQ \;,
\nonumber \\
i{\cal L}_{\rm ch}(Q) & = & -\frac{1}{2}\sum_c \trg\ln \left(
     {\opfone} + \frac{X_c}{\lambda}LQ\right) \;.
\end{eqnarray}
We note that
\mbox{$i{\cal L}_{\rm free}(Q)$}
coincides with the analogous quantity derived in Ref.~\citenum{AIE}.
At this stage, we implement the assumption of equivalent channels with ideal
couplings, as already discussed in Section~2, by setting
\mbox{$X_c=\lambda$}
for all $c$. Thus we obtain
\begin{equation}
i{\cal L}_{\rm ch}(Q) = -\frac{M}{2}\trg\ln ({\opfone} + LQ) \;,
\end{equation}
noting that
\mbox{$\sum_c 1 = M$}
by definition.
The source term can be treated similarly.

Letting
\mbox{$\omega\to 0$},
we finally arrive at the expression
\begin{eqnarray}
{\overline{|S_{ab}|^2}} & = & \int {\cal D}Q\,
     \exp\biggl[-\frac{M}{2}\,\trg\ln ({\opfone} + Q L) -
     \frac{\s}{4}\,\trg Q\tau^3Q\tau^3 \biggr]
\nonumber \\
& &  \quad\cdot
     \sum_{j=1}^2 \trg\left[\frac{1}{{\opfone} + Q L}{\opf I}_j(1)
     \frac{1}{{\opfone} + Q L}{\opf I}_j(2) \right] \;.
\label{qint}
\end{eqnarray}
The space of the supermatrices $Q$ can be parametrized by 16 real variables.
These can be chosen to comprise 2 `azimuthal' angles, 3 `Cayley-Klein'
parameters, 8 anticommuting variables and 3 eigenvalues (1 compact and 2
non-compact). All but the three eigenvalue parameters can be integrated out
analytically. The final threefold integral giving the exact result for
\mbox{$\overline{|S_{ab}|^2}$}
in its maximally reduced analytical form must be further evaluated numerically.
Our result in Eq.~(\ref{qint}) differs from that of Ref.~\citenum{VWZ}
by the appearance of the new source term containing
\mbox{${\opf I}_2(p)$}
and the symmetry-breaking term involving $\tau^3$ in the exponent.

\renewcommand{\Omega}{{\cal R}}
\chapter{Exact Results}
Using the method developed in Ref.~\citenum{AIMW}, we can perform
in Eq.~(\ref{qint}) the
integration over all the anticommuting as well as over several of the
commuting variables used in the parametrization of $Q$.
We are then left with a threefold
integration over real variables (the `eigenvalues' in Efetov's
parametrization).
A summary of the technical details involved in the derivation is presented in
Appendix~B.

For ease of subsequent discussion, it is convenient to separate the exact
result for
\mbox{$\overline{|S_{ab}|^2}$}
into two contributions:
\begin{equation}
\overline{|S_{ab}|^2} = \overline{|S_{1}|^2} + \overline{|S_{2}|^2} \;,
\label{exact1}
\end{equation}
where
\mbox{$\overline{|S_{1}|^2}$}
represents the part arising from the first source term
(\mbox{$j = 1$})
of Eq.~(\ref{qint}) and
\mbox{$\overline{|S_{2}|^2}$}
that from the second source term
(\mbox{$j = 2$}).
Then we have, for
\mbox{$a \neq b$},
\begin{eqnarray}
\overline{|S_{1}|^2} & = & \frac{1}{2}\int^{+1}_{-1}d\lambda
     \int^\infty_1 d\lambda_1 \int^\infty_1 d\lambda_2 \,
     \left(\frac{1+\lambda}{\lambda_1+\lambda_2}\right)^M
     e^{-2\s(\lambda^2_2-1)}
     \left[\frac{1-\lambda}{1+\lambda} +
     \frac{\lambda_1^2+\lambda_2^2-2}{(\lambda_1+\lambda_2)^2}
     \right]
\nonumber \\
& & \cdot\biggl\{\frac{1}{\Omega^2}\left[(1-\lambda^2)
     \bigl(1+e^{-2\s(1-\lambda^2)}\bigr) - (\lambda_1^2-\lambda_2^2)
     \bigl(1-e^{-2\s(1-\lambda^2)}\bigr)\right]
\nonumber \\
& & {}+\frac{4\s}{\Omega}\left[(1-\lambda^2)e^{-2\s(1-\lambda^2)} +
     \lambda_2^2\bigl(1-e^{-2\s(1-\lambda^2)}\bigr)\right]\biggr\} \;,
\label{exact2}
\end{eqnarray}
where
\begin{equation}
\Omega = \lambda^2+\lambda_1^2+\lambda_2^2-2\lambda\lambda_1\lambda_2-1 \;.
\end{equation}
And,
\begin{eqnarray}
\overline{|S_{2}|^2} & = & \frac{1}{2}\int^{+1}_{-1}d\lambda
     \int^\infty_1 d\lambda_1 \int^\infty_1 d\lambda_2 \,
     \left(\frac{1+\lambda}{\lambda_1+\lambda_2}\right)^M
     e^{-2\s(\lambda^2_2-1)} \frac{1}{\Omega}
\nonumber \\
& & \cdot\biggl\{
-\Bigl(1+e^{-2\s(1-\lambda^2)} -
     \frac{1-e^{-2\s(1-\lambda^2)}}{\s(1-\lambda^2)}\Bigr)
     \frac{1-\lambda}{(1+\lambda)(\lambda_1+\lambda_2)}
\nonumber \\
& & \cdot\biggl[
(\lambda_1-\lambda_2)\Bigl[
     3 + \frac{2}{\Omega}(1+\lambda)(\lambda_1\lambda_2-\lambda) \Bigr]
     + 4\s\left[
     (1+\lambda)(\lambda_2-\lambda\lambda_1)
     - \lambda_2^2(\lambda_1+\lambda_2)\right]
\biggr]
\nonumber \\
& &  {}+ \bigl(1-e^{-2\s(1-\lambda^2)}\bigr)
\biggl[
1 + \frac{2}{\Omega}(1-\lambda)(\lambda_1\lambda_2-\lambda)
\nonumber \\
& &  {}+ \frac{4\s}{\lambda_1+\lambda_2}\left[
     (1-\lambda)(\lambda_2-\lambda\lambda_1)
     + \lambda_2^2(\lambda_1-\lambda_2)\right]
\biggr]
\biggr\} \;.
\label{exact3}
\end{eqnarray}
We remark here that (i) for
\mbox{$\s = 0$},
\mbox{$\overline{|S_{2}|^2} = 0$},
so that only
\mbox{$\overline{|S_{1}|^2}$}
contributes to the GOE limit, while (ii) for
\mbox{$\s \to \infty$},
\mbox{$\overline{|S_{1}|^2} = \overline{|S_{2}|^2}$},
i.e.\ the two terms gives equal contributions to the GUE limit.

The S-matrix result above can be converted into a conductance by virtue of
Eq.~(\ref{avg}).
We define the quantity $\delta g$ to represent the positive deviation of the
mean conductance $\overline{g}$ from the GUE limit
\mbox{$\s \to \infty$}, i.e.
\begin{equation}
\delta g(\s,M) = \overline{g}(\s\to\infty,M) - \overline{g}(\s,M) \;.
\end{equation}

\section{Analytical Results for Special Cases}
For general values of $\s$, one must resort to a numerical evaluation of
Eqs.~(\ref{exact1})--(\ref{exact3}).
However, various special cases can be treated analytically.
Here, we shall quote the results, relegating details of the derivations to
Appendices~C and D.
(a) In the GUE limit
(\mbox{$\s\to\infty$}),
one integration drops out to yield
\begin{eqnarray}
\overline{|S_{ab}|^2} & = & \int^{+1}_{-1}d\lambda
     \int^\infty_1d\lambda_1 \, \left(\frac{1+\lambda}{1+\lambda_1}\right)^M
     \frac{1}{(\lambda_1-\lambda)^2}\left[\frac{1-\lambda}{1+\lambda} +
     \frac{\lambda_1-1}{\lambda_1+1}\right]
\nonumber \\
& = & \frac{1}{M} \;.
\label{GUE}
\end{eqnarray}
The corresponding value of the mean conductance in this limit is then
\begin{equation}
\overline{g}_{\rm GUE}(M) = \frac{M}{2} \;,
\end{equation}
in agreement with Eq.~(\ref{igue}).
The integral in Eq.~(\ref{GUE}) does indeed correctly represent the GUE
limit of
\mbox{$\overline{|S_{ab}|^2}$}.
This was worked out separately along the lines of Ref.~\citenum{VWZ},
replacing the GOE Hamiltonian ensemble used there by a GUE Hamiltonian
ensemble.
(b) In the GOE limit
(\mbox{$\s=0$}),
one has, for
\mbox{$a \neq b$},
\begin{eqnarray}
\overline{|S_{ab}|^2} & = & \frac{1}{2}\int^{+1}_{-1}d\lambda
     \int^\infty_1 d\lambda_1 \int^\infty_1 d\lambda_2 \,
     \left(\frac{1+\lambda}{\lambda_1+\lambda_2}\right)^M
     \frac{1-\lambda^2}{(\lambda^2+\lambda_1^2+\lambda_2^2-
     2\lambda\lambda_1\lambda_2-1)^2}
\nonumber \\
& &  \cdot\left[\frac{1-\lambda}{1+\lambda} +
     \frac{\lambda_1^2+\lambda_2^2-2}{(\lambda_1+\lambda_2)^2}
     \right]
\nonumber \\
& = & \frac{1}{M+1} \;.
\label{GOE}
\end{eqnarray}
This leads to a mean conductance for zero field that is given by
\begin{equation}
\overline{g}_{\rm GOE}(M) = \frac{M^2}{2(M+1)} \;.
\label{GGOE}
\end{equation}
We have also checked that, after a suitable change of integration variables,
the integral formula above agrees with Eq.~(8.10) of Ref.~\citenum{VWZ}
upon setting
\mbox{$T_c=1$}
for all $c$ and considering
\mbox{$a \neq b$}.
Details are provided in Appendix~C.
(c) One can also perform a large-$\s$ expansion of
Eqs.~(\ref{exact1})--(\ref{exact3}).
For values
\mbox{$M \ge 2$},
one obtains
\begin{equation}
\overline{|S_{ab}|^2} = \frac{1}{M} - \frac{M}{M^2-1} \frac{1}{8\s}
     + O\left(\frac{1}{\s^2}\right) \;.
\label{tail}
\end{equation}
The corresponding result for the large-$\s$ tail of the weak-localization term
is
\begin{equation}
\delta g(\s,M) = \frac{M^3}{M^2-1} \frac{1}{16\s}
     + O\left(\frac{1}{\s^2}\right) \;.
\label{slarge}
\end{equation}
The derivation of this result, which is not entirely straightforward, is
outlined in Appendix~D.

\section{General Magnetic Field}
For general values
\mbox{$0 < \s < \infty$},
the eigenvalue integrations in Eqs.~(\ref{exact2}, \ref{exact3})
can no longer be carried out analytically.
However, progress can made examining the implications of assuming that
\mbox{$\delta g(\s,M)$}
is indeed Lorentzian in the magnetic field:
\mbox{$B \sim \phi/\phi_0 \sim \sqrt{\s}$}.
Accordingly, we write
\begin{equation}
\delta g(\s,M) \quad = \quad \frac{1}{a+b\s} \quad = \quad
     \frac{1/a}{1+(b/a)\s} \;.
\label{fit}
\end{equation}
The unknown parameters $a$ and $b$ can be determined completely from the three
special cases discussed above.
To find $a$, we set
\mbox{$\s=0$}
in Eq.~(\ref{fit}). This gives
\begin{eqnarray}
\frac{1}{a} & = & \delta g(0,M)
\nonumber \\
& = & \overline{g}(\s\to\infty,M) - \overline{g}(0,M)
\nonumber \\
& = & \frac{M}{2(M+1)}
\label{avalue}
\end{eqnarray}
with the help of Eqs.~(\ref{GUE}) and (\ref{GUE}).
To find $b$, we expand Eq.~(\ref{fit}) in powers of $1/\s$, which yields
\begin{equation}
\delta g(\s,M) \quad = \quad \frac{1}{b\s}\left[\frac{1}{1+a/(b\s)}\right]
     \quad = \quad \frac{1}{b\s} + \cdots \;.
\end{equation}
Thus $1/b$ is the coefficient of $1/\s$ in
\mbox{$\delta g(\s,M)$}
as given in Eq.~(\ref{slarge}). Consequently, we see that
\begin{equation}
\frac{b}{a} = \frac{8(M-1)}{M^2} \;.
\label{bavalue}
\end{equation}
Substitution of Eqs.~(\ref{avalue}) and (\ref{bavalue}) into Eq.~(\ref{fit})
gives us the result for the expected Lorentzian:
\begin{equation}
\delta g(\s,M) = \frac{M}{2(M+1)}\frac{1}
     {1+{\displaystyle \frac{8(M-1)}{M^2}}\s} \;.
\label{lorentz}
\end{equation}

\chapter{Numerical Analysis}
In order to verify whether Eq.~(\ref{lorentz}) is indeed a good representation
of Eqs.~(\ref{exact1})--(\ref{exact3}), we should compute
\mbox{$\overline{|S_{ab}|^2}$}
numerically as a function of $\s$ and $M$.
This data is most easily digested if we plot the curves of reciprocal
weak-localiztion versus $\s$, i.e.
\begin{equation}
\delta g^{-1} = \frac{2}{M}\left[\frac{1}{M} - \overline{|S_{ab}|^2}
     \right]^{-1} \quad\mbox{vs}\quad \s \;,
\label{curve}
\end{equation}
for various values of $M$.
According to the Lorentzian hypothesis, we expect to find
\begin{equation}
\delta g^{-1}(\s,M) = a(M) + b(M)\s \;,
\label{line}
\end{equation}
i.e.\ a straight line in $\s$ for each value of $M$.

Unfortunately, the exact integral expression for
\mbox{$\overline{|S_{ab}|^2}$}
as given by Eqs.~(\ref{exact1})--(\ref{exact3})
is not directly amenable to numerical evaluation.
It must first be subjected to a number of changes of integration variable
in order to improve the behaviour of the integrands.
This procedure is outlined in Appendix~E.

The numerical results for the curves specified by Eq.~(\ref{curve}) are
presented in Figures~5 and 6.
One can make the following observations:
(i) For
\mbox{$M = 2,3,4,5,\ldots$}
and
\mbox{$\s \sml 100$},
the numerical plots agree with the `theoretical' straight lines to better than
$1$ part in $10^2$, and the degree of agreement improves as $M$ is increased.
(ii) Odd values of $M$ are unphysical, though the threefold integral
remains well-defined. This holds true even for
\mbox{$M=1$}.
We observe, however, that the
\mbox{$M=1$}
curve deviates significantly from a straight line.
Moreover, Eq.~(\ref{lorentz}) does not even make a sensible prediction for what
straight line could have been expected in this case.
To illustrate the extent of the deviation, we plot in Figure~7 the
weak-localization term $\delta g$ as a function of $\sqrt{t}$ for
\mbox{$M=1$},
and compare it with the corresponding Lorentzians obtained
from fitting to the large-$\sqrt{t}$ tail of the
\mbox{$M=1$}
curve, and also to its FWHM.
The inference that one draws from all this is that, viewing
\mbox{$\delta g(\s,M)$}
as a continuous function of the real variable $M$, a transition to almost
Lorentzian behaviour happens somewhere in the region
\mbox{$1 < M < 2$}.

One technical point, especially relevant to the case
\mbox{$M=1$},
that is worth mentioning is the following:
For given $M$, the quantity
\mbox{$\overline{|S_1|^2}$}
varies monotonically from the GOE value of $1/(M+1)$ at
\mbox{$\s=0$}
to half the GUE value, viz.\ $1/(2M)$, in the limit
\mbox{$\s\to\infty$}.
Thus for
\mbox{$M=1$},
it should remain constant with a value of $\half$ in spite of its complicated
structure in the variable $\s$. This is precisely what is observed numerically.

\chapter{Discussion}
To determine the magnetic-field scale on which the GOE~$\to$~GUE transition
occurs, let us look at the full width at half maximum (FWHM),
which we shall denote by
\mbox{$F_{1/2}(M)$},
of the $\delta g$ versus $\sqrt{\s}$ curve:
\mbox{$\delta g = 1/[a+b(\sqrt{\s})^2]$}.
{}From Eq.~(\ref{lorentz}), it is easy to see that
\begin{eqnarray}
F_{1/2}(M) & = & 2\sqrt{\frac{a}{b}} \quad=\quad
     \frac{M}{\sqrt{2(M-1)}}
\nonumber \\
& {\displaystyle \asym{M \gg 1}} & \sqrt{\frac{M}{2}} \;.
\label{fwhm}
\end{eqnarray}
Thus we conclude that the GOE~$\to$~GUE transition is delayed as the number of
open channels increases.
We should also observe from Eq.~(\ref{lorentz}) that the limits
\mbox{$\s\to\infty$} and
\mbox{$M\to\infty$}
do not commute:
\begin{eqnarray}
\lim_{\s\to\infty}\delta g(\s,M) & = & 0 \quad\mbox{for all~$M$} \;,
\nonumber \\
\lim_{M\to\infty}\delta g(\s,M) & = & \frac{1}{2} \quad\mbox{for all~$\s$} \;.
\end{eqnarray}
In the former case, one trivially remains in the GUE limit whatever the value
of $M$, while in the latter, no crossover is observed, with the GOE
weak-localization term holding for all values of $\s$.

This phenomenon can be related to two extreme cases in problems involving
quasi-one-dimensional rings.
When one considers an isolated quasi-one-dimensional ring (hence not coupled
to external channels) such as in the calculation of persistent currents
\cite{AIMW}, the transition from GOE to GUE sets in very rapidly with
increasing flux angle --- at a value around
\mbox{$\phi/\phi_0 \sim [d/(\pi E_c)]^{1/2}$},
where $E_c$ denotes the Thouless energy.
On the other hand, in the problem of AAS oscillations \cite{AAS} in the same
quasi-one-dimensional ring now coupled to two leads carrying a large number of
channels $M$, one computes the weak-localization term as an expansion in terms
of diffusons and Cooperons \cite{JZ}. Here, one does not observe a transition
to GUE behaviour for any value of the flux angle. The oscillations of the
magneto-conductance persist for all values of
\mbox{$\phi/\phi_0$},
as the Cooperon (a periodic function in
\mbox{$\phi/\phi_0$})
is never suppressed.
The hypothesis that the GOE~$\to$~GUE transtion is delayed with increasing
channel number clearly renders these two distinct observations consistent.

The broadening of the Lorentzian with increasing channel number implied by
Eq.~(\ref{lorentz}) can be simply understood in terms of the two relevant time
scales that enter the problem.
(i) We have $\tau_{\rm mix}$ --- the average time taken by the time-reversal
violating perturbation (due to $B$) to mix the GOE levels.
The width associated with this time scale is
\begin{eqnarray}
\Gamma_{\rm mix} & = & \frac{2\pi}{d}\left[\frac{\s}{N}
     \overline{\bigl|H_{\mu\nu}^{(A)}\bigr|^2}\right]
\nonumber \\
& = & \frac{2\pi \s}{d}{\cdot}\frac{\lambda^2}{N^2} \;,
\end{eqnarray}
in which case
\begin{equation}
\tau_{\rm mix} = \frac{\hbar}{\Gamma_{\rm mix}} =
     \frac{\pi\hbar}{2\s d} \;,
\end{equation}
since
\mbox{$d = \pi\lambda/N$}.
Thus, a stronger magnetic field requires a shorter time to mix the levels.
(ii) We also have $\tau_{\rm dec}$ --- the decay time of states in the internal
region due to the coupling with the leads. The associated width is given by
\begin{equation}
\Gamma_{\rm dec} = \frac{Md}{2\pi} \;,
\end{equation}
and this leads to
\begin{equation}
\tau_{\rm dec} = \frac{\hbar}{\Gamma_{\rm dec}} =
     \frac{2\pi\hbar}{Md} \;,
\end{equation}
from which we see that for more open channels, the electrons spend less time
within the stadium.
Therefore, an increase in the number of open channels implies a decrease in
$\tau_{\rm dec}$, which means that the same degree of time-reversal violation
will require a smaller $\tau_{\rm mix}$; and this is achieved by a stronger
magnetic field.
The ratio
\mbox{$\Gamma_{\rm mix}/\Gamma_{\rm dec} \sim \s/M$}
provides a measure of the time-reversal violation, and we note that for
\mbox{$M \gg 1$},
$\delta g$ depends on $\s$ only via the combination $\s/M$.
This feature is not explicit in the semiclassical result of
Ref.~\citenum{BJS} but can be inferred indirectly \cite{Jen,DSF}.

If one considers the problem of calculating
\mbox{$\delta g(\s,M)$}
in the framework of the asymptotic expansion of Refs.~\citenum{IWZ} and
\citenum{JZ}, which essentially proceeds in inverse powers of $M$, then one is
able to identify analogues of the diffuson and Cooperon as
\mbox{$\Pi_{\rm D}^{-1} = M$} and
\mbox{$\Pi_{\rm C}^{-1} = M + 8\s $},
respectively. Eq.~(\ref{lorentz}) has a very simple interpretation in terms of
these quantities: If one considers $M$ to be large (compared with unity) but
\mbox{$\s/M$}
not necessarily small, then one can write
\begin{equation}
\delta g(\s,M) \asym{M \gg 1} 2\left(\frac{M}{2}\right)
     \left(\frac{M}{2}\right)\Pi_{\rm D}\Pi_{\rm C} \;,
\end{equation}
where the first three factors represent the spin degeneracy and the summations
over the incoming and outgoing channels. In this approach, a diffuson factor is
always implied by the structure of the lowest-order contribution to the source
term, and reflects the coupling with the leads.
With increasing $\s$, the Cooperon contribution is damped out, and with it
$\delta g$.

We have treated the stadium billiard strictly as a one-particle system in
which classical chaos and magnetic flux combine to cause the
GOE~$\to$~GUE crossover transition. In actual fact, we should have
treated the stadium as a problem involving a large number (say $A$) of
non-interacting electrons. In such a system, the eigenvalues of the
many-body Hamiltonian
\mbox{$H = \sum_{j=1}^A H(j)$}
will {\it not} obey GOE
(or GUE) statistics, even if the eigenvalues of each of the single--particle
Hamiltonians
\mbox{$H(j)$},
\mbox{$j = 1,\ldots,A$}
do. Does this fact necessitate a
rethinking of our method and of our results? The answer is no. Indeed, using
the argument of Ref.~\citenum{Rea},
one can easily show that for non-interacting fermions, it is legitimate to
calculate the single-particle scattering matrix without paying attention to
the presence of the other fermions.


Finally, let us attempt a comparison of our theoretical result
(\ref{lorentz}) with experimental data.
There are several reasons why a direct comparison of our
Eq.~(\ref{lorentz}) with the results of Ref.~\citenum{Berry}
for $\overline g$ is not possible.
First, the stadium used in the experiment possesses a twofold
reflection symmetry.
Hence, in the closed stadium there exist classes of states pertaining to
different quantum numbers with respect to this symmetry. This fact
is not taken into account in our modelling of the Hamiltonian. Second, the
stadium is known \cite{Graf} to have a family of `bouncing ball'
orbits which might obscure the connection with our random-matrix model. In
view of the way the leads are attached to the stadium, it is likely that
these orbits affect the experiment only slightly. Thirdly, the data
of Ref.~\citenum{Berry} were obtained by averaging
over the gate voltage or, equivalently, over the number of channels and the
area of the stadium. Finally, the data show a marked dependence on
temperature which we do not take into account. For
purposes of orientation, we nonetheless take the data of
Ref.~\citenum{Berry} at face value.
For large $B$, the value of ${\overline g}$
is close to $3$. This suggests
\mbox{$M \approx 6$},
recalling that
\mbox{$\overline{g}_{\rm GUE}(M) = M/2$}.
Insertion of this value of $M$ into Eq.~(\ref{GGOE})
for the zero-field value of
$\delta g$ yields a theoretical prediction of $0.43$, to be compared with a
value of $1/3$ experimentally.
Also, one can write
\begin{equation}
\sqrt{\s} = k\phi/\phi_0 = (0.24)kab \;,
\end{equation}
where $a$ is the area of the stadium measured in
\mbox{$(\mu{\rm m})^2$}
and $b$ is the magnetic field in units of mT.
With the aid of Eq.~(\ref{fwhm}),
this gives the theoretical FWHM in $b$ to be
\begin{equation}
\Delta b_{\rm FWHM} \quad = \quad \frac{1}{0.24ka}\sqrt{\frac{M^2}{2(M-1)}}
     \quad \simeq \quad 2.7 \;,
\end{equation}
for
\mbox{$M = 6$},
\mbox{$a \simeq 0.5~(\mu {\rm m})^2$},
and taking
\mbox{$k \approx 5.8$}.
This value of $k$ results from
estimating the number $n$ of
electrons in the stadium to be about two thousand
and using the scaling relation
\begin{equation}
k_n = \left(\frac{n}{400}\right)^{1/4} k_{400} =
     3.87\left(\frac{n}{400}\right)^{1/4} \;,
\end{equation}
as discussed in Section~3.
Both the values for the height of the Lorentzian and the FWHM differ
somewhat from what we read off the experimentally fitted Lorentzian
shown as a dotted line in Figure~3(b) of Ref.~\citenum{Berry},
namely $1/3$ and $5$~mT, respectively. The discrepancy is consistent
with the presence in the experimental data of effects, such as finite
temperature and inelastic scattering, that will tend to wash out coherent
phenomena, and manifest themselves as a flattening of the peak and
broadening of the width.

\chapter{Conclusions}
We have derived a universal expression for the flux and channel-number
dependence of weak localization, $\delta g$, in classically chaotic ballistic
microstructures. For
\mbox{$M \ge 2$},
\mbox{$\delta g(\s,M)$}
is almost Lorentzian in
\mbox{$\sqrt{\s} \sim B$}.
The FWHM of
\mbox{$\delta g(\s,M)$}
versus $\sqrt{\s}$ grows with channel number essentially as $\sqrt{M/2}$,
so that the
\linebreak
GOE~$\to$~GUE crossover occurs at
\mbox{$\s\sim M/8$}.
This fact is arguably the most
interesting result of this paper: An increase in the number of channels
causes an increasing delay in the GOE ${\rightarrow}$ GUE crossover
transition.

It is remarkable that for values of $M$ as small as two,
\mbox{$\delta g(\s,M)$}
is almost Lorentzian in
\mbox{$\sqrt{\s} \sim B$}.
This is reminiscent of results for the S-matrix auto-correlation function
versus energy in the GOE limit.
In Ref.~\citenum{LW}, it was found that this function is close to Lorentzian
even for small $M$ provided all channels have
\mbox{$T_c = 1$}.
For `inequivalent' channels, i.e.\
channels with different $T_c$'s, marked deviations from the Lorentzian form
occur for $M$ as large as ten \cite{LW}.
We speculate that, similarly, gating one of the leads connected to the stadium
may significantly affect the functional dependence of
\mbox{$\delta g(\s,M)$}
on $\s$.

\section*{Acknowledgements}
We are grateful to M.R.~Zirnbauer for advice.
Z.P.\ and J.A.Z.\ would like to thank the members of the
Max--Planck--Institut f\"ur Kernphysik in Heidelberg for their hospitality
and support.
J.A.Z. also thanks P.J.~Forrester for useful discussions, and the
Mathematics Department at the University of Melbourne for its hospitality
during a visit when some of this work was done.
J.A.Z.\ acknowledges a research grant from NSERC (Canada),
Z.P.\ a grant CIPA--CT--92--2044/P2093 from the European Community,
and C.H.L. support by the NSF and DOE.

\appendix
\chapter{}
\renewcommand{\arraystretch}{0.5}
In this appendix, we collect the various constant matrices that appear
throughout the paper.
Matrices arising from the graded structure of the theory and the
symmetry-breaking due to the presence of advanced and retarded components are
given by
\begin{equation}
k^{\alpha\alpha'} = \left(
\begin{array}{rr}
1 &  0 \\
0 & -1
\end{array}
\right)_{\alpha\alpha'} \;, \quad
L_{pp'} = \left(
\begin{array}{rr}
{\opfone} &  0 \\
0 & -{\opfone}
\end{array}
\right)_{pp'} \;, \quad
s_{pp'} = \left(
\begin{array}{rr}
{\opfone} &  0 \\
0        & -k
\end{array}
\right)_{pp'} \;.
\end{equation}
Matrices associated with the doubling of dimensionality in the GOE problem
relative to the GUE are given by
\begin{equation}
\Lambda_{rr'} = \left(
\begin{array}{rr}
0 &  {\opfone} \\
k &  0
\end{array}
\right)_{rr'} \;, \quad
C = s \otimes \Lambda \;.
\end{equation}
Matrices arising from the presence of magnetic field and the source terms are
given by
\begin{equation}
\tau^3_{rr'} = \left(
\begin{array}{rr}
\opfone &  0 \\
0       & -\opfone
\end{array}
\right)_{rr'} \;, \quad
I_{pp'}(1) = \left(
\begin{array}{rr}
-k & 0 \\
 0 & 0
\end{array}
\right)_{pp'} \;, \quad
I_{pp'}(2) = \left(
\begin{array}{rr}
 0 & 0 \\
 0 & -k
\end{array}
\right)_{pp'} \;.
\end{equation}
In all cases, the explicit indices indicate the space to which the displayed
block structure pertains.

\chapter{}
In Efetov's parametrization, one writes
\mbox{$Q = UD\overline{U}$},
where
\begin{equation}
D = \left(
\begin{array}{rr}
\cos\hat\theta & i\sin\hat\theta \\
-i\sin\hat\theta & -\cos\hat\theta
\end{array}
\right) \;, \quad
\hat\theta = \left(
\begin{array}{rr}
\hat\theta_B & 0 \\
0            & \hat\theta_F
\end{array}
\right) \;,
\end{equation}
with
\begin{equation}
\hat\theta_B = i\left(
\begin{array}{rr}
\theta_1 & \theta_2 \\
\theta_2 & \theta_1
\end{array}
\right) \;, \quad
\hat\theta_F = \left(
\begin{array}{rr}
\theta & 0 \\
0      & \theta
\end{array}
\right) \;, \quad
U = \left(
\begin{array}{rr}
u & 0 \\
0 & v
\end{array}
\right) \;, \quad
\overline{U} = \left(
\begin{array}{rr}
\overline{u} & 0 \\
0 & \overline{v}
\end{array}
\right) \;,
\end{equation}
and we have the relations
\mbox{$\overline{u}= u^{\dagger}$},
\mbox{$\overline{v}= k v^{\dagger} k$}.
The rows and columns of matrices of dimensions 8, 4 and 2 are labelled by the
indices $(p,\alpha,r)$, $(\alpha,r)$, and $r$, respectively; the indices
follow in lexicographical order. (With $\alpha=0,1$, this order differs from
that of Efetov \cite{Efetov}.)
The $4\times4$ matrix $u$ is itself expressed as a product of two matrices
\mbox{$u = u_1u_2$}
(whence
\mbox{$\overline{u}=\overline{u}_2\overline{u}_1$})
where
\begin{equation}
u_{1} =\left(
\begin{array}{cc}
1-2\overline{\eta}\eta+6(\overline{\eta}\eta)^{2} &
-2(1-2\overline{\eta}\eta)\overline{\eta} \\
2\eta(1-2\overline{\eta}\eta) &
1-2\eta\overline{\eta}+6(\eta\overline{\eta})^{2}
\end{array}
\right) \;, \quad
u_{2} = \left(
\begin{array}{cc}
F_{1} & 0 \\
0     & F_{2}
\end{array}
\right) \;.
\label{zd7}
\end{equation}
Here,
\begin{equation}
\eta=\left(
\begin{array}{rr}
\eta_{1} & \eta_{2} \\
-\eta^{*}_{2} & -\eta^{*}_{1}
\end{array}
\right) \;, \quad
\overline{\eta}=\eta^{\dagger}=
\left(
\begin{array}{rr}
\eta^{*}_{1} & \eta_{2} \\
\eta^{*}_{2} & \eta_{1}
\end{array}
\right) \;,
\end{equation}
and
\begin{equation}
F_{1}=\exp(i\phi\tau^{3}) \;, \quad
F_{2}=\left(
\begin{array}{rr}
w          & z \\
-z^{*} & w^{*}
\end{array}
\right) \;,
\end{equation}
where
\mbox{$|w|^{2}+|z|^{2}=1$}.
We note that $F_{2}$ is an SU(2) matrix, which can be expressed in terms of
unconstrained parameters according to
\begin{equation}
w = \frac{(1-im)^{2}-|m_{1}|^{2}}{1+m^{2}+|m_{1}|^{2}} \;, \quad
z = \frac{-2 i m^*_{1}}{1+m^{2}+|m_{1}|^{2}} \;.
\end{equation}
Similarly, we write
\mbox{$v=v_{1}v_{2}$}
(whence
\mbox{$\overline{v}=\overline{v}_{2}\overline{v}_{1}$})
where
\begin{equation}
v_{1}=\left(
\begin{array}{cc}
1+2\overline{\rho}\rho+6(\overline{\rho}\rho)^{2} &
-2 i (1+2\overline{\rho}\rho)\overline{\rho} \\
2 i\rho(1+2\overline{\rho}\rho) &
1+2\rho\overline{\rho}+6(\rho\overline{\rho})^{2}
\end{array}
\right) \;, \quad
v_{2}=\left(
\begin{array}{cc}
\Phi_{1} & 0 \\
0        & \Phi_{2}
\end{array}
\right) \;.
\end{equation}
Here
\begin{equation}
\rho=\left(
\begin{array}{rr}
\rho_{1} & \rho_{2} \\
-\rho^{*}_{2} & -\rho^{*}_{1}
\end{array}
\right) \;, \quad
\overline{\rho}=\rho^{\dagger}= \left(
\begin{array}{rr}
\rho^{*}_{1} & \rho_{2} \\
\rho^{*}_{2} & \rho_{1}
\end{array}
\right) \;,
\label{zd13}
\end{equation}
and
\mbox{$\Phi_{1}=\exp(i\chi\tau^{3})$},
\mbox{$\Phi_{2}={\bf 1}$}.
The normalization of $u$'s and $v$'s is given by
\mbox{$\overline{u}u = \overline{v}v = 1$},
\mbox{$\overline{u}_{j}u_{j}=\overline{v}_{j}v_{j}=1$},
for
\mbox{$j=1,2$}.

In the parametrization given above,
\mbox{$\theta_{1},\theta_{2},\theta,m,\phi,\chi$}
are real commuting variables,
$m_1$ is a complex commuting variable,
while
\mbox{$\eta_{1},\eta^{*}_{1},\eta_{2},\eta^{*}_{2},
\rho_{1},\rho^{*}_{1},\rho_{2},\rho^{*}_{2}$}
are all anticommuting variables.
If we also introduce
\mbox{$\lambda_{1}=\cosh\theta_{1}$},
\mbox{$\lambda_{2}=\cosh\theta_{2}$} and
\mbox{$\lambda=\cos\theta$},
then in terms of these final variables, the integration measure on the
space of $Q$-matrices is given by
\begin{equation}
{\cal D}Q = \frac{2(1-\lambda^{2})}{(64\pi{\cal R})^{2}}
     d\lambda_{1}d\lambda_{2}d\lambda\,{\cal D}\!F_{2}\,
     d\phi d\chi\,d\eta_{1} d\eta^{*}_{1} d\eta_{2} d\eta^{*}_{2}
     d\rho_{1} d\rho^{*}_{1} d\rho_{2} d\rho^{*}_{2} \;,
\label{zd17}
\end{equation}
where
\mbox{${\cal R} = \lambda^{2}_{1}+\lambda^{2}_{2}+\lambda^{2}
     - 2\lambda\lambda_{1}\lambda_{2} - 1$}
and
\mbox{${\cal D}\!F_{2}$}
denotes the SU(2) integration measure
\begin{equation}
{\cal D}\!F_{2} = \frac{4}{\pi^{2}(1+m^{2}+|m_{1}|^{2})^3}
     dm\, d(\re m_{1})\, d(\im m_{1}) \;.
\end{equation}
One should note that Eq.~(\ref{zd17}) assumes Efetov's convention \cite{Efetov}
for integrals over the anticommuting variables,
\mbox{$\int d\chi\,\chi = 1$},
rather than the convention of Ref.~\citenum{VWZ}, namely
\mbox{$\int d\chi\,\chi = (2\pi)^{-1/2}$}.
The integration region corresponds to
\mbox{$1<\lambda_{1},\lambda_{2}<\infty$},
\mbox{$-1<\lambda<1$},
\mbox{$-\infty<m,\re m_{1},\im m_{1}<\infty$},
\mbox{$0<\phi,\chi<\pi$}.

To perform the Grassmann integration, we recall that only that part of the
integrand which is of the highest (eighth) order in the anticommuting
variables contributes \cite{AIMW}.
The $M$-dependent term of the integrand depends solely on the generalized
eigenvalues
\mbox{$\theta_{1},\theta_{2},\theta$}.
By diagonalizing the matrix $\hat\theta$, we find that
\begin{equation}
\exp\left\{-\half M\trg\ln(1+QL)\right\} = \detg^{-M/2} (1+QL)
= \left(\frac{1+\cos\theta}{\cosh\theta_{1}+\cosh\theta_{2}}
     \right)^{M} \;.
\label{zd21}
\end{equation}
After making the substitution
\begin{equation}
(1+QL)^{-1} = \half U \left(
\begin{array}{cc}
{\bf 1} & i\sin\hat\theta (1+\cos\hat\theta)^{-1} \\
i\sin\hat\theta (1+\cos\hat\theta)^{-1} & {\bf 1}
\end{array}
\right) \overline{U} \;,
\end{equation}
we find, for the symmetry-breaking and source terms, the explicit forms
\begin{eqnarray}
\lefteqn{\trg (Q\tau^{3}Q\tau^{3}) =
     \trg[(\cos\hat\theta\,\overline{u}_{2}{\cdot}\overline{u}_{1}
     \tau^{3}u_{1}{\cdot}u_{2})^{2} }
\nonumber \\
& & {}+ 2\sin\hat\theta\,\overline{u}_{2}{\cdot}\overline{u}_{1}\tau^{3}
     u_{1}{\cdot}u_{2}\sin\hat\theta\,\overline{v}_{2}
     {\cdot}\overline{v}_{1}\tau^{3} v_{1}{\cdot}v_{2} +
     (\cos\hat\theta\,\overline{v}_{2}{\cdot}\overline{v}_{1}
     \tau^{3} v_{1}{\cdot}v_{2})^{2}],
\label{zd24} \\
\lefteqn{\trg[(1+QL)^{-1}I_{1}(1)(1+QL)^{-1}I_{1}(2)] = }
\nonumber \\
& & -{\textstyle\frac{1}{4}}\trg [
     (1+\cos\hat\theta)^{-1}\sin\hat\theta\,\overline{u}_{2}
     {\cdot}\overline{u}_{1} k u_{1}{\cdot}u_{2}
     (1+\cos\hat\theta)^{-1}\sin\hat\theta\,\overline{v}_{2}
     {\cdot}\overline{v}_{1}k v_{1}{\cdot}v_{2}],
\label{zd25} \\
\lefteqn{\trg [(1+QL)^{-1}I_{2}(1)(1+QL)^{-1}I_{2}(2)] = }
\nonumber \\
& & -{\textstyle\frac{1}{4}}\trg [
     (1+\cos\hat\theta)^{-1}\sin\hat\theta\,\overline{u}_{2}
     {\cdot}\overline{u}_{1} k\tau^{3} u_{1}{\cdot}u_{2}
     (1+\cos\hat\theta)^{-1}\sin\hat\theta\,\overline{v}_{2}
     {\cdot}\overline{v}_{1}k\tau_{3}v_{1}{\cdot}v_{2}] \;.
\label{zd26}
\end{eqnarray}
The anticommuting variables enter the traces in
Eqs.~(\ref{zd24})--(\ref{zd26})
only via the matrices
$\overline{u}_{1}\mu u_{1}$ and
$\overline{v}_{1}\mu v_{1}$,
with
\mbox{$\mu=\tau^{3},k,k\tau^{3}$}.
By employing Eqs.~(\ref{zd7})--(\ref{zd13}),
we find that these matrices have the form
\begin{equation}
\overline{u}_{1}\mu u_{1}=U_{f}\mu_{f}U^{\dagger}_{f} \;,
\quad
\overline{v}_{1}\mu v_{1}=U_{g}\mu_{g}U^{\dagger}_{g} \;,
\label{zd27}
\end{equation}
where $U_{f}$, $U_{g}$ are the unitary matrices
(\mbox{$U_{f}U^{\dagger}_{f}=U_{g}U^{\dagger}_{g}=1$})
defined by
\begin{equation}
U_{f} = \left(
\begin{array}{cc}
{\bf 1} & 0 \\
0       & \exp(\gamma_{f})
\end{array}
\right) \;, \quad
U_{g} = \left(
\begin{array}{cc}
{\bf 1} & 0 \\
0       & \exp(\gamma_{g})
\end{array}
\right) \;,
\label{zd28}
\end{equation}
with
\begin{equation}
\gamma_{f} = 4\left(
\begin{array}{cc}
0 & -\eta_{1}\eta_{2} \\
\eta^{*}_{1}\eta^{*}_{2} & 0
\end{array}
\right) \;, \quad
\gamma_{g}=4\left(
\begin{array}{cc}
0 & \rho_{1}\rho_{2} \\
-\rho^{*}_{1}\rho^{*}_{2} & 0
\end{array}
\right) \;.
\label{zd29}
\end{equation}
We note that
(i) the matrix $U_{g}$ commutes with $\hat\theta$, $v_{2}$ and
$\overline{v}_{2}$,
(ii) the matrix $\tau^3_{f}$ depends solely on
\mbox{$\eta_{2},\eta^{*}_{2}$},
and
(iii) the matrix $\tau^3_{g}$ depends solely on
\mbox{$\rho_{2},\rho^{*}_{2}$}.
Making use of Eq.~(\ref{zd27}) and (i), we rewrite the traces standing on the
RHS of Eqs.~(\ref{zd24})--(\ref{zd26}) as follows:
\begin{eqnarray}
\lefteqn{\trg (Q\tau^{3}Q\tau^{3}) =
     \trg [(\cos\hat\theta\,U^{\dagger}_{g}\overline{u}_{2}U_{f}\tau^3_{f}
     U^{\dagger}_{f}u_{2}U_{g})^{2} }
\nonumber \\
& & {}+ 2\sin\hat\theta\,U^{\dagger}_{g}\overline{u}_{2}U_{f}\tau^3_{f}
     U^{\dagger}_{f}u_{2}U_{g}
     \sin\hat\theta\,\overline{v}_{2}\tau^3_{g}v_{2}
     + (\cos\hat\theta\,\overline{v}_{2}\tau^3_{g}v_{2})^{2}] \;,
\label{zd30} \\
\lefteqn{\trg[(1+QL)^{-1}I_{1}(1)(1+QL)^{-1}I_{1}(2)] = }
\nonumber \\
& & -{\textstyle \frac{1}{4}}\trg [(1+\cos\hat\theta)^{-1}\sin\hat\theta\,
     U^{\dagger}_{g}\overline{u}_{2}U_{f}k_{f} U^{\dagger}_{f}u_{2} U_{g}
     (1+\cos\hat\theta)^{-1}\sin\hat\theta\,\overline{v}_{2}k_{g}v_{2}] \;,
\label{zd31} \\
\lefteqn{\trg [(1+QL)^{-1}I_{2}(1)(1+QL)^{-1}I_{2}(2)] = }
\nonumber \\
& & -{\textstyle \frac{1}{4}}\trg [(1+\cos\hat\theta)^{-1}\sin\hat\theta\,
     U^{\dagger}_{g}\overline{u}_{2}U_{f}(k\tau_{3})_{f}
     U^{\dagger}_{f}u_{2}U_{g}
     (1+\cos\hat\theta)^{-1}\sin\hat\theta\,
     \overline{v}_{2}(k\tau_{3})_{g}v_{2}] \;.
\label{zd32}
\end{eqnarray}

The calculation can be greatly simplified by using a trick introduced in
Section~4.1 of Ref.~\citenum{AIMW}, which involves
passing from the SU(2) commuting
variables specifying $u_{2}$ to new (primed) SU(2) commuting variables
specifying $u'_{2}$ defined by
\begin{equation}
u'_{2} = U^{\dagger}_{f}u_{2}U_{g} \;,  \quad
\overline{u}'_{2} = U^{\dagger}_{g}\overline{u}_{2}U_{f} \;,
\label{zd33}
\end{equation}
with $u'_{2}$  being the same function of the variables
\mbox{$\phi, m', m'_{1}$}
as $u_{2}$ is of the variables
\mbox{$\phi,m,m_{1}$}.
The new commuting variables
\mbox{$m',m'_{1}$}
are functions of the old commuting variables
\mbox{$m,m_{1}$}
and of the products
\mbox{$\eta_{1}\eta_{2}$},
\mbox{$\eta^{*}_{1}\eta^{*}_{2}$},
\mbox{$\rho_{1}\rho_{2}$}
and
\mbox{$\rho^{*}_{1}\rho^{*}_{2}$}.
The Berezinian of this transformation of variables is equal to unity.
For simplicity, the primes will be suppressed in the sequel.
We note also that
the traces depend on the anticommuting variables only via the matrices
$\mu_{f}$ and $\mu_{g}$.
Furthermore,
the exponential symmetry breaking term depends only on the anticommuting
variables
\mbox{$\eta_{2},\eta^{*}_{2},\rho_{2},\rho^{*}_{2}$}:
\begin{eqnarray}
\trg (Q\tau^{3}Q\tau^{3}) & = &
     \trg[(\cos\hat\theta\,\overline{u}_{2}\tau^3_{f}u_{2})^{2}
     + 2\sin\hat\theta\,\overline{u}_{2}\tau^3_{f}u_{2}
     \sin\hat\theta\,\overline{v}_{2}\tau^3_{g}v_{2}
     + (\cos\hat\theta\,\overline{v}_{2}\tau^3_{g}v_{2})^{2}]
\nonumber \\
& = & 8(\sinh^{2}\theta_{2}+|z|^{2}\sin^{2}\theta)
+    64(\eta_{2}\eta^{*}_{2}-\rho_{2}\rho^{*}_{2})
     (\cosh^{2}\theta_{2} -
     \cosh\theta_{1}\cosh\theta_{2}\cos\theta-|z|^{2}\sin^{2}\theta)
\nonumber  \\
& & {}- 64[\eta_{2}\rho_{2} \cosh\theta_{1}\sinh\theta_{2}\sin\theta\,
     z^{*}\exp(-i(\phi+\chi))+ \mbox{c.c.}]
\nonumber  \\
& & {}+ 64[\eta_{2}\rho^{*}_{2} \sinh\theta_{1}\cosh\theta_{2}\sin\theta\,
     w^{*}\exp(-i(\phi-\chi))+ \mbox{c.c.}]
\nonumber  \\
& & {}+ 256\eta_{2}\eta^{*}_{2}\rho_{2}\rho^{*}_{2}
     [\cosh^{2}\theta_{1}-\cosh^{2}\theta_{2}+
     (|w|^{2}-|z|^{2})\sin^{2}\theta] \;.
\label{zd34}
\end{eqnarray}
Thus,
\begin{eqnarray}
\lefteqn{\exp\{-{\textstyle\frac{1}{4}}\s \trg(Q\tau^{3}Q\tau^{3})\}
     = \exp\{-2\s(\sinh^{2}\theta_{2} + |z|^{2}\sin^{2}\theta)\} }
\nonumber \\
& & \cdot\bigl\{1-16\s(\eta_{2}\eta^{*}_{2}-\rho_{2}\rho^{*}_{2})
     (\cosh^{2}\theta_{2}-\cosh\theta_{1}\cosh\theta_{2}\cos\theta
     - |z|^{2}\sin^{2}\theta)
\nonumber \\
& & {}+ 16\s [\eta_{2}\rho_{2}\cosh\theta_{1}\sinh\theta_{2}\sin\theta\,
     z^{*}\exp(-i(\phi+\chi)) + \mbox{c.c.}]
\nonumber \\
& & {}- 16\s [\eta_{2}\rho^{*}_{2}
     \sinh\theta_{1}\cosh\theta_{2}\sin\theta\,
     w^{*}\exp(-i(\phi-\chi)) + \mbox{c.c.}]
\nonumber \\
& & {}- 64\s\eta_{2}\eta^{*}_{2}\rho_{2}\rho^{*}_{2}
     [\cosh^{2}\theta_{1}-\cosh^{2}\theta_{2}
     + (|w|^{2}-|z|^{2})\sin^{2}\theta
\nonumber \\
& & {}+ 4\s(\cosh^{2}\theta_{2}-\cosh\theta_{1}\cosh\theta_{2}\cos\theta
     -|z|^{2}\sin^{2}\theta)^{2}
\nonumber \\
& & {}+ 4\s\,\sin^{2}\theta(|w|^{2}\sinh^{2}\theta_{1}\cosh^{2}\theta_{2}
     +|z|^{2}\cosh^{2}\theta_{1}\sinh^{2}\theta_{2})]\bigr\} \;.
\label{zd35}
\end{eqnarray}

For working out the integral, only those parts of the source terms which,
on multiplying Eq.~(\ref{zd35}), yield terms of the eighth order
in the anticommuting variables are of interest.
The explicit forms of these parts are as follows (cp = contributing part):
\begin{eqnarray}
\lefteqn{\trg [(1+QL)^{-1}I_{1}(1)(1+QL)^{-1}I_{1}(2)] }
\nonumber \\
& = & -{\textstyle\frac{1}{4}}
     \trg [(1+\cos\hat\theta)^{-1}\sin\hat\theta\,
     \overline{u}_{2}k_{f}u_{2}
     (1+\cos\hat\theta)^{-1}\sin\hat\theta\,
     \overline{v}_{2}k_{g}v_{2}]
\nonumber  \\
& \stackrel{\rm cp}{\longrightarrow} &
     -32\eta_{1}\eta^{*}_{1}\rho_{1}\rho^{*}_{1}
     (1-8\eta_{2}\eta^{*}_{2})(1+8\rho_{2}\rho^{*}_{2})
\nonumber \\ & & \cdot
     [(\cosh^{2}\theta_{1}+\cosh^{2}\theta_{2} - 2)
     (\cosh\theta_{1}+\cosh\theta_{2})^{-2}
     + \sin^{2}\theta\,(1+\cos\theta)^{-2}] \;,
\label{zd36} \\
\lefteqn{\trg [(1+QL)^{-1}I_{2}(1)(1+QL)^{-1}I_{2}(2)] }
\nonumber \\
& = & -{\textstyle\frac{1}{4}}
     \trg [(1+\cos\hat\theta)^{-1}\sin\hat\theta\,\overline{u}_{2}
     (k\tau^{3})_{f}u_{2}
     (1+\cos\hat\theta)^{-1}\sin\hat\theta\,\overline{v}_{2}
     (k\tau^{3})_{g}v_{2}]
\nonumber \\
& \stackrel{\rm cp}{\longrightarrow} &
     -32\eta_{1}\eta^{*}_{1}\rho_{1}\rho^{*}_{1}
     (\cosh\theta_{1}+\cosh\theta_{2})^{-1}(1+\cos\theta)^{-1}
\nonumber \\
& & \cdot \bigl\{(1-16\eta_{2}\eta^{*}_{2})(1+16\rho_{2}\rho^{*}_{2})
     (\cosh\theta_{1}+\cosh\theta_{2})(1-\cos\theta)(|w|^{2}-|z|^{2})
\nonumber \\
& & {}+ (\cosh\theta_{1}-\cosh\theta_{2})(1+\cos\theta)
     {}+ [16\eta_{2}\rho_{2}\sinh\theta_{2}\sin\theta\,
     z^{*}\exp(-i(\phi+\chi))+\mbox{c.c.}]
\nonumber \\
& & {}- [16\eta_{2}\rho^{*}_{2}\sinh\theta_{1}\sin\theta\,
     w^{*}\exp(-i(\phi-\chi))+\mbox{c.c.}]\bigr\} \;.
\label{zd37}
\end{eqnarray}
Using Eqs.~(\ref{zd35})--(\ref{zd37}), the desired leading part of the
integrand (i.e.\ eighth order in the anticommuting variables) can be found.
This part is independent of $\phi$ and $\chi$, and its dependence on the SU(2)
variables is carried by the factors
\mbox{$x^n \exp(ax)$},
where
\mbox{$a=\s\sin^{2}\theta$}
and
\mbox{$x=|w|^{2}-|z|^{2}$}.
(The index $n$ ranges over
\mbox{$n=0,1,2$}
for the source term of Eq.~(\ref{zd36}), and over
\mbox{$n=0,1,2,3$}
for the source term of Eq.~(\ref{zd37}).)
The integrals over the anticommuting variables and over the angles $\phi$
and $\chi$ can now be easily done.
In particular, the integration over the SU(2)
parameters can be carried out with the help of the formula (Section~4.2 of
Ref.~\citenum{AIMW})
\begin{equation}
\int {\cal D}F_2\, x^n e^{ax} = \frac{d^n}{da^n}\left(
     \frac{\sinh a}{a}\right)
\label{zd39}
\end{equation}
for
\mbox{$n = 0,1,2,3$}.
Having performed these integrations, we finally obtain
\mbox{$\overline{|S_{ab}|^2}$}
in the form of Eqs.~(\ref{exact1})--(\ref{exact3}).
\renewcommand{\arraystretch}{1.0}
\chapter{}
\vspace{-2ex}
\subsection*{GUE Limit}
We begin our discussion of the GUE limit
(\mbox{$\s\to\infty$})
by outlining the derivation of  the integral appearing in Eq.~(\ref{GUE}). For
\mbox{$\overline{|S_1|^2}$},
the contribution that survives the large-$\s$ limit
in Eq.~(\ref{exact2}) comprises the term whose coefficient is $4\s$, but
neglecting the exponentials
\mbox{$e^{-2\s(1-\lambda^2)}$},
i.e.
\begin{equation}
\overline{|S_{1}|^2} \asym{\s\to\infty} \frac{1}{2}\int^{+1}_{-1}d\lambda
     \int^\infty_1 d\lambda_1 \int^\infty_1 d\lambda_2 \,
     \left(\frac{1+\lambda}{\lambda_1+\lambda_2}\right)^M
     e^{-2\s(\lambda^2_2-1)}
     \left[\frac{1-\lambda}{1+\lambda} +
     \frac{\lambda_1^2+\lambda_2^2-2}{(\lambda_1+\lambda_2)^2}
     \right]
     \frac{4\s\lambda^2_2}{\Omega} \;.
\label{limit1}
\end{equation}
For
\mbox{$\s \gg 1$},
this integral is dominated by the region
\mbox{$\lambda_2 \sim 1$}.
Thus we write
\mbox{$\lambda_2 = 1+z$},
so that
\begin{equation}
e^{-2\s(\lambda_2^2-1)} = e^{-4\s z}(1 - 2\s z^2 + \cdots) \;.
\label{explim}
\end{equation}
The rest of the integrand should also be expanded in powers of $x$, retaining
only the lowest order contributions. This is equivalent to the replacement
\mbox{$\lambda_2 \to 1$}
everywhere except in the exponential on the RHS of Eq.~(\ref{explim}).
We note that
\begin{equation}
\Omega \asym{\lambda_2 \to 1} (\lambda_1 - \lambda)^2 \;.
\label{omegalim}
\end{equation}
The exponential integral over $z$ can now be trivially carried out, and we
obtain exactly half the integral expression of Eq.~(\ref{GUE}).
The contribution to
\mbox{$\overline{|S_2|}^2$}
in Eq.~(\ref{exact3}) that survives the large-$\s$ limit consists of the terms
with coefficient $4\s$, upon neglecting the exponentials
\mbox{$e^{-2\s(1-\lambda^2)}$}.
This can be simplified to
\begin{equation}
\overline{|S_{2}|^2} \asym{\s\to\infty} 4\s\int^{+1}_{-1}d\lambda
     \int^\infty_1 d\lambda_1 \int^\infty_1 d\lambda_2 \,
     \left(\frac{1+\lambda}{\lambda_1+\lambda_2}\right)^M
     e^{-2\s(\lambda^2_2-1)} \frac{1}{\Omega}
     {\cdot} \frac{\lambda_2^2(\lambda_1-\lambda\lambda_2)}
     {(1+\lambda)(\lambda_1+\lambda_2)} \;.
\label{limit2}
\end{equation}
To extract the large-$\s$ limit completely, the procedure used above for
\mbox{$\overline{|S_1|^2}$}
can be employed; namely, we write
\mbox{$\lambda_2 = 1+z$}
and expand in $z$ except for
\mbox{$e^{-4\s z}$}.
This also yields half the integral expression of Eq.~(\ref{GUE}), and hence the
result is established.

To now evaluate the GUE integral of Eq.~(\ref{GUE}) analytically, we first
cast it in the equivalent form
\begin{equation}
\overline{|S_{\rm GUE}|^2} = 2\int_{-1}^{+1}d\lambda \int_1^\infty
     d\lambda_1\, \left(\frac{1+\lambda}{1+\lambda_1}\right)^M
     \frac{1}{\lambda_1-\lambda} {\cdot}
     \frac{1}{(1+\lambda)(1+\lambda_1)} \;.
\end{equation}
Then we make the change of variables
\begin{eqnarray}
\lambda & = & \frac{1-w}{1+w} \;,
\nonumber \\
\lambda_1 & = & \frac{1-w}{1+w} + \frac{2x}{1+w} \;.
\label{XW}
\end{eqnarray}
The corresponding Jacobian is given by
\begin{equation}
\frac{\partial(\lambda,\lambda_1)}{\partial(x,w)} =
     \frac{4}{(1+w)^3} \;,
\end{equation}
and we also note that
\begin{eqnarray}
\frac{1}{\lambda_1-\lambda} & = & \frac{1+w}{2x} \;,
\nonumber \\
\frac{1}{(1+\lambda)(1+\lambda_1)} & = & \frac{(1+w)^2}{4(1+x)} \;,
\nonumber \\
\frac{1+\lambda}{1+\lambda_1} & = & \frac{1}{1+x} \;.
\end{eqnarray}
Putting all this together, we obtain
\begin{equation}
\overline{|S_{\rm GUE}|^2} \quad = \quad \int_0^\infty dx \int_0^x dw\,
     \frac{1}{x(1+x)^{M+1}} \quad = \quad \frac{1}{M} \;.
\end{equation}

\subsection*{GOE Limit}
To bring the integral in Eq.~(\ref{GOE}) into the form that appears
in Ref.~\citenum{VWZ}, we apply the transformation
\begin{eqnarray}
\lambda & = & 1-2\mu \;,
\nonumber \\
\lambda_1 & = & \left[(1+\mu_1)(1+\mu_2)+\mu_1\mu_2 +
     2\sqrt{\mu_1(1+\mu_1)\mu_2(1+\mu_2)}\right]^{1/2} \;,
\nonumber \\
\lambda_2 & = & \left[(1+\mu_1)(1+\mu_2)+\mu_1\mu_2 -
     2\sqrt{\mu_1(1+\mu_1)\mu_2(1+\mu_2)}\right]^{1/2} \;,
\end{eqnarray}
after re-expressing the non-compact integration in Eq.~(\ref{GOE}) as one
over the triangular region
\mbox{$\lambda_2 \le \lambda_1 < \infty$},
\mbox{$1 \le \lambda_2 < \infty$},
so that
\mbox{$\lambda_1 - \lambda_2 \ge 0$}
everywhere.
The associated Jacobian is given by
\begin{equation}
\frac{\partial(\lambda,\lambda_1,\lambda_2)}
     {\partial(\mu,\mu_1,\mu_2)} =
     \frac{\mu_1-\mu_2}{\sqrt{\mu_1(1+\mu_1)\mu_2(1+\mu_2)}} \;,
\end{equation}
and we note that
\begin{equation}
\Omega = 4(\mu+\mu_1)(\mu+\mu_2) \;.
\end{equation}
This allows us to write
\begin{eqnarray}
\overline{|S_{ab}|^2} & = & \frac{1}{8}\int_0^1 d\mu \int_0^\infty d\mu_1
     \int_0^\infty d\mu_2\,
     \frac{|\mu_1-\mu_2|}{\sqrt{\mu_1(1+\mu_1)\mu_2(1+\mu_2)}}
     \frac{(1-\mu)^M}{\left[(1+\mu_1)(1+\mu_2)\right]^{M/2}}
\nonumber \\
& & \frac{\mu(1-\mu)}{(\mu+\mu_1)^2(\mu+\mu_2)^2}
     \left[\frac{2}{1-\mu} - \frac{1}{1+\mu_1} - \frac{1}{1+\mu_2}\right] \;.
\end{eqnarray}
which agrees with Eq.~(8.10) of Ref.~\citenum{VWZ} for
\mbox{$a \neq b$},
having set
\mbox{$T_c = 1$}
for all $c$.

\newcommand{\lambar}{\overline{\lambda}}
\renewcommand{\Omega}{{\cal R}}
We now show how to evaluate the GOE integral exactly.
With the help of the identity
(\ref{ident}),
the GOE limit of
\mbox{$\overline{|S_{ab}|^2}$},
as given in Eq.~(\ref{GOE}),
can be written as
\begin{equation}
\overline{|S_{\rm GOE}|^2} =
     2\int_{-1}^{+1}d\lambda \int_1^\infty d\lambda_1
     \int_1^\infty d\lambda_2\,
     \left(\frac{1+\lambda}{\lambda_1+\lambda_2}\right)^{M}
     \frac{1-\lambda}{(\lambda_1+\lambda_2)^2} \frac{1}{\Omega} \left[
     1+\frac{(1+\lambda)(\lambda_1\lambda_2-\lambda)}{\Omega}\right] \;.
\end{equation}
Further simplification results from observing that
\begin{equation}
\frac{\partial}{\partial \lambda}\left(\frac{1}{\Omega}\right) =
     \frac{2(\lambda_1\lambda_2-\lambda)}{\Omega^2} \;,
\end{equation}
in which case
\begin{eqnarray}
\overline{|S_{\rm GOE}|^2} & = &
     \int_{-1}^{+1}d\lambda \int_1^\infty d\lambda_1
     \int_1^\infty d\lambda_2\,
     \left(\frac{1+\lambda}{\lambda_1+\lambda_2}\right)^{M+2}
     \frac{1-\lambda}{1+\lambda} \left[
     \frac{2}{1+\lambda}\frac{1}{\Omega} +
     \frac{\partial}{\partial \lambda}
     \left(\frac{1}{\Omega}\right)\right]
\nonumber \\
& = & \int_{-1}^{+1}d\lambda \frac{1}{1+\lambda}
     \left[1-(M-1)\frac{1-\lambda}{1+\lambda}\right]
     \int_1^\infty d\lambda_1 \int_1^\infty d\lambda_2\,
     \left(\frac{1+\lambda}{\lambda_1+\lambda_2}\right)^{M+2}
     \frac{1}{\Omega}
\end{eqnarray}
after an integration by parts.

Next, we make the change of integration variables
\mbox{$(\lambda_1,\lambda_2) \mapsto (\lambar, \alpha)$},
where
\begin{eqnarray}
\lambda_1 & = & 1 + \half (\lambar-1)(1+\alpha) \;,
\nonumber \\
\lambda_2 & = & 1 + \half (\lambar-1)(1-\alpha) \;,
\end{eqnarray}
the associated Jacobian being
\begin{equation}
\frac{\partial (\lambda_1,\lambda_2)}{\partial (\lambar,\alpha)} =
     \half (\lambar-1) \;.
\end{equation}
Then
\begin{equation}
\overline{|S_{\rm GOE}|^2} = \frac{1}{2}\int_{-1}^{+1}d\alpha
     \int_{-1}^{+1}d\lambda\,
     \left[1-(M-1)\frac{1-\lambda}{1+\lambda}\right]
     \int_1^\infty d\lambar\,
     \left(\frac{1+\lambda}{1+\lambar}\right)^{M+2}
     \frac{\lambar-1}{\Omega} \;,
\end{equation}
and we have
\begin{equation}
\Omega = (\lambar-\lambda)^2 -\half(1+\lambda)(\lambar-1)^2(1-\alpha^2) \;.
\end{equation}
We are now in a position to implement the change of variables on the pair
\mbox{$(\lambda,\lambar)$}
given in Eq.~(\ref{XW})
(with $\lambda_1$ replaced by $\lambar$ there), followed by setting
\mbox{$w = x(1-v)$}.
This leads to the representation
\begin{equation}
\overline{|S_{\rm GOE}|^2} = \int_0^\infty dx\,
     \frac{\ell_1(x)}{(1+x)^{M+2}} -
     (M-1)\int_0^\infty dx\, \frac{x\ell_2(x)}{(1+x)^{M+2}} \;,
\end{equation}
where
\begin{eqnarray}
\ell_1(x) & = & \int_0^1 d\alpha \int_0^1 dv\,
     \frac{v}{(1-v)(x+1+v)+\alpha^2v^2} \;,
\nonumber \\
\ell_2(x) & = & \int_0^1 d\alpha \int_0^1 dv\,
     \frac{v(1-v)}{(1-v)(x+1+v)+\alpha^2v^2} \;.
\end{eqnarray}
The $\alpha$-integration can be performed, to yield the expressions
\begin{eqnarray}
\ell_1(x) & = & \int_0^1 dv\,
     \frac{1}{\sqrt{(1-v)(x+1+v)}}\artan\frac{v}{\sqrt{(1-v)(x+1+v)}} \;,
\nonumber \\
\ell_2(x) & = & \int_0^1 dv\,
     \sqrt{\frac{1-v}{x+1+v}}\artan\frac{v}{\sqrt{(1-v)(x+1+v)}} \;.
\label{vint}
\end{eqnarray}

The integrals in Eqs.~(\ref{vint}) are easily done for the special case of
\mbox{$x=0$},
and give the results
\mbox{$\ell_1(0)=\pi^2/8$},
\mbox{$\ell_1(0) - \ell_2(0) = 1$}.
For general values of $x$, we can proceed by making the change of
integration variable
\begin{equation}
v = \left(1-\frac{x}{2}\right) -\frac{x}{2}u
\end{equation}
and introducing the quantity
\begin{equation}
\beta = \frac{x/2}{1+x/2} \;.
\end{equation}
Then we obtain
\begin{eqnarray}
\ell_1(x) & = & \int_\beta^1 du\, \frac{1}{\sqrt{1-u^2}}
     \artan\frac{u-\beta}{\sqrt{1-u^2}} \;,
\nonumber \\
\ell_2(x) & = & \left(1+\frac{x}{2}\right)
     \int_\beta^1 du\, \frac{1-u}{\sqrt{1-u^2}}
     \artan\frac{u-\beta}{\sqrt{1-u^2}} \;.
\end{eqnarray}
The derivatives with respect to $\beta$ have simple forms:
\begin{eqnarray}
\frac{d\ell_1}{d\beta} & = & -\frac{\ln (1+\beta)}{2\beta} \;,
\nonumber \\
\frac{d\tilde{\ell}_2}{d\beta} & = & -\frac{1-\beta}{2\beta}\left[
     1-\frac{\ln (1+\beta)}{\beta}\right] \;,
\end{eqnarray}
where we have introduced
\begin{equation}
\tilde{\ell}_2(x) \equiv \ell_2(x)/(1+x/2) \;.
\end{equation}

Now, after some integrations by parts, we can write
\begin{eqnarray}
\overline{|S_{\rm GOE}|^2} & = & \frac{\ell_1(0)-\ell_2(0)}{M+1} +
     \frac{1}{M+1}\int_0^\infty dx\, \frac{\ell'_1(x)}{(1+x)^{M+1}}
\nonumber \\
& &  {}- \frac{1}{2}\int_0^\infty dx\,
     \frac{\tilde{\ell}_2'(x)}{(1+x)^{M-1}}
     + \frac{M-1}{2(M+1)}\int_0^\infty dx\,
     \frac{\tilde{\ell}_2'(x)}{(1+x)^{M+1}}
\nonumber \\
& = & \frac{1}{M+1} + \int_0^1 d\beta\,
     \left(\frac{1-\beta}{1+\beta}\right)^M \frac{1}{1+\beta}
     \left[1 - \frac{\ln (1+\beta)}{\beta}\right]
\nonumber \\
& &  {}+ \frac{1}{2(M+1)}\int_0^1 \frac{d\beta}{\beta}\,
     \left(\frac{1-\beta}{1+\beta}\right)^{M+1}
     \left[1-\beta- \frac{\ln (1+\beta)}{\beta}\right] \;.
\label{final}
\end{eqnarray}
After noting that
\begin{equation}
\left(\frac{1-\beta}{1+\beta}\right)^M \frac{1}{(1+\beta)^2} =
     -\frac{1}{2(M+1)}\frac{d}{d\beta}
     \left(\frac{1-\beta}{1+\beta}\right)^{M+1}
\end{equation}
and
\begin{equation}
\frac{d}{d\beta}\biggl\{(1+\beta)\left[1-\frac{\ln (1+\beta)}{\beta}\right]
     \biggr\}
     = -\frac{1}{\beta}\left[1-\beta-\frac{\ln (1+\beta)}{\beta}\right] \;,
\end{equation}
and using this to perform a further partial integration on the first integral
on the RHS of Eq.~(\ref{final}),
it becomes clear that the two integrals in Eq.~(\ref{final}) cancel each
other, leaving us with the desired result of Eq.~(\ref{GOE}) in the main
text.

\chapter{}
\vspace{-2ex}
\subsection*{Preliminaries}
If we re-express $\Omega$ in terms of the variables
\mbox{$(x,w,z)$} introduced in the previous appendix,
and then make the further change of variables
\mbox{$(w,z) \mapsto (v',z')$},
where
\mbox{$w = vx$},
\mbox{$z = 2xz'$},
we obtain the form
\begin{equation}
\Omega = \frac{4x^2}{(1+vx)^2}D_v^x(z') \;,
\end{equation}
with
\begin{equation}
D^x_v(z') \equiv (z'+1+vxz')^2 - 4(1-v)z' \;.
\end{equation}
Now, in the large-$\s$ asymptotic expansion, due to the exponential factor
\mbox{$e^{-4\s z} = e^{-8\s xz'}$},
the significant domain of the integration is confined to values
\mbox{$xz' \ll 1$},
and hence
\mbox{$vxz' \ll 1$}
(since
\mbox{$0 \le v \le 1$}).
Since also
\mbox{$z' > 0$},
we can always make the replacement
\mbox{$1-vxz' \sim 1$} in our expression for
\mbox{$D_v^x(z')$},
valid for large $\s$.
Thus, we have
\begin{equation}
D_v^x(z') \quad {\displaystyle \asym{\s\gg 1}} \quad (z'+1)^2-4(1-v)z'
     \quad = \quad D^0_v(z') \;.
\end{equation}
We introduce the notation
\mbox{$D_v(z) \equiv D_v^0(z)$},
and also note that one can write
\mbox{$D_v(z) = (z-1)^2+4vz$}.

Likewise (in view of
\mbox{$xz' \ll 1$},
\mbox{$vxz' \ll 1$}),
we can make the following identifications and large-$\s$ replacements:
\begin{eqnarray}
\lambda_1\lambda_2 - \lambda & = & \frac{2x}{1+vx}[1+(2-v)xz'+z']
\nonumber \\
& {\displaystyle \asym{\s\gg 1}} & \frac{2x}{1+vx}(1+z') \;,
\label{L1} \\
\lambda\lambda_2 - \lambda_1 & = & -\frac{2x}{1+vx}[1+vxz'-z']
\nonumber \\
& {\displaystyle \asym{\s\gg 1}} & -\frac{2x}{1+vx}(1-z') \;,
\label{L2} \\
\lambda_2-\lambda\lambda_1 & = & \frac{2x}{(1+vx)^2}
     [2v(1+xz')+vx(1+vxz')-1+z']
\nonumber \\
& {\displaystyle \asym{\s\gg 1}} & \frac{2x}{(1+vx)^2}
     (2v-1+vx+z') \;.
\label{L3}
\end{eqnarray}
We should note that, in these combinations of the $\lambda$'s, it is not
justified to simply make the substitution
\mbox{$\lambda_2 = 1$}
(i.e.\
\mbox{$z'=0$})
in the asymptotic expansion, even when only leading order in $\s$ is
sought, because it is not necessarily true that the significant integration
region is confined to
\mbox{$z' \ll 1$}.

The following result will be used extensively in the subsequent discussion:
Suppose that $f(x)$ is a function such that
\mbox{$f(x\to\infty) \sim f_\infty(x) = Cx^\nu$}.
Then, (i) if
\mbox{$\nu < -1$},
we have
\begin{equation}
\int_0^\infty dx\, \frac{f(\s x)}{(1+x)^{M+1}} \asym{\s\gg 1}
     \frac{1}{\s}\int_0^\infty dx\, f(x) \;;
\label{R1}
\end{equation}
(ii) if
\mbox{$\nu > -1$},
we have
\begin{equation}
\int_0^\infty dx\, \frac{f(\s x)}{(1+x)^{M+1}} \asym{\s\gg 1}
     \s^\nu \int_0^\infty dx\, \frac{f_\infty(x)}{(1+x)^{M+1}} \;;
\label{R2}
\end{equation}
(iii) if
\mbox{$\nu = -1$},
we have
\begin{equation}
\int_0^\infty dx\, \frac{f(\s x)}{(1+x)^{M+1}} \asym{\s\gg 1}
     \frac{1}{s}\biggl\{\int_0^1 dx\, f(x) + \int_0^\infty
     \left[f(x)-\frac{C}{x}\right] + C\biggl[\ln\s - \sum_{j=1}^M\frac{1}{j}
     \biggr]\biggr\} \;.
\label{R3}
\end{equation}
It is useful to note that with
\mbox{$f(x\to\infty) \sim Cx^{-1}$}
and $f(x)$ of the form
\begin{equation}
f(x) = \int_0^\infty e^{-8xz} g(z) \;,
\end{equation}
in which case we must have
\mbox{$g(0) = 8C$},
Eq.~(\ref{R3}) reads
\begin{equation}
\int_0^\infty dx\, \frac{f(\s x)}{(1+x)^{M+1}} \asym{\s\gg 1}
     \frac{1}{8s}\biggl\{\int_1^\infty dz\, \frac{g(z)}{z} + \int_0^1
     \frac{g(z)-8C}{z} + 8C\biggl[\ln 8\s + \ln\gamma - \sum_{j=1}^M\frac{1}{j}
     \biggr]\biggr\} \;,
\label{RG3}
\end{equation}
where $\gamma$ denotes Euler's number.

It is also useful to define various functions for later convenience:
\begin{equation}
g_1(z) \equiv \int_0^1 dv\, \frac{1}{D_v(z)} =
     \frac{1}{4z}\ln\left(\frac{1+z}{1-z}\right)^2 \;,
\label{G1}
\end{equation}
and we have
\mbox{$g_1(0) = 1$},
\mbox{$g_1(z\to\infty) \sim z^{-2}$};
\begin{equation}
g_2(z) \equiv \int_0^1 dv\, \frac{2v}{D_v(z)} =
     \frac{1}{2z}\left[1- \frac{(1-z)^2}{4z}
     \ln\left(\frac{1+z}{1-z}\right)^2\right] \;,
\label{G2}
\end{equation}
and we have
\mbox{$g_2(0) = 1$},
\mbox{$g_2(z\to\infty) \sim z^{-2}$}.
Finally,
\begin{equation}
g_3(z) \equiv \int_0^1 dv\, \frac{2v}{D^2_v(z)} = \frac{1}{8z^2}\biggl[
     \ln\left(\frac{1+z}{1-z}\right)^2 + \left(\frac{1-z}{1+z}\right)^2
     -1 \biggr]\;.
\label{G4}
\end{equation}
Here we have
\mbox{$g_3(0) = 1$},
\mbox{$g_3(z\to\infty) \sim z^{-4}$}.
Various integrals involving these functions can be established. These
include
\begin{equation}
\int_0^\infty dz\, g_1(z) =
\int_0^\infty dz\, g_2(z) = \frac{\pi^2}{4} \;, \quad
\int_0^\infty dz\, \frac{g_1(z)-1}{z^2} = 0 \;.
\label{GINT1}
\end{equation}
Also,
\begin{equation}
\int_0^\infty dz\, g_3(z) = 1 \;, \quad
\int_0^1 dz\, \frac{g_3(z)-1}{z} + \int_1^\infty dz\, \frac{g_3(z)}{z}
     = -\half \;.
\label{GINT3}
\end{equation}

\subsection*{Large-$\s$ Expansion}
We isolate from the full expression (\ref{exact2}), four distinct
contributions to the $O(\s^{-1})$ term in
\mbox{$\overline{|S_1|^2}$}:
\begin{eqnarray}
L_1 & = & 4\s\int_{-1}^{+1}d\lambda \int_1^\infty d\lambda_1
     \int_1^\infty d\lambda_2\, e^{-2\s(\lambda_2^2-1)}
     \left(\frac{1+\lambda}{\lambda_1+\lambda_2}\right)^{M+2}
     \frac{\lambda_2^2}{(1+\lambda)^3} \left[
     1+\frac{(1+\lambda)(\lambda_1\lambda_2-\lambda)}{\Omega}\right] \;,
\nonumber \\
A_1 & = & \int_{-1}^{+1}d\lambda \int_1^\infty d\lambda_1
     \int_1^\infty d\lambda_2\, e^{-2\s(\lambda_2^2-1)}
     \left(\frac{1+\lambda}{\lambda_1+\lambda_2}\right)^{M+2}
     \frac{1}{(1+\lambda)^3}
     \left[1+\frac{(1+\lambda)(\lambda_1\lambda_2-\lambda)}{\Omega}\right]
\nonumber \\
& &  \cdot \frac{1-\lambda^2-\lambda_1^2+\lambda_2^2}{\Omega} \;,
\nonumber \\
B_1 & = & 4\s\int_{-1}^{+1}d\lambda \int_1^\infty d\lambda_1
     \int_1^\infty d\lambda_2\, e^{-2\s(\lambda_2^2-1)}
     \left(\frac{1+\lambda}{\lambda_1+\lambda_2}\right)^{M+2}
     \frac{1}{(1+\lambda)^3}
     \left[1+\frac{(1+\lambda)(\lambda_1\lambda_2-\lambda)}{\Omega}\right]
\nonumber \\
& &  \cdot (1-\lambda^2-\lambda_2^2)e^{-2\s(1-\lambda^2)} \;,
\nonumber \\
C_1 & = & \int_{-1}^{+1}d\lambda \int_1^\infty d\lambda_1
     \int_1^\infty d\lambda_2\, e^{-2\s(\lambda_2^2-1)}
     \left(\frac{1+\lambda}{\lambda_1+\lambda_2}\right)^{M+2}
     \frac{1}{(1+\lambda)^2}
     \frac{\lambda_1\lambda_2-\lambda}{\Omega^2}
\nonumber \\
& &  \cdot (1-\lambda^2+\lambda_1^2-\lambda_2^2) e^{-2\s(1-\lambda^2)} \;.
\label{LAB1}
\end{eqnarray}
The quantity $L_1$ contains the term of leading order in $\s^{-1}$, i.e.\
half the GUE limit, and can be seen to be equivalent to the RHS of
Eq.~(\ref{limit1}) upon making use of the identity
\begin{equation}
\frac{1-\lambda}{1+\lambda} +
     \frac{\lambda_1^2+\lambda_2^2-2}{(\lambda_1+\lambda_2)^2} =
     \frac{2\Omega}{(1+\lambda)(\lambda_1+\lambda_2)^2}\left[
     1+\frac{(1+\lambda)(\lambda_1\lambda_2-\lambda)}{\Omega}\right] \;.
\label{ident}
\end{equation}
We need to expand this term to first order is $\s^{-1}$.
The integrand of $A_1$ comprises the terms whose $\s$-dependence resides in
the common factor of
\mbox{$\exp[-2\s(\lambda_2^2-1)]$}.
This contribution is non-leading, and its large-$\s$ limit will yield a term
of order $O(\s^{-1})$.
The integrand of $B_1$ comprises the terms whose $\s$-dependence resides in
the common factor of
\mbox{$4\s \exp[-2\s(\lambda_2^2-1)]\exp[-2\s(1-\lambda^2)]$}.
This contribution is also non-leading, and its large-$\s$ limit will yield a
term of order $O(\s^{-1})$.
The term $C_1$ would appear to be of higher order in $\s^{-1}$. However,
because of the singular nature of the factor $\Omega^{-2}$, naive power
counting does not apply, and $C_1$ does in fact contribute at order
$O(\s^{-1})$.
There are four analogous contributions to the $O(\s^{-1})$ term in
the expression (\ref{exact3}) for
\mbox{$\overline{|S_2|^2}$},
given by
\begin{eqnarray}
L_2 & = & 4\s\int_{-1}^{+1}d\lambda \int_1^\infty d\lambda_1
     \int_1^\infty d\lambda_2\, e^{-2\s(\lambda_2^2-1)}
     \left(\frac{1+\lambda}{\lambda_1+\lambda_2}\right)^{M+1}
     \frac{\lambda_2^2}{(1+\lambda)^2}
     \frac{\lambda_1-\lambda\lambda_2}{\Omega} \;,
\nonumber \\
A_2 & = & \int_{-1}^{+1}d\lambda \int_1^\infty d\lambda_1
     \int_1^\infty d\lambda_2\, e^{-2\s(\lambda_2^2-1)}
     \left(\frac{1+\lambda}{\lambda_1+\lambda_2}\right)^{M+1}
     \frac{1}{(1+\lambda)^2} \frac{1}{\Omega}
\nonumber \\
& &  \cdot \left[4\lambda_2 - \lambda_1 - \lambda\lambda_2
     +\frac{2}{\Omega}(1-\lambda^2)(\lambda_1\lambda_2-\lambda)\lambda_2
     -\frac{2\lambda_2^2(\lambda_1+\lambda_2)}{1+\lambda}\right] \;,
\nonumber \\
B_2 & = & 4\s\int_{-1}^{+1}d\lambda \int_1^\infty d\lambda_1
     \int_1^\infty d\lambda_2\, e^{-2\s(\lambda_2^2-1)}
     \left(\frac{1+\lambda}{\lambda_1+\lambda_2}\right)^{M+1}
     \frac{1}{(1+\lambda)^2} \frac{1}{\Omega}
\nonumber \\
& &  \cdot \left[(\lambda_2-\lambda\lambda_1)
     (\lambda_2^2+\lambda^2-1)\right]
     e^{-2\s(1-\lambda^2)} \;,
\nonumber \\
C_2 & = & \int_{-1}^{+1}d\lambda \int_1^\infty d\lambda_1
     \int_1^\infty d\lambda_2\, e^{-2\s(\lambda_2^2-1)}
     \left(\frac{1+\lambda}{\lambda_1+\lambda_2}\right)^{M+1}
     \frac{1-\lambda}{1+\lambda}
     \frac{\lambda_1(\lambda_1\lambda_2-\lambda)}{\Omega^2} \;.
\label{LAB2}
\end{eqnarray}

One deals with $L_1$ and $L_2$ most easily by combining them into
\mbox{$L = L_1 + L_2$}.
Then
\begin{eqnarray}
L & = & 4\s\int_{-1}^{+1}d\lambda \int_1^\infty d\lambda_1
     \int_1^\infty d\lambda_2\, e^{-2\s(\lambda_2^2-1)}
     \left(\frac{1+\lambda}{\lambda_1+\lambda_2}\right)^{M+2}
     \frac{\lambda_2^2}{(1+\lambda)^3}
\nonumber \\
& & {}+ 4\s\int_{-1}^{+1}d\lambda \int_1^\infty d\lambda_1
     \int_1^\infty d\lambda_2\, e^{-2\s(\lambda_2^2-1)}
     \left(\frac{1+\lambda}{\lambda_1+\lambda_2}\right)^{M+2}
     \frac{\lambda_2^2}{(1+\lambda)^3} \frac{1}{\Omega}
\nonumber \\
& &  \cdot \Bigl[(\lambda_1-\lambda)^2 +
     2(\lambda_2+\lambda)(\lambda_1-\lambda) - \lambda(\lambda_2-1)^2
     \Bigr] \;.
\label{deltaL}
\end{eqnarray}
Next, we write
\mbox{$L = \overline{|S_{\rm GUE}|^2} + \delta L$}
which isolates the part that is zeroth order in $\s^{-1}$.
To extract
\mbox{$\delta L$},
which denotes the contribution of order
\mbox{$O(\s^{-1})$},
we must expand the integrand, save for
\mbox{$e^{-4\s z}$},
to first or second order in
\mbox{$z = \lambda_2 -1$}.
We shall split the result of the expansion into three components,
\mbox{$\delta L = \delta L' + \delta L'' + \delta L'''$},
where $\delta L'$ represents contribution from the first term on the RHS of
Eq.~(\ref{deltaL}), $\delta L''$ is the contribution coming from the
expression in square brackets in the second term on the RHS of
Eq.~(\ref{deltaL}),
and $\delta L'''$ denotes the rest.
Then we have
\begin{eqnarray}
\delta L' & = & 4\s\int_{-1}^{+1}d\lambda \int_1^\infty d\lambda_1\,
     \left(\frac{1+\lambda}{1+\lambda_1}\right)^{M+2}
     \frac{1}{(1+\lambda)^3} \int_0^\infty dz\, e^{-4\s z}\left[-2\s z^2
     + 2z -(M+2)\frac{z}{1+\lambda_1}\right]
\nonumber \\
& = & \frac{1}{16\s}\biggl[2\int_0^\infty dx\, \frac{x}{(1+x)^{M+2}}
     \int_0^1 dv\, - (M+2)\int_0^\infty dx\, \frac{x}{(1+x)^{M+3}}
     \int_0^1 dv\, (1+vx) \biggr]
\nonumber \\
& = & \frac{1}{16\s}\left[ \frac{1}{M} - \frac{2}{M+1}\right] \;.
\label{del1}
\end{eqnarray}
Next,
\begin{eqnarray}
\delta L'' & = & 4\s\int_{-1}^{+1}d\lambda \int_1^\infty d\lambda_1
     \int_0^\infty dz\, e^{-4\s z}
     \left(\frac{1+\lambda}{1+\lambda_1}\right)^{M+2}
     \frac{1}{(1+\lambda)^3} \frac{2z(\lambda_1-\lambda)-z^2\lambda}{\Omega}
\nonumber \\
& {\displaystyle \asym{\s\gg 1}} &
     4\s\int_0^\infty dx\, \frac{x^2}{(1+x)^{M+2}}
     \int_0^1 dv\, (1+vx) \int_0^\infty dz\, e^{-8\s xz}
     \frac{2z-z^2}{D_v(z)} \;.
\label{del2a}
\end{eqnarray}
The contribution involving $z^2$ is non-leading, being of order
\mbox{$O(\ln\s/\s^2)$},
while in the rest we can set
\mbox{$D_v(z) \sim D_v(0) = 1$}
by virtue of Eq.~(\ref{R1}), to obtain
\begin{eqnarray}
\delta L'' & {\displaystyle \asym{\s\gg 1}} &
     \frac{1}{8\s}\int_0^\infty dx\, \frac{1}{(1+x)^{M+2}}
     \int_0^1 dv\, (1+vx)
\nonumber \\
& = & \frac{1}{16\s}\left[ \frac{1}{M} + \frac{1}{M+1}\right] \;.
\label{del2b}
\end{eqnarray}
It is convenient to split $\delta L'''$ into a sum of two terms
\mbox{$\delta L''' = \delta L'''_1 + \delta L'''_2$}.
For the first term, we write
\begin{eqnarray}
\delta L'''_1 & = & 4\s\int_{-1}^{+1}d\lambda \int_1^\infty d\lambda_1
     \int_0^\infty dz\, e^{-4\s z}
     \left(\frac{1+\lambda}{1+\lambda_1}\right)^{M+2}
     \frac{1}{(1+\lambda)^3}\frac{1}{\Omega}
\nonumber \\
& &  \cdot \biggl[-2\s z^2 + 2z - (M+2)\frac{z}{1+\lambda_1}\biggr]
     [(\lambda_1-\lambda)^2 + 2(1+\lambda)(\lambda_1-\lambda)] \;.
\label{del31a}
\end{eqnarray}
Due to the presence of the factor
\mbox{$\lambda_1-\lambda$}
in the numerator, it is safe here to set
\mbox{$\Omega \asym{z\to 0} (\lambda_1-\lambda)^2$}
for large $\s$. This results in
\begin{eqnarray}
\delta L'''_1 & = & \frac{1}{4\s}\int_{-1}^{+1}d\lambda \int_1^\infty
     d\lambda_1\, \left(\frac{1+\lambda}{1+\lambda_1}\right)^{M+2}
     \frac{1}{(1+\lambda)^3}
     \left[1+\frac{2(1+\lambda)}{\lambda_1-\lambda}\right]
     \left[1-\frac{M+2}{1+\lambda_1}\right]
\nonumber \\
& = & \frac{1}{16\s}\int_0^\infty dx\, \frac{x+2}{(1+x)^{M+3}}
     \int_0^1 dv\, [2(1+x) - (M+2)(1+vx)]
\nonumber \\
& = & \frac{1}{16\s}\left[-2 + \frac{1}{M} + \frac{1}{M+1}\right] \;.
\label{del31b}
\end{eqnarray}
The remaining contribution to $\delta L'''$ is given by
\begin{eqnarray}
\delta L'''_2 & = & 4\s\int_{-1}^{+1}d\lambda \int_1^\infty d\lambda_1
     \int_0^\infty dz\, e^{-4\s z}
     \left(\frac{1+\lambda}{1+\lambda_1}\right)^{M+2}
     \frac{1}{(1+\lambda)^3}
     [(\lambda_1-\lambda)^2 + 2(1+\lambda)(\lambda_1-\lambda)]
\nonumber \\
& & \cdot \left[\frac{1}{\Omega} - \frac{1}{(\lambda_1-\lambda)^2}\right] \;.
\label{del32A}
\end{eqnarray}
Using the relation
\begin{equation}
\frac{(\lambda_1-\lambda)^2}{\Omega} -1 \asym{\s\gg 1}
     \frac{D_v(z')-1+2vxz'}{D_v(z')} \;,
\label{X1}
\end{equation}
we find
\begin{eqnarray}
\delta L'''_2 & {\displaystyle \asym{\s\gg 1}} & 4\s\int_0^\infty dx\,
     \frac{x(x+2)}{(1+x)^{M+2}} \int_0^1 dv \int_0^\infty dz\,
     e^{-8\s xz} \biggl[\left(\frac{1}{D_v(z)}-1\right)
     -\frac{2vxz}{D_v(z)}\biggr]
\nonumber \\
& = & \delta L'''_{\rm 2a}+\delta L'''_{\rm 2b} \;,
\label{X2}
\end{eqnarray}
where the two components pick out each of the two terms in the square
brackets, respectively.
In
\mbox{$\delta L'''_{\rm 2b}$},
we are allowed by Eq.~(\ref{R2}) to set
\mbox{$D_v(z) \sim D_v(0) =1$},
and so obtain
\begin{eqnarray}
\delta L'''_{\rm 2b} & {\displaystyle \asym{\s\gg 1}} &
     -\frac{1}{8\s}\int_0^\infty dx\,
     \frac{x+2}{(1+x)^{M+2}} \int_0^1 dv\, v
\nonumber \\
& = & -\frac{1}{16\s}\left[\frac{1}{M} + \frac{1}{M+1}\right] \;.
\label{del32b}
\end{eqnarray}
The component
\mbox{$\delta L'''_{\rm 2a}$}
can be expressed as
\begin{equation}
\delta L'''_{\rm 2a} = \int_0^\infty dx\,\frac{f(\s x)}{(1+x)^{M+1}} +
     \int_0^\infty dx\,\frac{f(\s x)}{(1+x)^{M+2}} \;,
\label{FINT}
\end{equation}
where
\begin{equation}
f(x) =  4x\int_0^\infty dz\, e^{-8xz} [g_1(z)-1]
     \asym{x\to\infty} -\frac{1}{3} \frac{1}{(8x)^2} \;.
\end{equation}
In this case, Eq.~(\ref{R1}) implies that
\begin{equation}
\delta L'''_{\rm 2a} \asym{\s\gg 1} \frac{2}{\s}\int_0^\infty dx\, f(x)
     = -\frac{1}{8\s}\int_0^\infty dz\, \frac{g_1(z)-1}{z^2} \;.
\end{equation}
We see from Eq.~(\ref{GINT1}) that
\mbox{$\delta L'''_{\rm 2a}$}
vanishes up to order $O(\s^{-1})$,
and consequently only
\mbox{$\delta L'''_{\rm 2b}$}
survives in
\mbox{$\delta L'''_2$}.
Collecting together Eqs.~(\ref{del1}), (\ref{del2b}) (\ref{del31b}) and
(\ref{del32b}), we obtain
\begin{equation}
\delta L \asym{\s\gg 1} \frac{1}{16\s}\left[-2 + \frac{2}{M}
     - \frac{1}{M+1}\right] \;.
\end{equation}

To extract the large-$\s$ limit of $A_1$, we proceed as above by writing
\mbox{$\lambda_2 = 1+z$}
and expanding the integrand to lowest order in $z$, except for
\mbox{$e^{-4\s z}$}
and the combinations in Eqs.~(\ref{L1})--(\ref{L2}) which are kept intact.
But first we re-express $A_1$ as
\begin{eqnarray}
A_1 & = & \int_{-1}^{+1}d\lambda \int_1^\infty d\lambda_1
     \int_1^\infty d\lambda_2\, e^{-2\s(\lambda_2^2-1)}
     \left(\frac{1+\lambda}{\lambda_1+\lambda_2}\right)^{M+2}
     \frac{1}{(1+\lambda)^3}
     \left[\frac{(1+\lambda)(\lambda_1\lambda_2-\lambda)}{\Omega}+1\right]
\nonumber \\
& &  \cdot \left[\frac{2\lambda_2(\lambda_2-\lambda\lambda_1)}{\Omega}
     -1\right] \;.
\end{eqnarray}
It is now convenient to write
\mbox{$A_1 = A_1^{(0)} + A_1^{(1)} + A_1^{(2)}$},
where $A_1^{(n)}$ denotes the contribution to $A_1$, as given above,
involving the factor
\mbox{$1/\Omega^n$}.
So, we have
\begin{eqnarray}
A_1^{(0)} & {\displaystyle \asym{\s\gg 1}} &
     -\int_{-1}^{+1}d\lambda \int_1^\infty d\lambda_1
     \int_0^\infty dz\, e^{-4\s z}
     \left(\frac{1+\lambda}{1+\lambda_1}\right)^{M+2}
     \frac{1}{(1+\lambda)^3}
\nonumber \\
& = & -\frac{1}{8s}\int_0^\infty \frac{x}{(1+x)^{M+2}}\int_0^1 dv
\nonumber \\
& = & -\frac{1}{8\s}\left[\frac{1}{M}-\frac{1}{M+1}\right] \;.
\label{X3}
\end{eqnarray}
Next, if we note that by Eqs.~(\ref{L1}) and (\ref{L3}) we can make the
replacement
\begin{equation}
2\lambda_2(\lambda_2-\lambda\lambda_1) -
     (1+\lambda)(\lambda_1\lambda_2-\lambda) \asym{\s\gg 1}
     \frac{4x}{(1+vx)^2}[2(v-1) + vx] \;,
\end{equation}
then we have
\begin{eqnarray}
A_1^{(1)} & {\displaystyle \asym{\s\gg 1}} &
     -\int_{-1}^{+1}d\lambda \int_1^\infty d\lambda_1
     \int_0^\infty dz\, e^{-4\s z}
     \left(\frac{1+\lambda}{1+\lambda_1}\right)^{M+2}
     \frac{1}{(1+\lambda)^3} \frac{1}{\Omega}
\nonumber \\
& & \cdot [2\lambda_2(\lambda_2-\lambda\lambda_1) -
     (1+\lambda)(\lambda_1\lambda_2-\lambda)]
\nonumber \\
& {\displaystyle \asym{\s\gg 1}} & \int_0^\infty dx\,
     \frac{x}{(1+x)^{M+2}} \int_0^1 dv \int_0^\infty dz\,
     e^{-8\s xz} \frac{2(v-1)+vx}{D_v(z)}
\nonumber \\
& {\displaystyle \asym{\s\gg 1}} & \int_0^\infty dx\,
     \frac{1}{(1+x)^{M+2}} \int_0^1 dv\, [2(v-1)+vx]
\nonumber \\
& = & \frac{1}{16\s}\left[\frac{1}{M}-\frac{3}{M+1}\right] \;.
\label{X4}
\end{eqnarray}
The observation that
\begin{equation}
\frac{2(\lambda_2-\lambda\lambda_1)}{\Omega} =
     -\frac{\partial}{\partial\lambda_2}\left(\frac{1}{\Omega}\right)
\end{equation}
allows us to express $A^{(2)}_1$ in a form that is amenable to integration
by parts. It is convenient to do so in order to avoid the emargence of
singular integrals (convergent only as principal values) in the process of
reducing to the large-$\s$ limit. Accordingly, we write
\begin{eqnarray}
A^{(2)}_1 & {\displaystyle \asym{\s\gg 1}} &
     -\int_{-1}^{+1} d\lambda \int_1^\infty d\lambda_1 \int_0^\infty dz\,
     e^{-4\s z} \left(\frac{1+\lambda}{1+\lambda_1}\right)^{M+2}
     \frac{1}{(1+\lambda)^2} (\lambda_1\lambda_2-\lambda)
     \frac{\partial}{\partial z}\left(\frac{1}{\Omega}\right)
\nonumber \\
& {\displaystyle \asym{\s\gg 1}} &
     -\int_{-1}^{+1} d\lambda \int_1^\infty d\lambda_1 \int_0^\infty dz\,
     e^{-4\s z} \left(\frac{1+\lambda}{1+\lambda_1}\right)^{M+2}
     \frac{1}{(1+\lambda)^2}
\nonumber \\
& &  \cdot \biggl\{4\s\left[\frac{\lambda_1-\lambda}{\Omega}
     - \frac{1}{\lambda_1-\lambda}\right]
     - (1-4\s z)\frac{\lambda_1}{\Omega}\biggr\}
\nonumber \\
& = & A^{\rm (2a)}_1 + A^{\rm (2b)}_1 \;.
\end{eqnarray}
Let us first deal with
\mbox{$A^{\rm (2b)}_1$}.
We can write this component as
\begin{equation}
A^{\rm (2b)}_1 = \left(1+\s\frac{d}{d\s}\right)\tilde{A}^{\rm (2b)}_1
\label{A2btilde}
\end{equation}
with
\begin{eqnarray}
\tilde{A}^{\rm (2b)}_1 & = &
     \int_{-1}^{+1} d\lambda \int_1^\infty d\lambda_1 \int_0^\infty dz
     e^{-4\s z} \left(\frac{1+\lambda}{1+\lambda_1}\right)^{M+2}
     \frac{\lambda_1}{(1+\lambda)^2} \frac{1}{\Omega}
\nonumber \\
& {\displaystyle \asym{\s\gg 1}} &
     \half\int_0^\infty dx\, \frac{1}{(1+x)^{M+2}} \int_0^1 dv\,
     [(2-v)x+1] \int_0^\infty dz\, e^{-8\s xz}\frac{1}{D_v(z)} \;.
\end{eqnarray}
Now, the leading contribution to the term involving
\mbox{$(2-v)$},
which is of order $O(\s^{-1})$, is annihilated by the differential operator
in Eq.~(\ref{A2btilde}). In the rest, only the logarithmic term, whose
presence follows from Eq.~(\ref{R3}), survives.
Thus, we obtain
\begin{eqnarray}
\tilde{A}^{\rm (2b)}_1  & {\displaystyle \asym{\s\gg 1}} &
     \frac{1}{2}\left(1+\s\frac{d}{d\s}\right)
     \int_0^\infty dx\, \frac{1}{(1+x)^{M+2}}
     \int_0^\infty dz\, e^{-8\s xz} g_1(z)
\nonumber \\
& {\displaystyle \asym{\s\gg 1}} &
     \frac{1}{16}\left(1+\s\frac{d}{d\s}\right)\frac{\ln\s}{\s}
\nonumber \\
& = & \frac{1}{16\s} \;.
\end{eqnarray}
Finally, we have
\begin{equation}
A^{\rm (2a)}_1 {\displaystyle \asym{\s\gg 1}}
     -4\s\int_{-1}^{+1} d\lambda \int_1^\infty d\lambda_1 \int_0^\infty dz\,
     e^{-4\s z} \left(\frac{1+\lambda}{1+\lambda_1}\right)^{M+2}
     \frac{\lambda_1-\lambda}{(1+\lambda)^2}
     \left[\frac{1}{\Omega} - \frac{1}{(\lambda_1-\lambda)^2}\right] \;.
\end{equation}
With the help or Eq.~(\ref{X1}), this can be reduced to
the form
\begin{equation}
A^{\rm (2a)}_1 {\displaystyle \asym{\s\gg 1}}
     -4\s\int_0^\infty dx\, \frac{x}{(1+x)^{M+2}} \int_0^1 dv \int_0^\infty
     dz\, e^{-8\s xz} \biggl[\left(\frac{1}{D_v(z)}-1\right)
     -\frac{2xvz}{D_v(z)}\biggr] \;.
\end{equation}
One should note that this expression for
\mbox{$A^{\rm (2a)}_1$}
has a structure very similar to that of
\mbox{$\delta L'''_2$}
in Eq.~(\ref{X2}), and an almost identical analysis leads to the result
\begin{equation}
A^{\rm (2a)}_1 \asym{\s\gg 1} \frac{1}{16\s} \frac{1}{M+1} \;.
\end{equation}
Adding together all contributions to $A^{(2)}_1$ and combining thus with
the results (\ref{X3}) for $A^{(0)}_1$ and
(\ref{X4}) for $A^{(1)}_1$, we see that
\begin{equation}
A_1 \asym{\s\gg 1} \frac{1}{16\s}\left(1-\frac{1}{M}\right) \;.
\end{equation}

To extract the large-$\s$ limit of $A_2$, we also split it into a sum of
regular and singular parts,
\linebreak
\mbox{$A_2 = A_2^{(1)} + A_2^{(2)}$},
where $A_2^{(n)}$ denotes the contribution to $A_2$ in Eq.~(\ref{LAB2})
that involves $1/\Omega^{n}$. Then
\begin{eqnarray}
A_2^{(1)} & = & \int_{-1}^{+1} d\lambda \int_1^\infty d\lambda_1
     \int_1^\infty d\lambda_2\, e^{-2\s(\lambda_2^2-1)}
     \left(\frac{1+\lambda}{\lambda_1+\lambda_2}\right)^{M+1}
     \frac{1}{(1+\lambda)^3}\frac{1}{\Omega}
\nonumber \\
& & \cdot \bigl\{2\lambda_2[(1-\lambda^2)-(\lambda_1-\lambda)] +
     (1+\lambda)(\lambda_1\lambda_2-\lambda)
     + 2\lambda_2(1-\lambda_2)\bigr\} \;,
\nonumber \\
& = & A_2^{\rm (1a)} + A_2^{\rm (1b)} + A_2^{\rm (1c)},
\end{eqnarray}
where the three components pick out each of the three terms in the braces,
respectively. The first component yields
\begin{eqnarray}
A_2^{\rm (1a)} & {\displaystyle \asym{\s \gg 1}} & 2\int_{-1}^{+1}
     d\lambda \int_1^\infty d\lambda_1 \int_0^\infty dz\, e^{-4\s z}
     \left(\frac{1+\lambda}{1+\lambda_1}\right)^{M+1}
     \frac{1}{(1+\lambda)^3}
     \frac{(1-\lambda^2) - (\lambda_1-\lambda)}{\Omega}
\nonumber \\
& {\displaystyle \asym{\s \gg 1}} & \int_0^\infty dx\, \frac{x}{(1+x)^{M+1}}
     \int_0^1 dv\, \left[(2v-1) + vx\right]
     \int_0^\infty dz\, e^{8\s xz} \frac{1}{D_v(z)}
\nonumber \\
& {\displaystyle \asym{\s \gg 1}} & -\frac{1}{16\s}\left[\frac{1}{M-1}
     - \frac{1}{M}\right] \;,
\end{eqnarray}
having set
\mbox{$D_v(z) \sim D_v(0) =1$}
by virtue of Eq.~(\ref{R2}).
With the aid of Eq.~(\ref{L2}), the second component gives us
\begin{eqnarray}
A_2^{\rm (1b)} & {\displaystyle \asym{\s \gg 1}} & \int_{-1}^{+1}
     d\lambda \int_1^\infty d\lambda_1 \int_0^\infty dz\, e^{-4\s z}
     \left(\frac{1+\lambda}{1+\lambda_1}\right)^{M+1}
     \frac{1}{(1+\lambda)^2}
     \frac{\lambda\lambda_2 - \lambda_1}{\Omega}
\nonumber \\
& {\displaystyle \asym{\s \gg 1}} & -\int_0^\infty dx\, \frac{x}{(1+x)^{M+1}}
     \int_0^\infty dz\, e^{8\s xz} (1-z)
     \int_0^1 dv\, \frac{1}{D_v(z)} \;.
\end{eqnarray}
The term involving $z$ is non-leading, being of order
\mbox{$O(\ln\s/\s^2)$},
while in the rest we can set
\linebreak
\mbox{$D_v(z) \sim D_v(0) =1$}
to obtain
\begin{equation}
A_2^{\rm (1b)} \asym{\s \gg 1}  -\frac{1}{8\s}\frac{1}{M} \;.
\end{equation}
The contribution from the third component $A_2^{\rm (1c)}$ is of higher order
\mbox{$O(\ln\s/\s^2)$},
and hence discarded.
On combining the three components, we obtain
\begin{equation}
A_2^{(1)} \asym{\s \gg 1}  -\frac{1}{16\s}\left[\frac{1}{M} +
     \frac{1}{1+M}\right] \;.
\label{A(1)2}
\end{equation}
Taking account of Eq.~(\ref{L1}), we have for $A^{(2)}_2$,
\begin{eqnarray}
A_2^{(2)} & {\displaystyle \asym{\s \gg 1}} &
     2\int_{-1}^{+1}d\lambda \int_1^\infty d\lambda_1\,
     \int_0^\infty dz\, e^{-4\s z}
     \left(\frac{1+\lambda}{1+\lambda_1}\right)^{M+1}
     \frac{1-\lambda}{1+\lambda}
     \frac{\lambda_1\lambda_2-\lambda}{\Omega^2}
\nonumber \\
& {\displaystyle \asym{\s \gg 1}} &
     \int_0^\infty dx\, \frac{1}{(1+x)^{M+1}}
     \int_0^\infty dz\, e^{-8\s xz} (1+z) \int_0^1 dv\,
     \frac{2v}{D^2_v(z)} \;.
\label{A2SV}
\end{eqnarray}
We express this as
\begin{equation}
A_2^{(2)} \asym{\s \gg 1} \int_0^\infty dx\, \frac{f_1(\s x)}{(1+x)^{M+1}}
     + \int_0^\infty dx\, \frac{f_2(\s x)}{(1+x)^{M+1}} \;,
\label{.}
\end{equation}
where
\begin{eqnarray}
f_1(x) & = & \int_0^\infty dz\, e^{-8xz} g_3(z)
     \quad {\displaystyle \asym{x\to\infty}} \quad \frac{1}{8x} \;,
\nonumber \\
f_2(x) & = & \int_0^\infty dz\, e^{-8xz} zg_3(z)
     \quad {\displaystyle \asym{x\to\infty}} \quad \frac{1}{(8x)^2} \;.
\end{eqnarray}
Then, Eq.~(\ref{R1}) applied to $f_2(x)$ and Eq.~(\ref{RG3}) applied to
$f_1(x)$, together with the integral identities (\ref{GINT3}), imply that
\begin{equation}
A^{(2)}_2 \asym{\s\gg 1} \frac{1}{8\s}\biggl[\frac{1}{2} + \ln 8\s +
     \ln\gamma -\sum_{j=1}^M\frac{1}{j}\biggr] \;.
\label{A(2)2}
\end{equation}
The full result for $A_2$ is obtained by adding Eqs.~(\ref{A(1)2}) and
(\ref{A(2)2}).

The sum
\mbox{$B=B_1+B_2$}
can be simplified to yield
\begin{eqnarray}
B & = &  4\s\int_{-1}^{+1}d\lambda \int_1^\infty d\lambda_1
     \int_1^\infty d\lambda_2\, e^{-2\s(\lambda^2-1)}
     e^{-2\s(1-\lambda^2)}
     \left(\frac{1+\lambda}{\lambda_1+\lambda_2}\right)^{M+2}
     \frac{\lambda_1^2-1}{(1+\lambda)^2}
     \frac{1-\lambda^2-\lambda_2^2}{\Omega}
\nonumber \\
& {\displaystyle \asym{\s \gg 1}} &
     -2\s\int_{-1}^{+1}d\lambda \int_1^\infty d\lambda_1
     \int_0^\infty dz\, e^{-4\s z} e^{-2\s(1-\lambda^2)}
     \left(\frac{1+\lambda}{1+\lambda_1}\right)^{M+1}
     \frac{\lambda_1-1}{(1+\lambda)^2}
     \frac{1}{\Omega}
\nonumber \\
& {\displaystyle \asym{\s \gg 1}} & -8\s\int_0^\infty dx\,
     \frac{x}{(1+x)^{M+1}} \int_0^1 dv\, e^{-8\s xv} (1-v)
     \int_0^\infty dz\, e^{-8\s xz}\frac{1}{D_v(z)}
\nonumber \\
& = & \sum_{\alpha=0}^2 \int_0^\infty dx\,
     \frac{f_\alpha(\s x)}{(1+x)^{M+1}} \;,
\end{eqnarray}
where we have used the fact that only values
\mbox{$w = vx \ll 1$}
are significant for large $\s$, and we have found it convenient to
introduce
\begin{eqnarray}
f_0(x) & = & -\int_0^\infty dv\, e^{-8xv} \quad \asym{x\to\infty} \quad
     -\frac{1}{8x} \;,
\nonumber \\
f_1(x) & = & -8x\int_0^1 dv\, e^{-8xv} \int_0^\infty dz\, e^{-8xz}
     \left[\frac{1}{D_v(z)}-1\right] \quad \asym{x\to\infty} \quad
     -\frac{2}{(8x)^2} \;,
\nonumber \\
f_2(x) & = & 8x\int_0^1 dv\, ve^{-8xv} \int_0^\infty dz\, e^{-8xz}
     \frac{1}{D_v(z)} \quad \asym{x\to\infty} \quad \frac{1}{(8x)^2} \;.
\end{eqnarray}
Application of Eq.~(\ref{R1}) to $f_1(x)$ and $f_2(x)$, followed by the
$x$-integration, results in
\begin{eqnarray}
\int_0^\infty dx\,\frac{f_1(\s x)}{(1+x)^{M+1}}
     & {\displaystyle \asym{\s\gg 1}} &
     -\frac{1}{8\s}\int_0^1 dv \int_0^\infty dz\, \frac{1}{(v+z)^2}
     \left[\frac{1}{D_v(z)}-1\right]
     \quad = \quad -\frac{1}{16\s}\int_0^\infty dz\, g_2(z) \;,
\nonumber \\
\int_0^\infty dx\,\frac{f_2(\s x)}{(1+x)^{M+1}}
     & {\displaystyle \asym{\s\gg 1}} &
     \frac{1}{8\s}\int_0^1 dv \int_0^\infty dz\,
     \frac{v}{(v+z)^2} \frac{1}{D_v(z)} \quad = \quad
     \frac{1}{16\s}\int_0^\infty dz\, g_2(z) \;.
\label{DB1}
\end{eqnarray}
These two contributions cancel out, and leave us with
\begin{equation}
B  \quad {\displaystyle \asym{\s \gg 1}} \quad
     \int_0^\infty dx\,\frac{f_0(\s x)}{(1+x)^{M+1}}
     \quad {\displaystyle \asym{\s\gg 1}} \quad
     -\frac{1}{8\s}\biggl[\ln 8\s + \ln \gamma - \sum_{j=1}^M \frac{1}{j}
     \biggr]
\end{equation}
according to Eq.~(\ref{RG3}), having noted that in this application
\mbox{$g(z) = \Theta(1-z)$},
with $\Theta$ denoting the Heaviside step function.
We note that $B$ cancels the non-analytic term in $A_2$.

We combine
\mbox{$C = C_1 + C_2$}
and note that
\mbox{$1-\lambda^2+\lambda_1^2-\lambda_2^2 = -\Omega
     + 2\lambda_1(\lambda_1-\lambda\lambda_2)$}.
Then we decompose
\mbox{$C=C^{(1)}+C^{(2)}$}
according to the power of $\Omega^{-1}$ that the contributions involve.
Thus, we have
\begin{eqnarray}
C^{(1)} & = & \int_{-1}^{+1}d\lambda \int_1^\infty d\lambda_1
     \int_1^\infty d\lambda_2\, e^{-2\s(\lambda_2^2-1)} e^{-2\s(1-\lambda^2)}
     \left(\frac{1+\lambda}{\lambda_1+\lambda_2}\right)^{M+2}
     \frac{1}{(1+\lambda)^2} \frac{\lambda_1\lambda_2-\lambda}{\Omega} \;,
\label{C012} \\
C^{(2)} & = & \int_{-1}^{+1}d\lambda \int_1^\infty d\lambda_1
     \int_1^\infty d\lambda_2\, e^{-2\s(\lambda_2^2-1)} e^{-2\s(1-\lambda^2)}
     \left(\frac{1+\lambda}{\lambda_1+\lambda_2}\right)^{M+2}
     \frac{\lambda_1}{(1+\lambda)^2}
     (\lambda_1\lambda_2-\lambda)
     \frac{\partial}{\partial \lambda_2}\left(\frac{1}{\Omega}\right) \;.
\nonumber \\
\end{eqnarray}
The term $C^{(1)}$ is clearly of higher order in $1/\s$.
To evaluate $C^{(2)}$, we first integrate by parts in order to avoid the
appearance of singular integrals, as in $A^{(2)}_1$. This yields
\begin{eqnarray}
C^{(2)} & {\displaystyle \asym{\s\gg 1}} &
     \int_{-1}^{+1}d\lambda \int_1^\infty d\lambda_1
     \int_0^\infty dz\, e^{-4\s z} e^{-2\s(1-\lambda^2)}
     \left(\frac{1+\lambda}{1+\lambda_1}\right)^{M+2}
     \frac{\lambda_1}{(1+\lambda)^2}
     (\lambda_1\lambda_2-\lambda)
     \frac{\partial}{\partial z}\left(\frac{1}{\Omega}\right)
\nonumber \\
& {\displaystyle \asym{\s\gg 1}} &
     \int_{-1}^{+1}d\lambda \int_1^\infty d\lambda_1
     \int_0^\infty dz\, e^{-4\s z} e^{-2\s(1-\lambda^2)}
     \left(\frac{1+\lambda}{1+\lambda_1}\right)^{M+2}
     \frac{\lambda_1}{(1+\lambda)^2}
\nonumber \\
& & \cdot \biggl\{4\s\left[\frac{\lambda_1-\lambda}{\Omega}
     - \frac{1}{\lambda_1-\lambda}\right] -
     (1-4\s z)\frac{\lambda_1}{\Omega}\biggr\}
\nonumber \\
& = & C^{\rm (2a)} + C^{\rm (2b)} \;.
\end{eqnarray}
Making use of Eq.~(\ref{X1}) gives us
\begin{eqnarray}
C^{\rm (2a)} & {\displaystyle \asym{\s\gg 1}} &
     4\s\int_0^\infty dx\, \frac{x(1+2x)}{(1+x)^{M+2}} \int_0^1 dv\,
     e^{-8\s xv} \int_0^\infty dz\, e^{-8\s xz}
     \left[\frac{1}{D_v(z)}-1\right]
\nonumber \\
& {\displaystyle \asym{\s\gg 1}} &
     \int_0^\infty dx\, \frac{f(\s x)}{(1+x)^{M+1}}
     -\half \int_0^\infty dx\, \frac{f(\s x)}{(1+x)^{M+2}} \;,
\end{eqnarray}
with
\begin{equation}
f(x) = 8x\int_0^1 dv\, e^{-8xv}\int_0^\infty dz\, e^{-8xz}
     \left[\frac{1}{D_v(z)}-1\right]
     \asym{x\to\infty} \frac{2}{(8x)^2} \;.
\end{equation}
Then, using Eq.~(\ref{R1}) and performing the $x$-integration, we
have
\begin{equation}
C^{\rm (2a)} \asym{\s\gg 1} \frac{1}{16\s}\int_0^1 dv \int_0^\infty dz\,
     \frac{1}{(v+z)^2} \left[\frac{1}{D_v(z)}-1\right]
     =  \frac{1}{32\s}\int_0^\infty dz\, g_2(z) \;.
\label{C1}
\end{equation}
For the component $C^{\rm (2b)}$, we can write
\begin{equation}
C^{\rm (2b)} \asym{\s\gg 1} \half\int_0^\infty dx\,
     \biggl[\frac{1}{(1+x)^{M+2}} + \frac{4x}{(1+x)^{M+1}}\biggr]
     \int_0^1 dv\, e^{-8\s xv} \int_0^\infty dz\, (1-8\s xz)e^{-8\s xz}
     \frac{1}{D_v(z)} \;.
\end{equation}
The second term in the square brackets is non-leading, leaving us with
\begin{equation}
C^{\rm (2b)} \asym{\s\gg 1} \int_0^\infty dx\,
     \frac{f(\s x)}{(1+x)^{M+2}} \;,
\end{equation}
where
\begin{equation}
f(x) = -\half\int_0^1 dv\, e^{-8xv} \int_0^\infty dz\, (1-8xz)e^{-8xz}
     \frac{1}{D_v(z)} \asym{x\to\infty} \frac{1}{(8x)^3} \;.
\end{equation}
Hence, after appealing to Eq.~(\ref{R1}) and carrying out the $x$-integration,
we obtain
\begin{equation}
C^{\rm (2b)} \asym{\s\gg 1} -\frac{1}{16\s}\int_0^1 dv \int_0^\infty dz\,
     \frac{v}{(v+z)^2} \frac{1}{D_v(z)}
     =  -\frac{1}{32\s}\int_0^\infty dz\, g_2(z) \;.
\label{C2}
\end{equation}
We see from Eqs.~(\ref{C1}) and (\ref{C2}) that $C^{(2)}$ vanishes up to
order $O(\s^{-1})$, and therefore $C$ provides no contribution to the final
result.
All the separate components can now be combined to give
\begin{equation}
\delta L + \sum_{j=1}^2 (A_j+B_j+C_j) \asym{\s\gg 1} -\frac{1}{8\s}
     \cdot \frac{M}{M^2-1} \;,
\end{equation}
which leads directly to Eqs.~(\ref{tail}) and (\ref{slarge}) in the main text.

\subsection*{Additional Remarks}
The double integrals which appear in Eqs.~(\ref{DB1}), (\ref{C1}) and
(\ref{C2}) can be easily performed by making the change of integration
variables
\mbox{$(v,z) \mapsto (\tilde{v},\tilde{z})$}
defined by
\begin{equation}
v = \frac{\tilde{v}\tilde{z}}{1+\tilde{v}\tilde{z}} \;, \quad
z = \frac{(1-\tilde{v})\tilde{z}}{1+\tilde{v}\tilde{z}} \;.
\end{equation}
Then we have
\begin{eqnarray}
\int_0^1 dv \int_0^\infty dz & = & \int_0^1 d\tilde{v}\int_0^\infty
     d\tilde{z}\,\frac{\tilde{z}}{(1+\tilde{v}\tilde{z})^3} \;,
\nonumber \\
D_v(z) & = & \frac{D_{\tilde{v}}(\tilde{z})}{(1+\tilde{v}\tilde{z})^2} \;,
\nonumber \\
\frac{1-v}{v+z} & = & \frac{1}{\tilde{z}} \;.
\end{eqnarray}
The results used in the text follow immediately.

\chapter{}
A numerical treatment of the integrals in Eqs.~(\ref{exact2},\ref{exact3})
cannot be implemented directly as their integrands exhibit an
algebraic singularity on the surface
\begin{equation}
\Omega \equiv
     \lambda^2+\lambda_1^2+\lambda_2^2-2\lambda\lambda_1\lambda_2-1=0 \;,
\end{equation}
which, of course, cancels out between numerator and denominator in the end,
once all relevant contributions are taken into account.
This apparent singularity can be eliminated altogether via the following
change of integration variables:
\begin{eqnarray}
\lambda & = & 1-2q \;,
\nonumber \\
\lambda_1 & = & 1+2qq_1 \;,
\nonumber \\
\lambda_2 & = & 1+2qq_2 \;.
\end{eqnarray}
Then
\mbox{$q, q_1, q_2$}
span the ranges
\mbox{$0 \le q_1,q_2 < \infty$}
and
\mbox{$0 \le q \le 1$}.

Let us now consider the case
\mbox{$\s=0$}
and write
\begin{equation}
\overline{|S_1|^2} = \int_0^1 dq\, I_M(q) \;.
\end{equation}
Then one can show, for example taking
\mbox{$M=1$},
that
\mbox{$I_1(q) \asym{q\to\infty} \mbox{const.}\times q^{-1/2}$},
so that there is an integrable singularity at
\mbox{$q=0$}.
The divergence of
\mbox{$I_M(q\to 0)$}
does not disappear for larger values of $M$.
It can be removed by the transformation
\mbox{$q = 1/(p+1)$}.
The resulting integral is regular in $p$ but displays a consequent
slow fall-off as
\mbox{$p\to\infty$},
going as $p^{-3/2}$ for
\mbox{$M=1$}.
So, to achieve an accuracy of $10^{-4}$, the numerical cut-off in the
$p$-integral would have to be taken as $10^8$.

Nonetheless, we thus arrive at the expressions
\begin{eqnarray}
\overline{|S_1|^2} & = &
     \int_0^\infty dp \int_0^\infty dq_1 \int_0^\infty dq_2\,
     \left(\frac{p}{1+p+q_1+q_2}\right)^{M-1}
     \frac{\exp[-8\s q_2(1+p+q_2)/(p+1)^2]}{(1+p+q_1+q_2)^2}
     \frac{1}{\Omega_p}
\nonumber \\
& & \left[1+q_1+q_2 + \frac{p}{p+1}\frac{2q_1q_2}{1+p+q_1+q_2}\right]
\biggl\{
\frac{p+1}{\Omega_p}\left[(1+{\cal E})p -
     (1-{\cal E})(q_1-q_2)(1+p+q_1+q_2)\right]
\nonumber \\
& & {}+ \frac{4\s}{(p+1)^2}\left[4{\cal E}p + (1-{\cal E})(1+p+2q_2)^2\right]
\biggr\}
\end{eqnarray}
and
\begin{eqnarray}
\overline{|S_2|^2} & = &
     \int_0^\infty dp \int_0^\infty dq_1 \int_0^\infty dq_2\,
     \left(\frac{p}{1+p+q_1+q_2}\right)^{M-1}
     \frac{\exp[-8\s q_2(1+p+q_2)/(p+1)^2]}{(1+p+q_1+q_2)^2}
     \frac{1}{\Omega_p}
\nonumber \\
& & \cdot \biggl\{
-\frac{1+{\cal E}-{\cal F}}{p+1}\biggl[ (q_1-q_2)\left[ 3 +
     \frac{2p}{\Omega_p}\left((p+1)(1+q_1+q_2) +2q_1q_2\right)\right]
\nonumber \\
& & {}+ \frac{4\s}{(p+1)^2}\left[ 2p\left((p+1)(1-q_1+q_2) +2q_2\right)
     - (1+p+q_1+q_2)(1+p+q_2)^2\right]\biggr]
\nonumber \\
& & {}+ \frac{(1-{\cal E})p}{p+1}\biggl[\left[
     1+\frac{2}{\Omega_p}\left((p+1)(1+q_1+q_2) + 2q_1q_2\right)\right]
     (1+p+q_1+q_2)
\nonumber \\
& & {}+ \frac{4\s}{(p+1)^2}\left[2(p+1)(1-q_1+q_2) + 4q_1 +
     (q_1-q_2)(1+p+2q_2)^2\right]\biggr]
\biggr\} \;.
\end{eqnarray}
We have set
\mbox{$\Omega_p \equiv (p+1)\Omega$}
so that
\begin{equation}
\Omega_p = (p+1)(1+q_1+q_2)^2 - 4pq_1q_2 \;,
\end{equation}
and here
\mbox{${\cal E} = \exp[-8\sp/(p+1)^2]$}.
In this form, the integration is completely free of singularities;
but, for small $\s$, the integrand decays slowly (with a power law)
in all directions.
Consequently, all three upper integration limits must be taken very large,
and this makes accurate computation slow and difficult.

To circumvent this, we introduce two compact coordinates by transforming to
two angle variables $\alpha$, $w$ and one
radial parameter $r$ according to the definitions
\begin{eqnarray}
q_1 & = & \half r(1-\alpha)(1+w) \;,
\nonumber \\
q_2 & = & \half r(1-\alpha)(1-w) \;,
\nonumber \\
p & = & r\alpha \;.
\label{xay}
\end{eqnarray}
The ranges of the new variables are given by
\mbox{$0 \le r \le \infty$},
\mbox{$0 \le \alpha \le 1$},
\mbox{$-1 \le w \le 1$}.
However, because of the very asymmetric form that the exponential now
acquires, there exist both rapidly and very slowly decaying directions for the
integrand.
Indeed, the upper limit of the $r$-integral must still be taken extremely
large in order to achieve sufficient accuracy for the slowly decaying parts.
However, this means that the integration region also extends far along the
rapidly decaying directions where the integrand is virtually zero;
and, in fact, the integrand turns out to have support
only in a tiny fraction of the entire integration volume.
The integration routine is then unable to properly determine when the desired
accuracy has been achieved. This is because, at some stage in the iteration,
further subdivision of the grid is likely to create new points essentially
only in the pervading insignificant region, and hence not alter the previous
value of the integral to within the requisite accuracy.

The easisest way to remedy this
problem is by compactifying the $x$-integration through the
variable change
\begin{equation}
y = \frac{r}{1+r} \;,
\label{y}
\end{equation}
so that
\mbox{$0 \le y \le 1$}.
The transformations of Eq.~(\ref{xay},\ref{y})
result in a simple bounded integration over a
hyper-rectangular region and an
integrand that has a significant value over an appreciable fraction of this
region.
The change of integration measure and region that corresponds to the variable
transformation
\mbox{$(p,q_1,q_2) \to (y,\alpha,w)$}
can be summarized by the equation
\begin{eqnarray}
& & \int_0^\infty dp \int_0^\infty dq_1 \int_0^\infty dq_2\,
     \left(\frac{p}{1+p+q_1+q_2}\right)^{M-1}
     \frac{1}{(1+p+q_1+q_2)^2} \; [\ldots]
\nonumber \\
& & =
     \frac{1}{2}\int_0^1 dy\, \frac{y^{M+1}}{(1-y)^2}
     \int_0^1 d\alpha\, (1-\alpha)\alpha^{M-1}
     \int_{-1}^{+1} dw\; [\ldots] \;.
\end{eqnarray}

Then we can write for
\mbox{$\overline{|S_j|^2}$}
\mbox{$(j=1,2)$},
\begin{eqnarray}
\overline{|S_j|^2} & = & \frac{1}{2}\int_0^1 dy\, y^{M+1}
     \int_0^1 d\alpha\, (1-\alpha)\alpha^{M-1}
     \int_{-1}^{+1} dw\, \frac{1-y}{\Omega_y}
\nonumber \\
& & \cdot
\exp\left\{-2\s\frac{y(1-\alpha)(1-w)}{1-y(1-\alpha)}\left[ 2 +
     \frac{y(1-\alpha)(1-w)}{1-y(1-\alpha)}\right]\right\}
F_j(y,\alpha,w) \;,
\end{eqnarray}
where
\begin{eqnarray}
F_1(y,\alpha,w) & = & \frac{1}{1-y} \left[1-y\alpha -
     \frac{1}{2}\frac{y^3\alpha(1-\alpha)^2(1-w^2)}{1-y(1-\alpha)}\right]
\nonumber \\
& &
\cdot \biggl\{
\frac{y\left(1-y(1-\alpha)\right)}{\Omega_y}\left[
     (1+{\cal E})(1-y)\alpha -(1-{\cal E})(1-\alpha)w\right]
\nonumber \\
& &  {}+ 4\s\biggl[ 4{\cal E}\frac{y(1-y)\alpha}{[1-y(1-\alpha)]^2}
     + (1-{\cal E})\left(1+\frac{y(1-\alpha)(1-w)}{1-y(1-\alpha)}
     \right)^2\biggr]
\biggr\}
\end{eqnarray}
and
\begin{eqnarray}
F_2(y,\alpha,w) & = &
-(1+{\cal E}-{\cal F})\biggl\{
\frac{y(1-\alpha)w}{1-y(1-\alpha)}\left[3+\frac{2y\alpha}{\Omega_y}
     \left( (1-y) + y^2\alpha(1-\alpha) + \half y^2(1-\alpha)^2(1-w^2)
     \right)\right]
\nonumber \\
& & {}+
\frac{4\s}{[1-y(1-\alpha)]^3}\left[2y\alpha
     \left( 1-y - y^2\alpha(1-\alpha)w\right) -
     \left( 1 - y(1-\alpha)w\right)^2 \right]
\biggr\}
\nonumber \\
& & {}+ (1-{\cal E})\frac{y\alpha}{1-y(1-\alpha)}\biggl\{
\frac{1}{1-y} + \frac{1}{\Omega_y}\left[ 2(1-y)
     + 2y^2\alpha(1-\alpha) + y^2(1-\alpha)^2(1-w)^2\right]
\nonumber \\
& & {}+
\frac{4\s}{1-y}\frac{1}{[1-y(1-\alpha)]^2}\biggl[
     2(1-y)\left(1-y - y^2\alpha(1-\alpha)w\right)
\nonumber \\
& &  {}+ y(1-\alpha)w\left(1-y(1-\alpha)w\right)^2\biggr]
\biggr\} \;,
\end{eqnarray}
with
\begin{eqnarray}
\Omega_y & = & y^2(1-y)(1-\alpha)^2
     + (1-y)[1-y(1-\alpha)][1-y + 2y(1-\alpha)]
     + y^3\alpha(1-\alpha)w^2 \;,
\nonumber \\
{\cal E} & = &  \exp\left\{-8\s\frac{y(1-y)\alpha}
     {[1-y(1-\alpha)]^2}\right\} \;,
\nonumber \\
{\cal F} & = & \frac{1 - e^{-8\s x}}{4\s x} \;,\quad
     x = \frac{y(1-y)\alpha}{[1-y(1-\alpha)]^2} \;.
\end{eqnarray}
In the numerical analysis, we represent ${\cal F}$ as the integral
\begin{equation}
{\cal F}(4\s x) = 2\int_0^1 du\, e^{-8\s xu}
\end{equation}
when the argument is small.
It is also useful to note the relation
\begin{equation}
\frac{1}{\Omega_p} = \frac{(1-y)^3}{\Omega_y} \;.
\end{equation}
In the parameterization above, a mild algebraic singularity appears along the
line
\mbox{$y=1,w=0$}.
This can be easily dealt with by setting the upper limit of the $y$-integration
to $1-\delta$ with $\delta$ an arbitrarily small positive number.

\newpage
\section*{Figure captions}
{\bf Figure 1:}
Representation of a typical lead--stadium configuration. The size of the
stadium $L$ is large compared with the elastic mean free path $\ell$,
ensuring that the electron motion is ballistic. A total of $M$ external
channels couple to the stadium through the two leads.

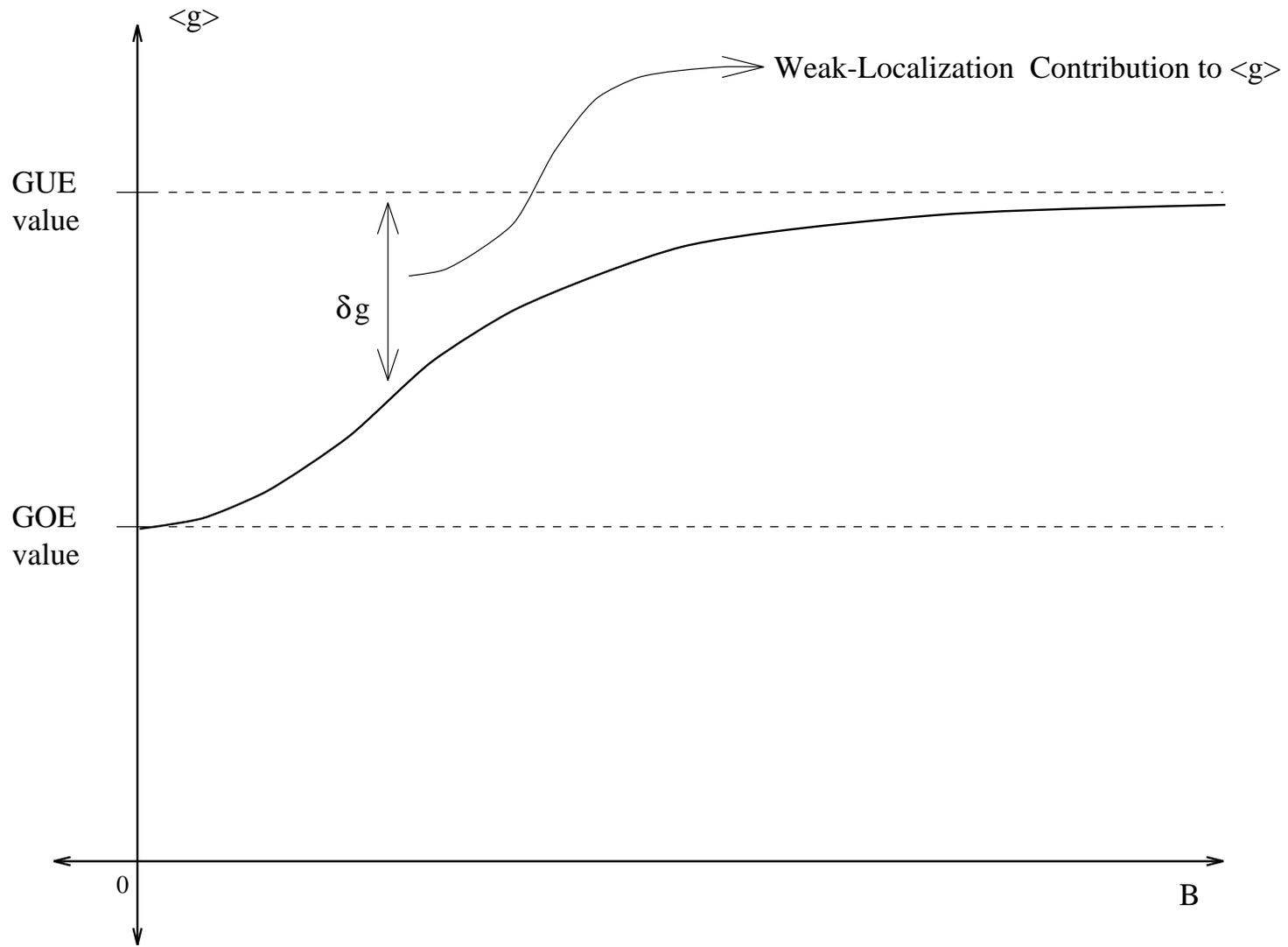
\noindent {\bf Figure 2:}
Schematic depiction of the zero-field dip in the average magnetoconductance
associated with weak localization. The curve interpolates between values
associated with pure GOE and GUE for low and high magnetic fields $B$,
respectively.

\noindent {\bf Figure 3a:}
Graph of the $\Delta_3$ statistic versus energy interval $L$ for flux angle
\mbox{$\phi/\phi_0 = 0$}.
The solid line corresponds to setting
\mbox{$t=0$}
in Eq.~(\ref{Y2PM}) (i.e.\ the GOE result). The dashed line is the GUE
result.

\noindent {\bf Figure 3b:}
Graph of the $\Delta_3$ statistic versus energy interval $L$ for flux angle
\mbox{$\phi/\phi_0 = 0.1$}.
The solid line shows the best fit to the data points furnished by
Eq.~(\ref{Y2PM}), obtained by adjusting
\mbox{$t=0.15$}
The upper and lower dashed lines correspond to the GOE and GUE results,
respectively.

\noindent {\bf Figure 3c:}
Graph of the $\Delta_3$ statistic versus energy interval $L$ for flux angle
\mbox{$\phi/\phi_0 = 0.2$}.
The solid line shows the best fit to the data points furnished by
Eq.~(\ref{Y2PM}), obtained by adjusting
\mbox{$t=0.6$}
The upper and lower dashed lines correspond to the GOE and GUE results,
respectively.

\noindent {\bf Figure 4:}
The correlator $\kappa(t)$ (defined in Eq.~(\ref{kappa})) as a function of the
time $t$ computed as a numerical average over phase space, taking $10^7$
trajectories. The
\mbox{$t\to\infty$}
asymptote is
\mbox{$\kappa \simeq 0.06$}.

\noindent {\bf Figure 5:}
Graph of the reciprocal of the weak-localization term $\delta g^{-1}$ versus
the magnetic-field parameter $\s$ for
\mbox{$M = 1,2,4,6,8,10$}
on a large scale
(\mbox{$0 \le \s \le 20$}).

\noindent {\bf Figure 6:}
Graph of the reciprocal of the weak-localization term $\delta g^{-1}$ versus
the magnetic-field parameter $\s$ for
\mbox{$M = 1,2,4,6,8,10$}
on a smaller scale
(\mbox{$0 \le \s \le 4$}).

\noindent {\bf Figure 7:}
Plot of the weak-localization term $\delta g$ as a function of $\sqrt{t}$
for the case
\mbox{$M=1$},
compared with the corresponding Lorentzians obtained from fitting to the
large-$\sqrt{t}$ tail of the
\mbox{$M=1$}
curve, and to its FWHM.
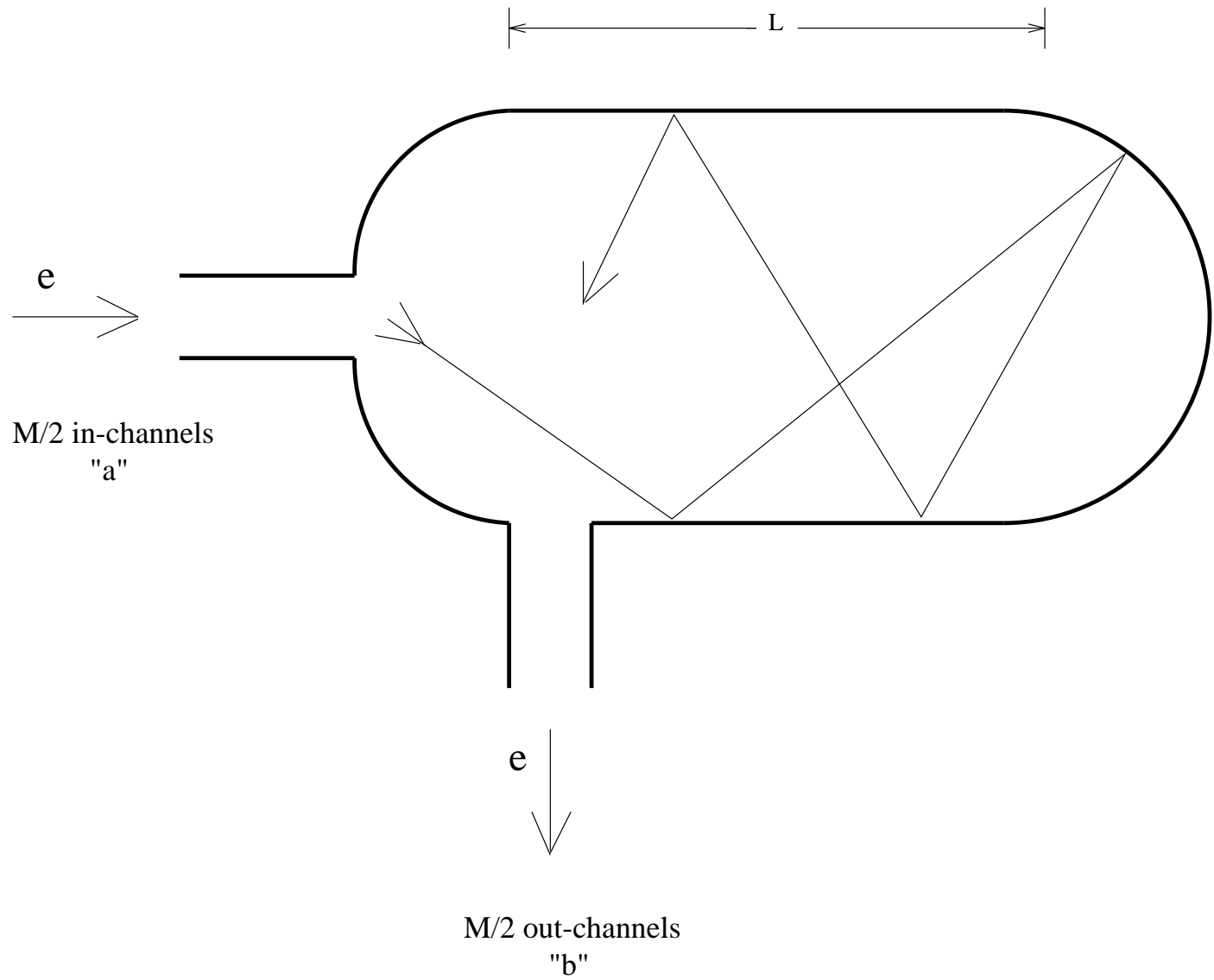
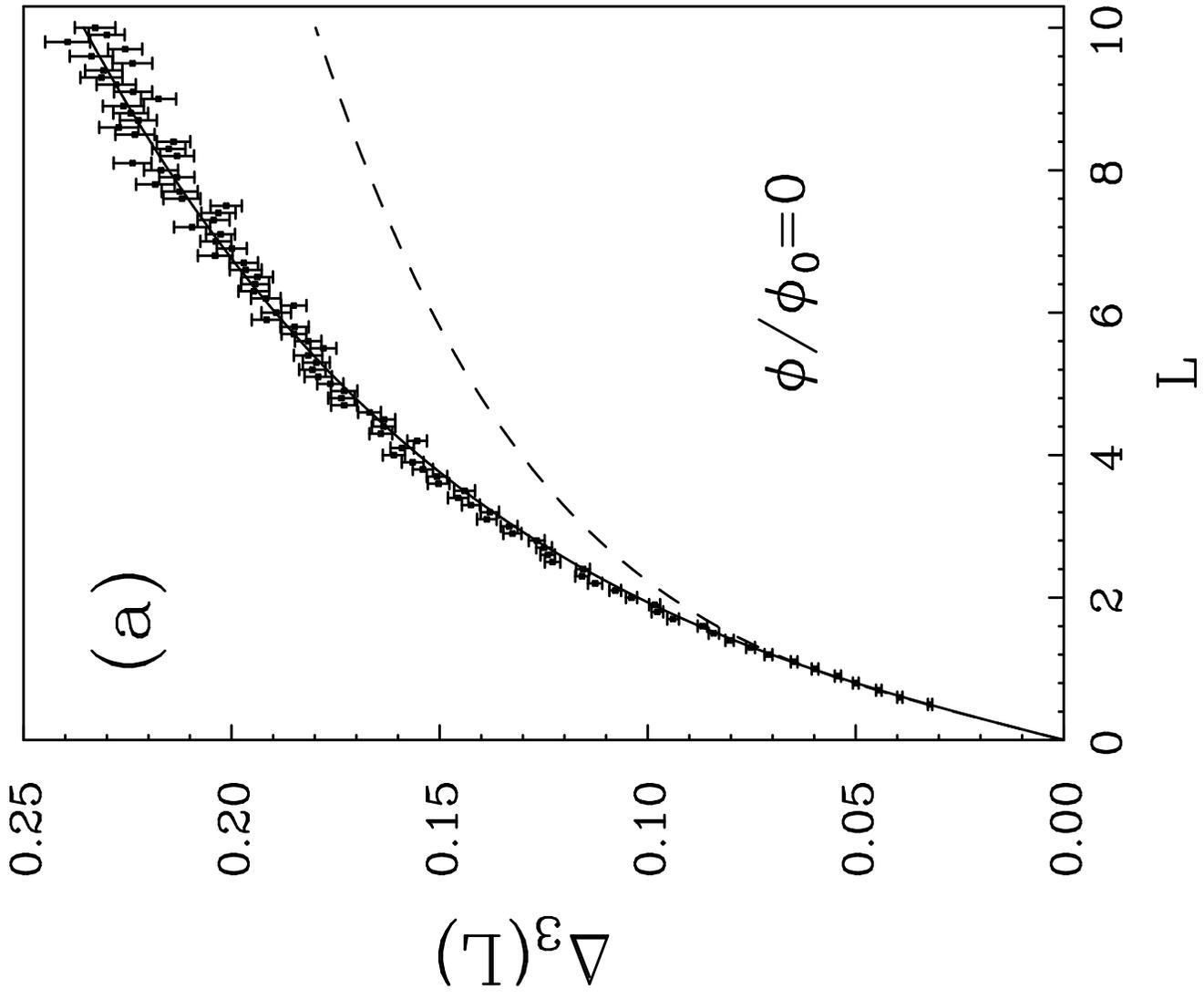
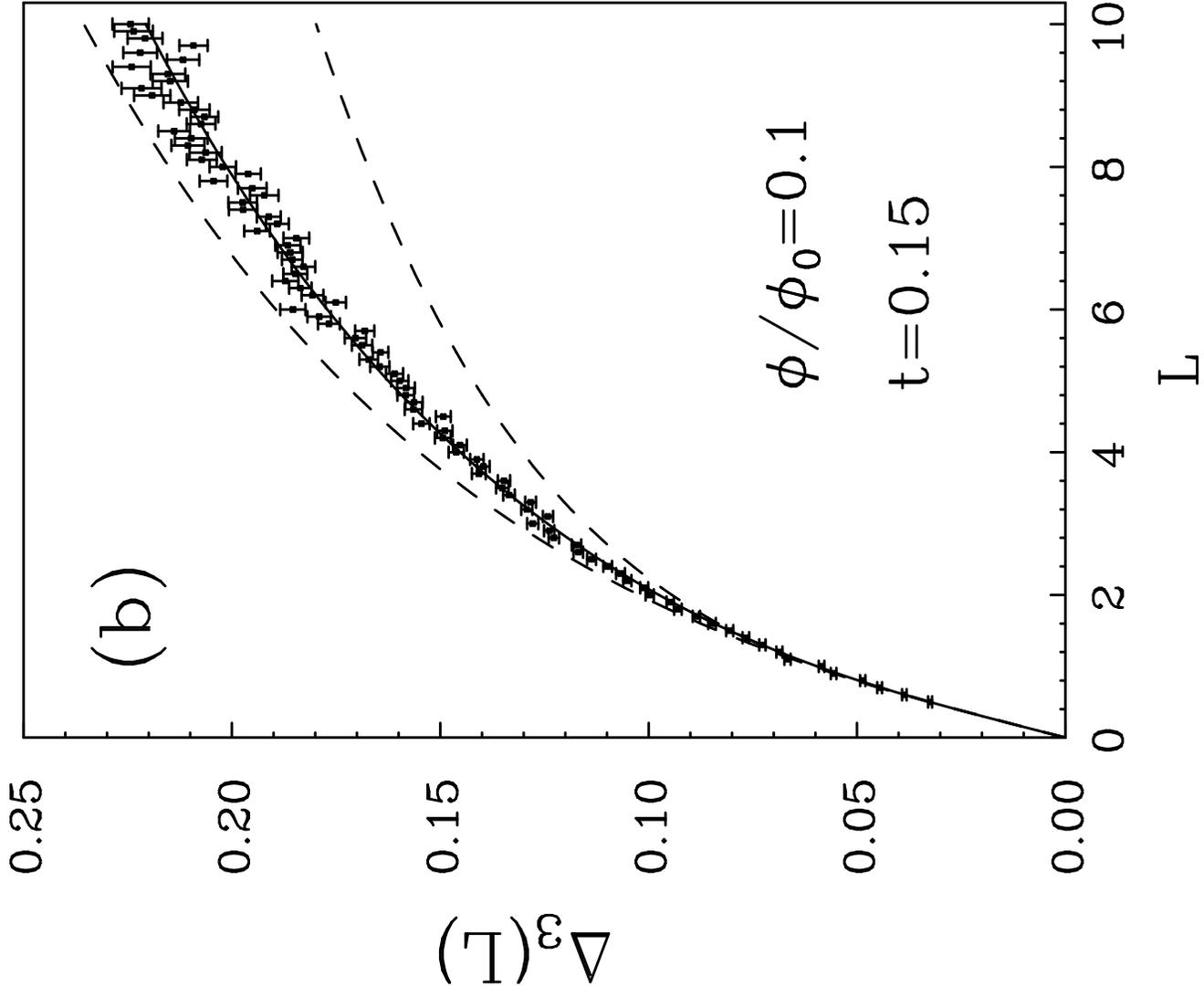
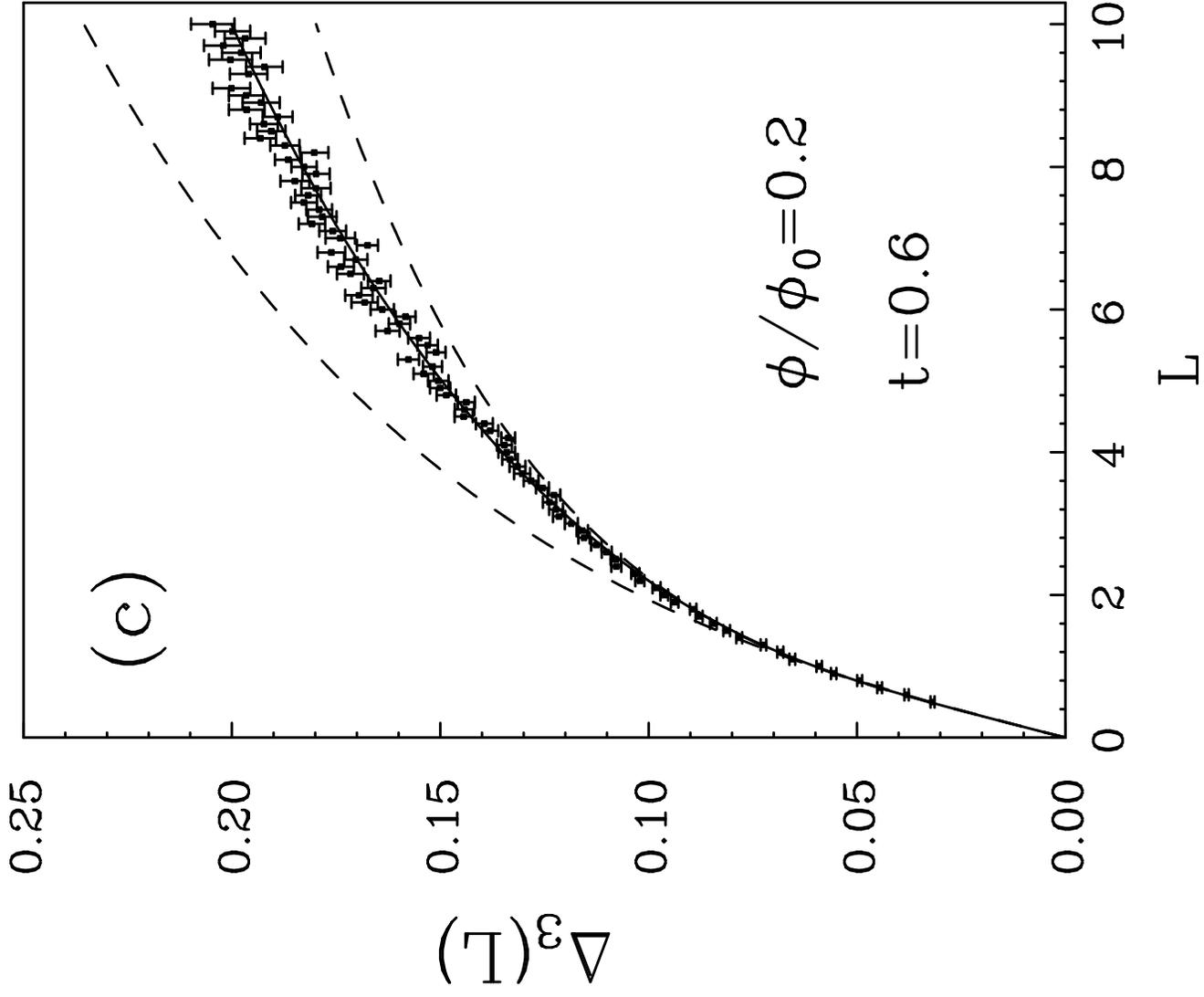
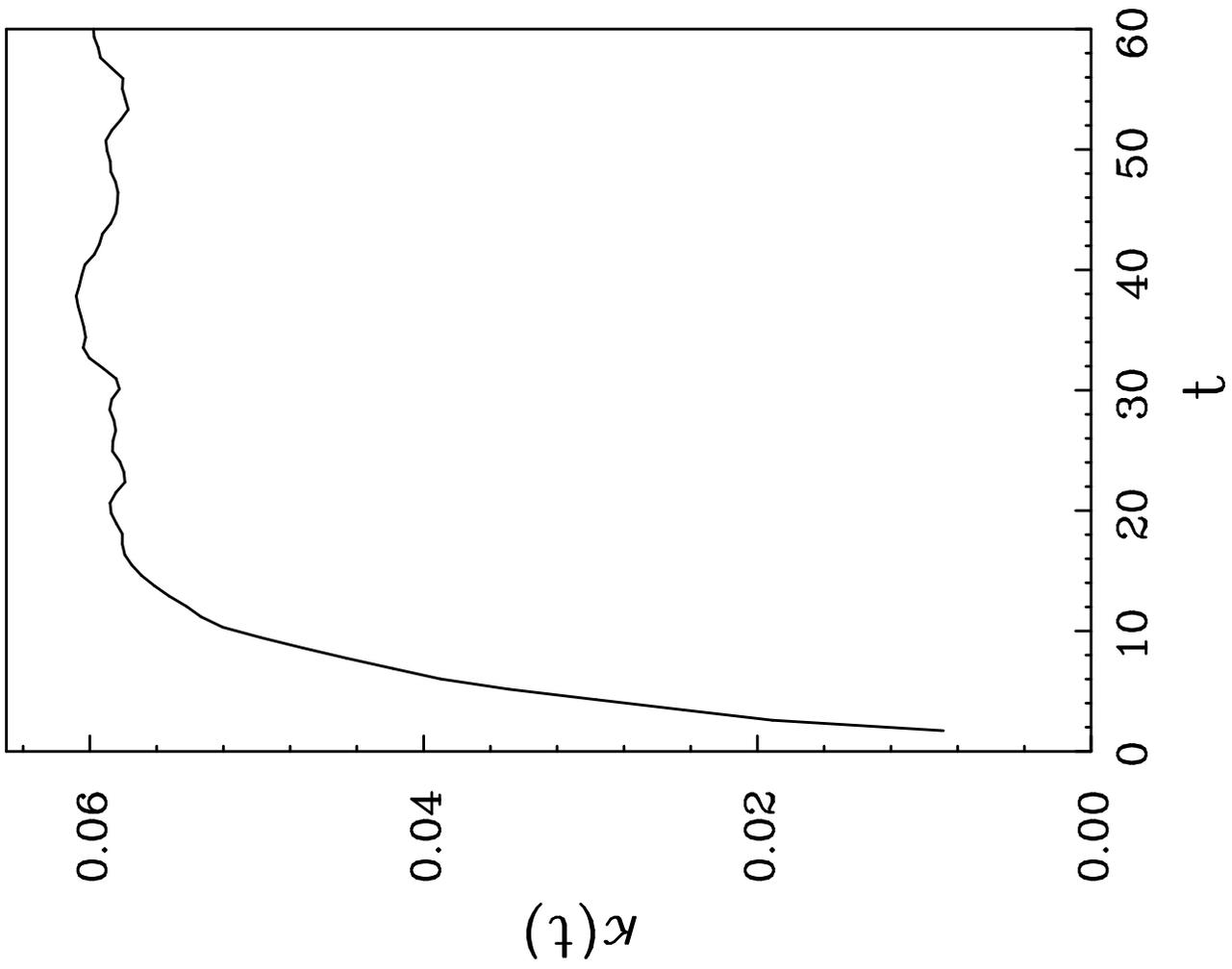
\end{document}